\documentclass[superscriptaddress,twocolumn,10pt,pra,aps,showpacs,longbibliography]{revtex4-1}
\usepackage[latin9]{inputenc}
\usepackage{mathtools}
\usepackage{amsbsy}
\usepackage{amstext}
\usepackage{amsthm}
\usepackage{float}
\usepackage{latexsym}
\usepackage[dvipsnames]{xcolor}
\usepackage[unicode=true,pdfusetitle,
 bookmarks=false,
 breaklinks=false,pdfborder={0 0 1},backref=false,colorlinks,urlcolor=Black, linkcolor=Red,citecolor=CornflowerBlue]
 {hyperref}
\hypersetup{
 bookmarksnumbered=false,bookmarksopen=false}

\makeatletter

\providecommand{\bra}[1]{\langle #1 \rvert}
\providecommand{\ket}[1]{\lvert #1 \rangle}

\def\aop{\hat{a}}
\def\xop{\hat{x}}
\def\pop{\hat{p}}
\def\Xop{\hat{X}}
\def\adag{\hat{a}^{\dagger}}

\def\sigz{\hat{\sigma}_z}
\def\sigx{\hat{\sigma}_x}
\def\sigy{\hat{\sigma}_y}

\def \Uop{\hat{U}_x(0\rightarrow t)}

\def \Oop{\hat{O}}
\def \Of{\omega}
\def \Oq{\Omega}
\def\rop{\hat{\rho}}

\newcommand{\Hop}{\hat{H}}
\newcommand{\Aop}{\hat{A}}

\newcommand{\Nop}{\hat{N}}
\newcommand{\Sop}{\hat{S}}
\newcommand{\Pop}{\hat{P}}

\newcommand{\Jop}{\hat{J}}

\newcommand{\resub}[1]{\textcolor{Blue}{#1}}

\newtheorem{thm}{Heuristic}

\usepackage{amsmath}
\usepackage{graphicx}
\usepackage{amssymb}
\usepackage{txfonts,color}
\makeatother

\begin{document}
\title{Critical Quantum Metrology with Fully-Connected Models:\\ 
From Heisenberg to Kibble-Zurek Scaling}
\author{Louis Garbe}
\affiliation{Vienna Center for Quantum Science and Technology, Atominstitut, TU Wien, 1040 Vienna, Austria}
\author{Obinna Abah}
\affiliation{School of Mathematics, Statistics and Physics, Newcastle University, Newcastle upon Tyne NE1 7RU, United Kingdom}
\affiliation{Centre for Theoretical Atomic, Molecular and Optical Physics, Queen's University Belfast, Belfast BT7 1NN, United Kingdom}
\author{Simone Felicetti}
\affiliation{Istituto di Fotonica e Nanotecnologie, Consiglio Nazionale delle Ricerche (IFN-CNR), \\ Via Cineto Romano 42, 00156 Rome, Italy}
\author{Ricardo Puebla}
\affiliation{Instituto de F{\'i}sica Fundamental, IFF-CSIC, Calle Serrano 113b, 28006 Madrid, Spain}
\affiliation{Centre for Theoretical Atomic, Molecular and Optical Physics, Queen's University Belfast, Belfast BT7 1NN, United Kingdom}
\begin{abstract}
Phase transitions represent a compelling tool for classical and quantum sensing applications. It has been demonstrated that quantum sensors can in principle saturate the Heisenberg scaling, the ultimate precision bound allowed by quantum mechanics, in the limit of large probe number and long measurement time. Due to the critical slowing down, the protocol duration time is of utmost relevance in critical quantum metrology. However, how the long-time limit is reached remains in general an open question. So far, only two dichotomic approaches have been considered, based on either static or dynamical properties of critical quantum systems.
Here, we provide a comprehensive analysis of the scaling of the quantum Fisher information for different families of protocols that create a continuous connection between static and dynamical approaches. In particular, we consider fully-connected models, a broad class of quantum critical systems of high experimental relevance. Our analysis unveils the existence of universal precision-scaling regimes. These regimes remain valid even for finite-time protocols and finite-size systems. We also frame these results in a general theoretical perspective, by deriving a precision bound for arbitrary time-dependent quadratic Hamiltonians.
\end{abstract}

\maketitle

\section{Introduction}

Critical systems, i.e. those undergoing a phase transition, represent a valuable resource for metrology and sensing applications. Indeed, in proximity of the critical point of a phase transition, a small variation of physical parameters can lead to dramatic changes in equilibrium and dynamical properties~\cite{Huang,Tauber}. In turn, when one or more system parameters depend on an external field, this diverging susceptibility can be exploited to obtain a very precise estimation of the field intensity. Such criticality-based sensing has already found  applications in current technological devices, such as transition-edge detectors~\cite{irwin_transition-edge_2005} and bolometers~\cite{bolometer_review}. These kind of sensors are based on a classical working principle, that is, they do not follow optimal sensing strategies from the quantum mechanical point of view, even when quantum models are required to describe their physical behavior. In this context, the aim of critical quantum metrology is to exploit quantum fluctuations in proximity of a quantum phase transition (QPT)~\cite{Sachdev} to achieve quantum advantage in sensing protocols.
In the last few years, a series of theoretical studies~~\cite{zanardi_quantum_2008, invernizzi2008Optimal,ivanov_adiabatic_2013, bina_dicke_2016, fernandez-lorenzo_quantum_2017, rams_at_2018, frerot_quantum_2018, heugel2020_quantum, Mirkhalaf2020, Wald2020, Ivanov_2020steady, Salado2021, Niezgoda2021, mishra2021integrable,tsang_quantum_2013,macieszczak_dynamical_2016} have shown that quantum critical sensors can achieve the so-called Heisenberg limit, where the squared signal-to-noise ratio scales like $N^2 T^2$, where $N$ is the number of probe systems, and $T$ is the protocol duration.

Only very recently, it has been shown~\cite{garbe2020} that the Heisenberg scaling can also be achieved using finite-component QPTs~\cite{bakemeier2012quantum,Ashhab2013, Hwang:15,Puebla:16,Puebla:17,Hwang:18,Zhu:20,Puebla:20b}. In these systems, we have only a finite numbers of components interacting; the usual thermodynamic limit is then replaced by a scaling of the system parameters~\cite{Casteels2017,Bartolo2016Exact,Minganti2018,peng_unified2019,felicetti2020universal}. A variety of protocols based on finite-component QPTs have been proposed considering light-matter interaction models~\cite{Ivanov2020,Chu2021,gietka2021,hu2021,liu2021,ilias2021criticality} and quantum nonlinear resonators~\cite{di2021critical}. A critical quantum sensor can then be realized using small-scale atomic or solid-state devices, circumventing the complexity of implementing and controlling many-body quantum systems.
Finite-component critical systems belong to fully-connected models, whose low-energy physics can effectively be  described in terms of non-linear quantum oscillators in the thermodynamic or parameter-scaling limit~\cite{Lambert:04,Lambert:05,Ribeiro:07,Ribeiro:08,Hwang:15,Puebla:16}. The class of fully-connected models is of high interest for two main reasons: 1) It provides a very convenient theoretical testbed to derive fundamental results with both analytical and numerical techniques; 2) It includes models of immediate experimental relevance for different quantum platforms, such as the quantum Rabi (QR), the Dicke and the Lipkin-Meshkov-Glick (LMG) models.

Critical quantum metrology protocols can be categorized in two main approaches, which we will label as \textit{static} and \textit{dynamical}. A) The static approach~\cite{zanardi_quantum_2008, invernizzi2008Optimal,ivanov_adiabatic_2013, bina_dicke_2016, fernandez-lorenzo_quantum_2017, rams_at_2018, heugel2020_quantum, Mirkhalaf2020, Wald2020, Ivanov_2020steady, Salado2021, Niezgoda2021, mishra2021integrable,garbe2020} exploits the susceptibility of equilibrium properties of critical systems. In a Hamiltonian settings, the static approach consists  an adiabatic sweep that brings the system in close proximity of the critical point, to then measure an observable on the system \textit{ground state}. Similarly, in a driven-dissipative setting, the static approach consists in exploiting the critical properties of the system \textit{steady-state}. The static approach is simpler to realize in practical implementations but it is limited by the critical slowing down: As the critical point is approached, the estimation precision diverges, but also the time required to prepare such equilibrium state. B) The dynamical approach~\cite{tsang_quantum_2013,macieszczak_dynamical_2016,Chu2021} typically refers to a sudden quench that brings the system close to the critical point, i.e., to the QPT, to then monitor the \textit{dynamical evolution} of the system, which may also have a critical dependence on the system parameters.  

In general, however, we can interpolate between these two approaches, considering protocols that bring the system close to the QPT in a continuous and time-dependent fashion. This naturally establishes a bridge between two distinct research fields, namely, quantum metrology and the study of non-equilibrium critical dynamics triggered by a QPT~\cite{Polkovnikov:11,Eisert:15}.  One example is the emergence of universal scaling laws as predicted by the Kibble-Zurek  mechanism~\cite{Zurek:96,Zurek:05,Dziarmaga:05,Damski:05,delCampo:14,rams_at_2018}. However, these intermediate protocols have hitherto rarely been considered from a metrological perspective.

Understanding the scaling of the estimation precision with respect to the protocol duration time is of utmost relevance in critical quantum metrology.
Recent results suggest that, for a large class of spin systems, the dynamical and equilibrium approaches have a similar scaling of the estimation precision in the thermodynamic limit~\cite{rams_at_2018}. 
For fully-connected models, either under thermodynamic~\cite{Lambert:04,Lambert:05,Ribeiro:07,Ribeiro:08} or parameter-scaling limit~\cite{Hwang:15,peng_unified2019,felicetti2020universal} it was shown that dynamical protocols have a constant factor advantage over static protocols due to the critical slowing down~\cite{Chu2021}. It has also been shown that a direct application of shortcuts-to-adiabaticity~\cite{Torrontegui:13} can not improve the scaling of the estimation precision of critical quantum sensors~\cite{gietka2021}. However, a unifying treatment is still missing.
Indeed, so far the scaling of the estimation precision has only been analyzed considering either sudden quenches or strictly adiabatic evolutions, and focused on specific models and observables in the thermodynamic limit.

In this article, we present a theoretical analysis of different families of finite-time metrological protocols that allows us to establish a connection between static and dynamic approaches to critical quantum metrology. 
In particular, we provide a comprehensive analysis of the metrological power of fully-connected models displaying a QPT. The critical properties of these Hamiltonians can be described by a single unifying model, made of a non-linear oscillator. We evaluate the quantum Fisher information (QFI) achievable with protocols based on sudden quenches, adiabatic sweeps and finite-time ramps towards the critical point. 
We also derive a precision bound for protocols involving Gaussian states under time-dependent Hamiltonians. This bound accurately reproduces our findings in most parameter regimes and thus put them in a more general perspective.  Our analysis unveils the existence of different time-scaling regimes for the QFI, such as the emergence of a Kibble-Zurek scaling law in the QFI under finite-time ramps. We show that these scalings are not limited to a certain model or a certain regime of parameters, but describe the vast majority of critical estimation protocols with fully-connected models. Importantly, these results are valid both in and outside of the thermodynamic limit.

In the following subsection we provide a summary of the results, while the rest of this article is organized as follows. In Sec.~\ref{sec:QPT} we show how the critical properties of fully-connected models can be captured by a non-linear oscillator. We discuss this mapping in details with two examples, namely, the LMG model~\cite{Lipkin:65,Ribeiro:07,Ribeiro:08} and the QR model~\cite{Hwang:15,Puebla:16}.
In Sec.~\ref{sec:QCM}, we introduce our metrological protocol, briefly recalling the definition of the QFI, and several important bounds to it which can be found in the literature. In Sec.~\ref{sec:bound}, we introduce our bound for metrology with time-dependent Hamiltonian and Gaussian states. In Sec.~\ref{sec:Quench} and \ref{sec:Ramp}, we discuss the metrological properties of three different protocols in the vicinity of the QPT. We show how these protocols allow to draw a connection between the static and dynamic approach. Finally, in Sec.~\ref{sec:Conclusion}, we present the main conclusions of the article and an outlook.

\subsection{Summary of results}
\label{sec:Summary}
We start by introducing and putting into context fully-connected models and quantum metrological protocols. In fully-connected models such as the QR and LMG models, we can define an effective "system size", $\eta$, which controls the non-linearity of the system; in the case of the Rabi model, it is given by the frequency ratio of the qubit and the field, while in the LMG or Dicke model, it corresponds to the number of qubits. In the so-called thermodynamic or scaling limit, $\eta\rightarrow\infty$, all of these models can be effectively described as $\Hop=\Hop_0+\Hop_1$, with ($\hbar=1$)
\begin{align}
 	\Hop_0&=\Of\left[\frac{\pop^2}{2}+(1-g^2)\frac{\xop^2}{2}\right], \label{eq:H0} \\
 	\Hop_1&=\Of \frac{f(g)}{\eta}\xop^4,
 	\label{potquartic}
 \end{align}
up to first order in $1/\eta$, where $\xop$ and $\pop$ are the quadratures of a bosonic field, $\xop=\frac{\aop+\adag}{\sqrt{2}}$ and $\pop=\frac{i(\adag-\aop)}{\sqrt{2}}$. Here $g$ is an effective and dimensionless coupling strength, $\Of$ is the typical frequency scale of the system, while $f(g)$ is a function of the dimensionless coupling that depends on the considered model (but is typically of order $1$). In the thermodynamic limit $\eta\rightarrow\infty$, this model undergoes a QPT at $g_c=1$ (cf. Sec.~\ref{sec:QPT}). In this study, we will always remain in the normal phase, for $g<g_c$. This stands in contrast with other approaches where one actually crosses  the critical point, typically to exploit symmetry-breaking effects \cite{ivanov_adiabatic_2013,fernandez-lorenzo_quantum_2017,heugel2020_quantum,Salado2021}.\\

The working principle of the considered families of protocols is as follows. We assume that all parameters are known, except the one to be estimated, such as for example $g$ or $\omega$. Notice that this framework is relevant in the design of practical sensors~\cite{di2021critical}. First, the system is initialized in its ground state, far from the critical point. Then the value of a controllable parameter is changed in order to push the system in proximity of the phase transition. Finally, we perform a measurement on the final state, profiting from the high critical susceptibility  to gain information about the parameter we want to estimate.

For the sake of clarity, let us compare the working principle of the protocols here considered with the standard interferometric approach to quantum metrology. In the latter, a given probe system is initialized, then the phase to be estimated is imparted on the probe, and finally a measurement is performed. On the contrary, in the protocols here considered the information about the parameter to be estimated is encoded in the probe system during the sweep or quench of the controllable parameter (which is varied to bring the system close to the critical point). In this sense, the quantum criticality  is not used only as a mean to generate a quantum state that is subsequently used in a parameter estimation protocol: the critical nature of the system is exploited to efficiently encode information about the parameter to be estimated onto the probe system itself.

The precision achievable by a given estimation protocol is upper-bounded by the QFI $I_x=4[\langle \partial_x \psi|\partial_x \psi\rangle -|\langle \partial_x\psi|\psi\rangle|^2]$, where $\ket{\psi}$ is the system state at the end of the protocol, which depends on the unknown $x$. The QFI is a figure of merit of theoretical relevance, which corresponds to the estimation precision when the optimal measurement is performed, a task which can be highly nontrivial. In practice, one rather considers the squared signal-to-noise ratio (SNR) $Q_x=x^2 I_x$, which gives the estimation precision obtained with a specific measurement setup.

\subsubsection{Scaling regimes}
\begin{figure}[!h] 
    \centering
    \includegraphics[width=\linewidth]{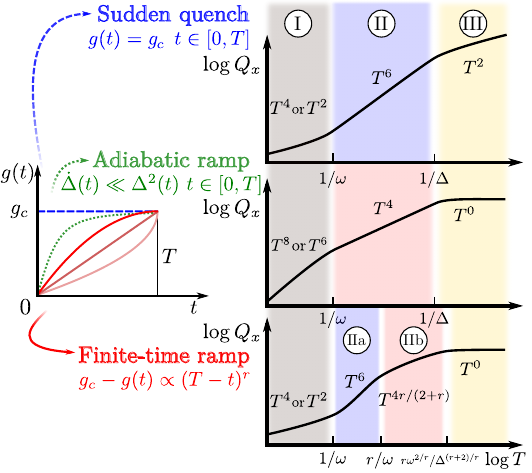}
    \caption{\small{Schematic illustration of the main results regarding the scaling of the QFI with respect to the parameter $x$ of interest and versus the time $T$. Left: Different protocols considered here, namely, a sudden quench (blue dashed line), adiabatic (green dotted line) and finite-time ramps (red solid lines). Right: Sketch of the three different regimes (I, II, and III) for the different dynamical protocols, $\log Q_x$ vs $\log T$. On top, we show the scaling regimes for these three protocols. The first regime I corresponds to very fast dynamics $T\ll 1/\omega$. In this regime, the QFI scales in a non-universal fashion, and depends on the specific parameter $x$ considered. We also have a small signal-to-noise ratio $Q_x=x^2 I_x\ll 1$. The second regime II is within the range $1/\omega \ll T\ll 1/\Delta$, where here $\Delta$ is the energy gap at the critical point (for finite $\eta$). In this regime, the sudden quench leads to a remarkable $T^6$ scaling, while the adiabatic ramp achieves a $T^4$ scaling. For the finite-time ramps we find two different behaviors depending on $r$. On the one hand, for $1/\omega\ll T\ll r/\omega$, the finite-time ramp mimics the dynamics of a sudden quench, and thus one recovers the $T^6$ scaling (IIa). On the other hand, for $r/\omega\ll T\ll 1/\Delta$, the QFI adopts a Kibble-Zurek scaling, which here corresponds to $T^{4r/(2+r)}$ (IIb). This scaling is dominated by the ground-state properties of the system; in the limit $r\gg1$, we recover the adiabatic $T^4$ scaling. Therefore, the finite ramp draws a connection between the "dynamic" and "static" scalings (which we depict in blue and red, respectively). Finally, for times $T\gg 1/\Delta$, the QFI becomes quadratic in time for the sudden quench, while for adiabatic the QFI saturates to a constant value given by the ground state at $g_c$. For the finite-time ramp, the QFI saturates to the same value, but after a time $\frac{r}{\Delta}\left(\frac{\omega}{\Delta}\right)^{2/r}$. Note that in the thermodynamic limit, the third regime is pushed to $T\rightarrow \infty$ as $\Delta\rightarrow 0$. Comparing the three strategies, we find that the sudden quench always yields the best scaling for $\omega T\gtrsim 1$. The same is true if we compare the absolute value of the precision, instead of the scaling.}}
    \label{fig_sketchmainresults}
\end{figure}

We consider three different preparation protocols, as illustrated in Fig.~\ref{fig_sketchmainresults}. First, sudden quenches in which the coupling is abruptly increased from $g=0$ to its final value $g_f\sim g_c=1$, followed by an evolution for a time $T$. Second, adiabatic ramps in which the coupling varies in time $g(t)$ towards the QPT always fulfilling the adiabatic condition $\dot{\Delta}(t)\ll \Delta^2(t)$~\cite{Messiah,Chandra,Rigolin:08,deGrandi:10}, where $\Delta$ denotes the energy gap between ground and first-excited states, which vanishes at the QPT as $\Delta\sim |g-g_c|^{z\nu}$, where here $z\nu=1/2$~\cite{Hwang:15,Ribeiro:07,garbe2020}. We found that if $g$ evolves according to
\begin{equation}
g(t)=\left(1-\frac{1}{1+(t/\tau_Q)^{2}}\right)^{1/2},
\label{adiabevo_general}
\end{equation}
with $\tau_Q$ some time constant verifying $\tau_Q\gg1/\omega$, then the evolution remains adiabatic at all time. However, the critical point is only approached in the long-time limit, but never reached. These two families of protocols (sudden quenches and adiabatic ramps) epitomize the "dynamic" and "static" approaches, and constitute the two poles of our analysis. The third family of protocols is given by finite-time ramps in which
\begin{equation}
    g(t)=1-\left(\frac{T-t}{T}\right)^r,
\label{finiteramp_general}
\end{equation}
where $r>0$ is an exponent which describes the non-linearity of the ramp close to the QPT~\cite{Barankov:08}. Contrary to the adiabatic ramp, this protocol allows one to reach the critical point in finite time, but it does not ensure perfect adiabaticity. By tuning $r$ and $T$, we can make the evolution more or less adiabatic, and thus draw a connection between the two previous protocols.\\

As sketched in Fig.~\ref{fig_sketchmainresults}, we found three different scaling regimes for the QFI depending on the total duration of the protocol $T$. The first one, I, concerns fast evolutions, in which the scaling of the QFI depends on the parameter to be estimated in a non-universal manner. Note that this regime is characterized by a small signal-to-noise ratio $Q_x=x^2 I_x\ll 1$. 
The second and third regimes, II and III, are universal in the sense that we obtain the same QFI scaling for almost any parameter $x$, and any fully-connected model. The regime II is valid for $\frac{1}{\omega}\ll T\ll1/\Delta$, where here  $\Delta$ represents the gap at $g=g_c$ for finite $\eta$. In this regime, the finite-size effects are not yet relevant, and thus one can ignore the quartic term~\eqref{potquartic}, so that the system behaves as in the thermodynamic limit. For a sudden quench in this regime, we found that the QFI scales as $(\omega T)^6$. For the adiabatic preparation, the precision is dominated by the ground-state properties of the Hamiltonian. The QFI in proximity of the QPT diverges as $I_x\sim |g-g_c|^{-\tilde{\gamma}}$ with $\tilde{\gamma}=2$~\footnote{Note that the critical exponent of the susceptibility is typically denoted by $\gamma$. Here, $\tilde{\gamma}$ denotes the critical exponent of the QFI, which is proportional to the fidelity susceptibility.}. Plugging in the profile \eqref{adiabevo_general}, we find that the QFI scales as $(\omega T)^{{\tilde{\gamma}}/(z\nu)}=(\omega T)^4$, as previously reported in~\cite{garbe2020}.

The finite-time ramp leads to a scaling which depends on the behavior of $g(t)$ close to the critical point, i.e. on the non-linear exponent $r$. 
For $r\gg 1$ and $T \lesssim r/\omega$, the finite-time ramp mimics a sudden quench, and hence the characteristic $T^6$ scaling is recovered. We refer to this regime as IIa. On the contrary, for $r/\omega\!\ll\!T\!\ll\! 1/\Delta$, one enters another domain, which we call IIb. In this region, the QFI obeys a Kibble-Zurek scaling law, such that  $Q_x\!\sim\!(\omega T)^{\tilde{\gamma} r/(z\nu r+1)}$, which in this case gives a $(\omega T)^{4r/(2+r)}$ scaling. In the limit $r\gg 1$ and $T>\frac{r}{\omega}$, i.e., for a very slow and non-linear evolution, the adiabatic scaling $(\omega T)^4$ is retrieved. Therefore, the finite-time ramp draws a connection between the two extreme cases studied before, that is, for short enough $T$ the ramp behaves as a sudden quench, while for long $T$ and high $r$, it becomes a fully adiabatic evolution.

Finally, if the protocol time surpasses $1/\Delta$, we enter the regime III, where finite-size effects can no longer be neglected. In this regime, we find that a sudden quench leads to $(\omega T)^2$, restoring an Heisenberg-like behavior. To the contrary, for the adiabatic and finite-time ramps, the QFI saturates to its maximum ground-state value $I_x\sim \Delta^{-\tilde{\gamma}/(z\nu)} $, 
and thus it displays a $T^0$ scaling. In the thermodynamic limit, the energy gap at the critical point vanishes, $\Delta\rightarrow 0$, and the third regime is pushed to $T\rightarrow \infty$. It is worth mentioning that comparing the three strategies, we find that a sudden quench to $g=1$ always gives the highest QFI for a given $T$, both with and without finite-size effects.
 
\subsubsection{Precision bound}
In addition to this, we have derived an upper bound to the QFI which allowed us to re-derive the various scaling regimes, with little to no calculations. This bound is valid if the system state is a squeezed state evolving under a quadratic, time-dependent Hamiltonian of the form
\begin{equation}
    \Hop_x(t)=(\xop, \pop) h_x(t) (\xop, \pop)^T,
\end{equation}
 with $h_x(t)$   being a (in general time-dependent) two-by-two matrix which encodes the parameter $x$ to be evaluated.  This simple model captures most of the metrological protocols with fully-connected systems, in the thermodynamic limit. In general, this evolution does not conserve the average number of excitations; hence it belongs to the category of \textit{active} interferometry. For such an evolution, we found that the QFI is bounded by 
 \begin{equation}
  I_x(T)\leq 8 \left[\int_0^T dt \sqrt{\chi(t)^2+\phi(t)^2} \Big( 2N(t)+1\Big)\right]^2,   
  \label{genebound_general}
 \end{equation}
with $T$ the duration of the protocol, $N(t)$ the average number of excitations at time $t$, $N(t)=\bra{\psi(t)}\adag\aop\ket{\psi(t)}$, and $\phi(t)$ and $\chi(t)$ are the eigenvalues of the matrix $M_x(t)=\partial_x h_x(t)$. Note that $N(t)$ is a very coarse-grained description of the system, while $\chi$ and $\phi$ can be obtained just from $\Hop_x$. Therefore, this expression allows to bound the QFI with minimal information about the state of the system. Notice that, contrary to previous bounds \cite{pinel_quantum_2013,sparaciari_bounds_2015,sparaciari_gaussian-state_2016,safranek_optimal_2016,safranek_gaussian_2016}, this expression takes explicitly into account the fact that the number of excitations varies in time. We stress that this bound efficiently reproduces the scaling in $T$ and general features of regimes II and III. Furthermore, although this bound has been derived for squeezed states and purely quadratic Hamiltonian, we showed that a similar expression can be found when the state is coherently displaced, which could be used to describe more general protocols.

\subsubsection{Saturability of the QFI}
The QFI provides an upper bound on the best precision allowed by quantum mechanics. However, in a practical implementation, we need to measure a specific observable, and use the measurement results to infer the value of the unknown parameter. Depending on the choice of observable, the actual precision may saturate or be very far away from the QFI. We have analyzed the accuracy that can be reached by performing homodyne or photon-counting measurements, or more complex observables. Let us summarize our main findings in the following.
The first insight is that, in general, the relevant information is encoded in the \textit{noise} of the bosonic mode. The quadrature always have zero mean value $\langle\xop\rangle=\langle\pop\rangle=0$. Instead, the \textit{fluctuations} $\langle\xop^2\rangle$, $\langle\pop^2\rangle$ become very sensitive to the parameter to be measured, as the critical point is approached. 
Then we notice that in all cases a given quadrature shows a much larger noise than the other ones, and in general the antisqueezed quadrature provides the best precision. However, we found that homodyne measurement of either of the two will almost always give the \textit{same} precision scaling, except in one specific case where measuring the anti-squeezed quadrature provides a significant advantage. The reason why the anstisqueezed quadrature is convenient (although it shows larger fluctuations) is that these fluctuations themselves are highly sensitive to the parameter of interest. This is in contrast with the traditional paradigm of quantum metrology, where noise of some observable is reduced in order to obtain an effective probe to sense small displacements or phase-shifts. Critical quantum metrology provides an alternative framework, in which noise is not an hindrance blurring the signal, but instead is the signal itself.

Let us now focus on the differences between different families of critical quantum protocols. For the adiabatic ramp, or the finite-time ramp with small $r$, we show that measuring the fluctuations of a single quadrature is  enough to saturate the QFI. For the sudden quench, measuring a single quadrature yields a precision scaling like $T^4$, less favourable than the QFI $T^6$. We showed that reaching in practice the $T^6$ scaling of the QFI of sudden-quench protocols is possible, but it requires implementing non-standard measurement setups. Accordingly, the sudden quench allows reaching a higher QFI than a finite-time ramp, but more complex measurements are required to fully exploit this advantage.

\subsubsection{Effect of decoherence}
Finally, we studied how our findings are affected by decoherence. We analyzed the dynamics near $g=1$ in the presence of boson loss with a rate $\kappa$. For the adiabatic (non-linear) ramp, we found two distinct cases: As long as the protocol duration is smaller than the dissipation timescale, i.e. for $T\ll\frac{1}{\kappa}$, the system does not have time to equilibrate. The dynamics is then very similar to what we have in the absence of decay, and the QFI still follows the $T^4$ (Kibble-Zurek $T^{4r/(2+r)}$) scaling. However, if $T\gtrsim\frac{1}{\kappa}$, the system reaches the steady-state before the end of the protocol. Then, we observe that the QFI saturates at a value $\sim \frac{\omega^4}{\kappa^4}$. Further increasing $T$ does not improve the precision beyond this point.

For the sudden quench, the situation is more intricate. We find the appearance of another timescale, $T_0=\omega^{-2/3}\kappa^{-1/3}$. This timescale corresponds to the moment where the purity of the system starts dropping. Then the QFI behaves as follows: for $T\ll T_0$, the system is still essentially pure and behaves as it did in the absence of decay, so that one recovers the $T^6$ scaling. For $T_0\lesssim T\lesssim \frac{1}{\kappa}$, the system is no longer pure, but it does not yet reach the steady-state. The QFI shows a scaling $T^3$. Finally, for $T\gtrsim \frac{1}{\kappa}$, the system reaches the steady-state, and the QFI saturates at a value $\sim \frac{\kappa^4}{\omega^4}$.

\section{Quantum phase transitions in fully-connected models}\label{sec:QPT}

As aforementioned, here we focus on a family of fully-connected models that undergo a QPT. In the following we provide the general details of an effective model (cf. Sec.~\ref{ss:eff}), namely, a non-linear oscillator, that captures the critical features of this family of models, such as the QR (cf. Sec.~\ref{ss:QRM}) and the LMG (cf. Sec.~\ref{ss:LMG}), among others. This effective description is also valid to describe other relevant systems~\cite{felicetti2020universal}, such as a driven Kerr resonator~\cite{Bartolo2016Exact} or other long-range interacting systems~\cite{Mottl:12}. 

\subsection{Effective model: Non-linear oscillator}\label{ss:eff}

The effective model that describes the low-energy physics of fully-connected models consists in a non-linear oscillator, whose Hamiltonian is given by Eqs.~\eqref{eq:H0}-\eqref{potquartic}. We are particularly interested in the regime where $\eta\rightarrow \infty$, i.e. in $\Hop_0$ given in Eq.~\eqref{eq:H0}, which we will call in the following the \textit{scaling} or \textit{thermodynamic} limit. In this limit, the Hamiltonian exhibits a second-order QPT at the critical point $g_c=1$. The ground state of $\Hop_0$ is a squeezed vacuum state, which can be expressed in two equivalent ways, that is,
 \begin{align} \nonumber
\ket{\psi_b}&={\rm exp}\left[\frac{1}{2}\left(\lvert z\rvert e^{-i\theta}\aop^2-\lvert z\rvert e^{i\theta}a^{\dagger 2}\right)\right]\ket{0}\\
&=(1-4\lvert b\rvert^2)^{1/4}{\rm exp}\left[e^{i\theta}|b| a^{\dagger 2} \right]\ket{0},
\label{squeezedstate}
\end{align}
where we have defined $z=\lvert z\rvert e^{i\theta}$ and $b=\frac{1}{2}\tanh(\lvert z\rvert) e^{i\theta}$. The direction of squeezing is encoded in $\theta$. The squeezing norm can be expressed either through $\lvert z\rvert$ or $\lvert b\rvert$. Each of these two conventions can be more or less convenient depending on the calculation to make; in the following, we will use both. The number of excitations is related to the squeezing parameter via  $N=\sinh^2(\lvert z\rvert)$.  In the ground-state of \eqref{eq:H0}, we have $b=-1/2+(1+\sqrt{1-g^2})^{-1}$ and $\theta=0$ for $0\leq g\leq 1$. The quadrature fluctuations change with the coupling as

\begin{align}
    \label{Eq:quadflu_GS}
    \langle\xop^2\rangle\propto\frac{1}{\sqrt{1-g^2}},\\ \nonumber
    \langle\pop^2\rangle\propto\sqrt{1-g^2}.
\end{align} 
The spectrum is harmonic, and the energy gap is equal to

\begin{equation}
    \Delta=\Of\sqrt{1-g^2},
    \label{Eq:gap_GS}
\end{equation}

so that $\Delta\sim |g-g_c|^{z\nu}$ with $z\nu=1/2$ for $|g-g_c|< 1$. When the system approaches the critical point, the squeezing diverges and the number of excitations becomes infinite.

For $\eta<\infty$, however, the quartic term~\eqref{potquartic} is no longer negligible near the critical point, and stabilizes the system. Although the model can then no longer be solved exactly,  one can still resort to a variational approach to extract scaling arguments, supported by exact numerical simulations~\cite{Hwang:15}. 
 The quartic term will be non-negligible when $\langle(1-g^2)\xop^2\rangle\sim\langle\frac{f(g)}{\eta}\xop^4\rangle$, where the average is taken over the ground state in the $\eta\rightarrow\infty$ limit. Since the thermodynamic-limit ground state is Gaussian, Wick's theorem gives $\langle\xop^4\rangle\sim\langle\xop^2\rangle^2$ up to an irrelevant prefactor. Hence, we find that the quartic term starts to play a role for $1-g^2\sim \eta^{-2/3}$. 
When the quartic term becomes important, the quadrature variance can no longer diverge. Instead, it saturates  $\langle\xop^2\rangle\sim\eta^{1/3}$, and becomes weakly dependent on $g$. Similarly, the gap stabilizes around a finite value, $\Delta\sim\omega\eta^{-1/3}$. Finally, above the critical point $g>1$, the effective $\xop$ potential assumes a double-well structure~\cite{Hwang:15,peng_unified2019,felicetti2020universal}, and the ground state becomes degenerate, which marks the phase transition. This structure corresponds to the usual Landau potential for second-order phase transitions.
 
 To summarize, the region $g<1$ can be divided into two regimes: On the one hand, for $g^2<1-\eta^{-2/3}$, the system remains harmonic, and the quantities describing the system scale with $g$.  On the other hand, for $1-\eta^{-2/3}<g^2<1$, the variance and gap saturate at certain values which scales with $\eta$. This region is denoted as the \textit{critical region}. Its typical width, denoted by $\Gamma$, scales as $\Gamma\sim \eta^{-1/\nu}$, with $\nu=3/2$. 
 As $\eta$ increases, the critical region shrinks, and the saturation value of the gap becomes smaller.
 
This behavior can be retrieved by using the concept of scale-invariance and critical exponents \cite{Fisher:72,Fisher:74,Botet:82,Botet:83,Hwang:15,Puebla:17,Wei:17qrm,PueblaPhD}. We give here a brief sketch of the argument, and refer the reader to App.~\ref{App_criexp} for more details. Even though the model~\eqref{potquartic} has no spatial structure, we can still define a renormalization group-like approach~\cite{Botet:82,Botet:83}, in which the field quadratures, rather than the position in space, are rescaled. More precisely, performing a transformation $\pop \to \alpha\pop$, $\xop \to\frac{1}{\alpha}\xop$, $\eta \to \frac{1}{\alpha^6}\eta$, $\omega \to \frac{1}{\alpha^2} \omega$, $1-g^2 \to \alpha^4(1-g^2)$, the Hamiltonian~\eqref{eq:H0}-\eqref{potquartic} remains approximately invariant; the invariance becomes exact at the critical point. Thus, quantities such as the energy gap must also be invariant. Now, in the thermodynamic limit, the gap is given by $\Delta=\Of(1-g^2)^{1/2}$. For general $g$ and $\eta$, we can write $\Delta\sim \Of(1-g^2)^{1/2}f(g,\eta,\omega)$, with $f$ some scaling function~\cite{Botet:82,Botet:83}. Such scaling function must satisfy several constraints, since $\Delta$ must be invariant under the scaling transformation and must become independent of $\eta$ in the limit $\eta\rightarrow\infty$, and independent of $g$ in the critical region, in the limit $1-g^2\ll\Gamma$. The simplest expression satisfying these constraints is $\Delta\sim \Of(1-g^2)^{1/2}f\left(\frac{1-g^2}{\Gamma}\right)\sim\Of(1-g^2)^{1/2}f\left(\frac{1-g}{\eta^{1/\nu}}\right)$, with $f(x)\rightarrow1$ for $x\gg1$, and $f(x)\rightarrow x^{-1/2}$ for $x\ll1$. In general, for any quantity $A$ that behaves as $A\propto |g-g_c|^\alpha$ in the $\eta\rightarrow\infty$ limit, one can write $A\propto |g-g_c|^\alpha h_A\left((g_c-g)\Gamma^{-1}\right)=|g-g_c|^\alpha h_A\left((g_c-g)\eta^{1/\nu}\right)$ for $\eta<\infty$, with $h_A(x)\rightarrow 1$ for $x\gg1$, and $h_A(x)\rightarrow x^{-\alpha}$ for $x\ll1$. This means that, within the critical region $1-g^2<\Gamma$, the quantity $A$ will saturate at a value that scales as $\Gamma^{\alpha}=\eta^{-\alpha/\nu}$. Note that this corresponds to the standard finite-size scaling in spatially extended systems. Therefore, knowing how $A$ scales with $g$ in the thermodynamic limit, we can infer how it scales with $\eta$ in the critical region. This can be summarized by the following heuristic:

\begin{thm}
 The scaling of a quantity in the critical region can be obtained by taking its scaling with $g$ in the thermodynamic limit, and substituting $1-g^2$ with $\Gamma\sim\eta^{-2/3}$. Or said differently, a quantity for $\eta$ finite and $1-g^2\leq\eta^{-2/3}$ is equal to the same quantity in the thermodynamic limit, at a coupling $g^{*2}=1-\eta^{-2/3}$: the value inside the critical zone is equal to the value near the edge of the zone.
 \label{heur1}
\end{thm}
 
 We can readily verify that this heuristic is satisfied for both $\Delta$ and $\langle\xop^2\rangle$. In the thermodynamic limit, we have $\Delta\sim\omega\sqrt{1-g^2}$ and $\langle\xop^2\rangle\sim(1-g^2)^{-1/2}$; near the critical point, we have $\Delta\sim \omega\eta^{-1/3}\sim \omega\sqrt{\Gamma}$ and $\langle\xop^2\rangle\sim\eta^{1/3}\sim \Gamma ^{\,-1/2}$. Exact numerical simulations~\cite{Hwang:15} confirm that this behavior holds in general. 
 
\subsection{Quantum Rabi model}\label{ss:QRM}

Let us consider a single two-level system, or a spin, interacting with a bosonic mode according to the QR model Hamiltonian, 
\begin{equation}
    \Hop_{\rm QR}=\Oq \sigz +\Of \adag\aop + 2\lambda(\adag+\aop)\sigx,
    \label{HRabi}
\end{equation}
where $\hat{\sigma}_i$ are Pauli operators describing the qubit (we take the convention $[\sigma_x,\sigma_y]=i\sigma_z$), and $[a,\adag]=1$. This model can describe a large number of physical systems, such as a single atom interacting with a photonic mode in the context of cavity QED~\cite{Walther:12}, the interaction of internal degrees of freedom of a trapped ion with the vibrational motion~\cite{Leibfried:03}, or artificial atoms interacting with an LC resonator in superconducting circuits~\cite{Blais:21}. We are interested in evaluating one of the parameters $\Of$, $\Oq$, or $\lambda$, assuming the other two are known. This task can be mapped to several estimation problems using the above-mentioned platform; for example, in trapped ions, the qubit frequency $\Oq$ can be sensitive to an external magnetic field. Therefore, the estimation of $\Oq$ could be useful for space-resolved quantum magnetometry.

In the limit $\Of\!\ll\!\Oq$, this model can be mapped to the non-linear oscillator \eqref{eq:H0}-\eqref{potquartic}, by the mean of a perturbative Schrieffer-Wolff transformation~\cite{Hwang:15}. Let us define the frequency ratio $\eta\!=\!\frac{\Oq}{\Of}$. We apply a unitary operator $\Sop\!=\!e^{i\frac{\lambda}{\lambda_c\sqrt{\eta}}\sigy(\aop+\adag)-\frac{\lambda^3}{3\lambda_c^3\eta\sqrt{\eta}}\sigy(\aop+\adag)^3}$, where we have defined $\lambda_c=\frac{\sqrt{\omega\Oq}}{2}$. This leads to $e^{i\Sop} \Hop_{\rm QR} e^{-i\Sop}=\Oq\sigz+\Of\frac{\lambda^2}{2\lambda_c^2}\sigz(\aop+\adag)^2+\Of\adag\aop -\Of\frac{\lambda^4}{8\lambda_c^4\eta}\sigz(\aop+\adag)^4$ 
plus higher-order terms of order $\frac{1}{\eta\sqrt{\eta}}$ and smaller. Such transformation is valid for $0\leq \lambda\leq \lambda_c$, where the spin and the boson are now decoupled, up to second order in perturbation theory. We can now project the spin in its low-energy subspace $\ket{\downarrow_z}$, and obtain the effective Hamiltonian
\begin{equation}
 	\Hop_{\rm QR,eff}=\Of\left\{\frac{\pop^2}{2}+\left[1-\left(\frac{\lambda}{\lambda_c}\right)^2\right]\frac{\xop^2}{2}+\frac{\lambda^4}{4\lambda_c^4\eta}\xop^4\right\}-\frac{\Oq+\Of}{2},
 	\label{potquarticRabi}
 \end{equation}
 which is in the same form of the Hamiltonian $\Hop$ given in Eqs.~\eqref{eq:H0}-\eqref{potquartic}, up to a constant term, and with $g\!=\!\frac{\lambda}{\lambda_c}\!=\!\frac{2\lambda}{\sqrt{\omega\Oq}}$, $\eta=\frac{\Oq}{\omega}$, and $f(g)=g^4/4$. Therefore, the physics of the QR model can be captured by the non-linear oscillator model, where $g$ is the physical spin-boson coupling, normalized by the frequencies,  and $\eta$ is the ratio between the qubit and boson frequencies. Previous numerical simulations confirm that this is a faithful description of the system close to the critical point~\cite{Hwang:15}.

For $g\leq 1$, the system finds itself in the so-called normal phase. To first order in $1/\eta$ in the $\eta\rightarrow\infty$ limit, the ground state reads as $\ket{\downarrow_z}\ket{\psi_b}$ where the field is in a vacuum squeezed state. As one gets closer to the critical point $g_c=1$, the fluctuations and the number of bosons increase, while the spin remains unperturbed due to the large energy difference. Beyond the critical point, for $g>1$, the system enters the so-called superradiant phase, with a doubly-degenerate ground state, which feature a bosonic population $\propto \eta$ and a coherent state in the spin degree of freedom~\cite{Hwang:15}. 
A direct correspondence can be established between this phenomenology and the superradiant QPT of the Dicke~\cite{Emary:03,Kirton2019,peng_unified2019,GarbePhD} and LMG model (see below, and for example Ref.~\cite{Puebla:17}), with the frequency ratio $\eta=\frac{\Oq}{\Of}$ playing the role of a large number of spins. Therefore, we see here that the parameter $\eta$ can be interpreted as an effective system \textit{size}, even in this model with no spatial extension. It is worth mentioning that the existence of such superradiant QPT in light-matter systems has been subject to a vibrant theoretical debate~\cite{Rzazewski:75,nataf2010no,Vukics:14,de_bernardis_cavity_2018,de_bernardis_breakdown_2018,Andolina2019}. Such fundamental limitation can however be sidestep relying on effective implementations of ultrastrongly-coupled systems~\cite{braumuller2017analog,langford2017experimentally,Markovic2018,Lv2018,Peterson2019}, as demonstrated by recent experimental observations of this QPT in distinct platforms~\cite{Baumann:10,Zhiqiang:17,Cai:21}.

\subsection{Lipkin-Meshkov-Glick model}\label{ss:LMG}
The LMG model~\cite{Lipkin:65} describes a system of $N$ spins coupled through an all-to-all interaction, whose Hamiltonian can be written as
\begin{equation}
    \Hop_{\rm LMG}=h\Jop_z-\frac{\Lambda}{N}\Jop^2_x.
\end{equation}
Here $\Jop_z=\sum_i^N\hat{\sigma}^i_z$ is a collective operator describing the excitations of the chain of spins. The term $\Jop^2_x$ accounts for the all-to-all interaction,  $\sum_{i,j}\hat{\sigma}^i_x\hat{\sigma}^j_x$, while the parameter $\Lambda$ controls its strength and so the nature of the ground state of the LMG. The LMG can be considered as a limiting case of the Ising model with long-range interactions, and thus has been proven very useful to test different aspects of critical quantum dynamics and the role of long-range interactions ~\cite{Porras:04,PerezFernandez:11,Koffel:12,Hauke:13,Santos:16,Jaschke:17,Defenu:18,Zunkovic:18,Halimeh:18,Puebla:19,Zakaria:21}, which has been experimentally realized in  trapped-ion setup~\cite{Islam:11,Richerme:14,Jurcevic:17,Zhang:17b} and with cold gases~\cite{Zibold:10,Hoang:16b,Anquez:16}.

In the thermodynamic limit $N\!\rightarrow\!\infty$, the system undergoes a QPT at $\Lambda_c\!=\!h$, as shown in~\cite{Ribeiro:07,Ribeiro:08,Dusuel:04,Dusuel:05}. For $|\Lambda|\!\leq\! \Lambda_c$ the system is in the normal paramagnetic phase, while it enters in the symmetry-broken ferromagnetic phase for $\Lambda>\Lambda_c$ where the ground state is two-fold degenerate, and $\langle J_x\rangle\neq 0$ acts as a good order parameter. In the thermodynamic limit, this system can be mapped to the non-linear oscillator model \eqref{eq:H0}-\eqref{potquartic}, by performing a Holstein-Primakoff transformation~\cite{Holstein:40,Dusuel:04}. The LMG Hamiltonian commutes with the total spin operator $\Jop^2=\Jop^2_x+\Jop^2_y+\Jop^2_z$, and therefore, the Hilbert space can be split into sectors corresponding to the value of $\Jop^2$. In each spin sector, the interaction term can be decomposed as $\Jop_x=\frac{1}{2}(\Jop_++\Jop_-)$, where the operators $\Jop_{\pm}$ describe raising and lowering within the spin ladder. In the subspace with largest angular momentum $J=N/2$, and for $0\leq\Lambda\leq \Lambda_c$, the Holstein-Primakoff transformation maps the spin states to a bosonic field according to $\Jop_z=-\frac{N}{2}+\adag\aop$, and $\Jop_+=\adag\sqrt{N-\adag\aop}$. Intuitively, the operator $\Jop_+$ is mapped on a bosonic creation operator, plus some extra term which encodes the non-linearity of the spin operator. In the limit $N\rightarrow\infty$, this non-linearity becomes negligible, and we have $\Jop_+\sim\sqrt{N}\adag$. Then we can expand the spin operator as $\Jop_+=\sqrt{N}\adag-\frac{\aop^{\dag 2 }\aop}{2\sqrt{N}}$, plus higher-order terms. In this manner, one finds 
\begin{align}
    \Hop_{\rm LMG,eff}=&h\left(\frac{\pop^2}{2}+\left(1-\frac{\Lambda}{h}\right)\frac{\xop^2}{2}\right)\nonumber\\&+\frac{\Lambda}{4N}\left(\xop^4+\frac{\xop^2\pop^2+\pop^2\xop^2}{2}-2\xop^2+\frac{1}{2}\right)-\frac{h}{2},
\end{align}
which is very similar to the effective non-linear oscillator, Eqs.~\eqref{eq:H0}-\eqref{potquartic}, upon the identification $\omega=h$, $g^2=\frac{\Lambda}{h}$, $\eta=N$ and $f(g)=g^2/4$. Note that the constant term $-\frac{h}{2}$ can be safely ignored. In the limit $N\rightarrow\infty$, the quartic term vanish; the ground state is a squeezed state with $\langle\xop^2\rangle\propto(1-g^2)^{-1/2}$, and $\langle\pop^2\rangle\propto(1-g^2)^{1/2}$, which leads to a diverging entanglement among the $N$ coupled spins as $g\rightarrow 1$~\cite{Latorre:05}. The higher-order terms start to become relevant close to the point $g=1$. At this stage, the fluctuations of $\xop$ become dominant over $\pop$. Hence, $\langle\pop^4\rangle\ll\langle\pop^2\xop^2\rangle\ll\langle\xop^4\rangle$, and $\langle\xop^2\rangle\ll\langle\xop^4\rangle$. Hence, finite-size effects appear primarily as a quartic potential $\xop^4$ as in the non-linear oscillator. Therefore, the phenomenology of the LMG reduces to that of the non-linear oscillator \eqref{eq:H0}-\eqref{potquartic}.

\section{Quantum critical metrology}
\label{sec:QCM}

\subsection{Protocol}
\label{Protocol}

We  now discuss how a metrological protocol exploiting quantum critical effects can be implemented. For concreteness, we will consider  a system described by the QR model~\eqref{HRabi}; the discussion would be exactly the same for other fully-connected models. Let us assume, for instance, that we want to evaluate the frequency $\Of$, assuming that the other parameters are known and controllable. We prepare the system in its ground-state at $\lambda\!=\!0$, with the boson and qubit decoupled. Then we change the coupling constant from $0$ to a certain target value $\lambda$, within a time $T$. As discussed in Sec.~\ref{sec:Summary}, we consider three families or protocols, i.e., quenches, adiabatic ramps and finite-time ramps. These different profiles are sketched in the left side of Fig.~\ref{fig_sketchmainresults}. In all cases, the system at the end of the evolution can be written as
\begin{equation}
\ket{\psi_{\Of}(\lambda,T)}=\ket{\psi_f}.
\end{equation}
In the left side, we have indicated explicitly that the final state depends both on the final coupling value $\lambda$, the protocol duration $T$, and the (unknown) bosonic frequency $\Of$. On the right side, we have used a short-hand notation to lighten the equations. Then we measure some observable on the system, such as the number of bosonic excitations, spin state, etc. Finally, the measurement results are used to reconstruct the value of $\Of$. Depending on the choice of observable, the evaluation may be more or less precise. A standard result in quantum metrology~\cite{paris_quantum_2011} states that if the choice of observable is optimized,  the maximum achievable precision is bounded by the quantum Fisher information (QFI) $I_\Of$ according to
\begin{align}\nonumber
\delta \omega&\geq \frac{1}{\sqrt{I_\Of(\lambda,t)}},\\
\end{align}
where the QFI reads as
\begin{align}
 I_\Of&=4\left[\langle \partial_\Of \psi_f|\partial_\Of \psi_f\rangle -\lvert\langle \partial_\Of\psi_f|\psi_f\rangle\rvert^2\right].
 \label{QFI_general}
\end{align}
Here $\delta\omega$ denotes the standard deviation of the estimated $\Of$, and $\ket{\partial_\Of \psi_f}$ is the derivative of the state $\ket{\psi_f}$ with respect to $\Of$. Under certain conditions \cite{paris_quantum_2011}, it is guaranteed that there exists a choice of observable which allow to saturate this bound. For example, in the case of the evaluation of $\Of$ in the critical Rabi model, some of us have shown in~\cite{garbe2020} that quadrature or photon-number measurements on the bosonic field are optimal. In the case of pure state, the QFI can also be identified with the susceptibility~\cite{Hauke:16}, a quantity commonly used in the condensed matter community. The same reasoning can be applied if, instead of $\Of$, one aims at evaluating the coupling $\lambda$, or any other parameter $x$. This reasoning can also be extended to mixed states, for which the QFI has a more involved expression~\cite{paris_quantum_2011}. Here we will only consider pure states.

To compute the QFI, we can now apply the mapping which we discussed in the Sec.~\ref{sec:QPT}. For fully-connected models, the system can be described in terms of the non-linear oscillator model, Eqs.~\eqref{eq:H0}-\eqref{potquartic}. We consider first the thermodynamic limit, when the quartic term is negligible. In this case, whether through the quench or ramp process, the bosonic mode evolves under an effective quadratic Hamiltonian. At the end of the evolution, the system is in a squeezed vacuum state $\ket{\psi_{b(x,T)}}$, where the squeezing parameter $b$ depends both on the unknown parameter $x$ and on the protocol duration $T$. For squeezed states, the expression~\eqref{QFI_general} can be rewritten in terms of the derivative of the squeezing parameter (see App.~\ref{App:QFI_squeezed} for details of the derivation)
\begin{align}
\nonumber
I_x&=2\left(\left(\frac{\partial|z|}{\partial x}\right)^2+\cosh^2(|z|)\sinh^2(|z|)\left(\frac{\partial\theta}{\partial x}\right)^2\right)\\ \nonumber
&=\frac{8}{(1-4|b|^2)^2}\bigg|\frac{\partial b}{\partial x}\bigg|^2\\
&=\frac{8}{(1-4|b|^2)^2}\left(\left(\frac{\partial|b|}{\partial x}\right)^2+\lvert b\rvert^2\left(\frac{\partial\theta}{\partial x}\right)^2\right).
	\label{QFI_squeezed}
\end{align}
Recall that $b=\lvert b\rvert e^{i\theta}=\frac{1}{2}\text{tanh}(\lvert z\rvert)e^{i\theta}$. Therefore, if we want to evaluate a parameter $x$ using the Rabi or LMG model, we can find analytically the expected precision by mapping the model to the effective bosonic model, express the squeezing as a function of $x$, then use Eq.~\eqref{QFI_squeezed} to compute the QFI. Rather than the QFI itself, we will mostly focus on the quantity $Q_x=x^2 I_x$, which gives the squared signal-to-noise ratio (SNR) of the estimation protocol.

For finite $\eta$, the evolution is no longer quadratic, and the state cannot be expressed analytically. However, we can still obtain the QFI relying on numerical simulations truncating the Fock state basis, which complements the analytical results previously obtained.

\subsection{General bounds, Heisenberg and super-Heisenberg scaling}  

Although the achievable precision can be obtained exactly by computing the QFI, this computation is, in general, a difficult task. Even for the very simple non-linear oscillator model Eqs.~\eqref{eq:H0}-\eqref{potquartic}, numerical simulations have to be used unless $\eta\rightarrow\infty$. The computation becomes even more challenging when mixed states are considered. Therefore, there has been considerable efforts to derive bounds which are insensitive to the specifics of the evolution. Several bounds can be found in the literature \cite{demkowicz-dobrzanski_chapter_2015,giovannetti_quantum_2006,boixo_generalized_2007,pang_optimal_2017}, although there is sometimes some confusion about their range of validity. For the sake of clarity, we provide here a short review of these bounds, and when they can or cannot be used. 
Let us consider a probe system evolving for a time $T$ under an Hamiltonian $\Hop_x(t)$, which depends on the unknown parameter $x$. $\Hop$ may be in general time-dependent and/or depend on $x$ is a non-trivial way. 
At the end of the evolution, the parameter is now encoded in the final state $\ket{\psi_x(T)}$. By measuring this state, we can now evaluate $x$ with a precision bounded by the QFI: $I_x=4\left[\langle \partial_x \psi_x(T)|\partial_x \psi_x(T)\rangle +(\langle \partial_x\psi_x(T)|\psi_x(T)\rangle)^2\right]$. 
Without computing this expression exactly, we can find useful bounds by making several assumptions about the evolution. In particular, let us consider the following set of assumptions:

1) The probe system is composed of a fixed number of probes $N$. 2) The Hamiltonian $\Hop_x$ is bounded, and acts independently on each probe $\Hop_x=\sum_i \Hop_x^i$. 3) $\Hop_x$ depends linearly in the unknown parameter $x$: $\Hop_x=x\Aop$, with $\Aop$ some operator independent of $x$. 4) $\Hop_x$ is time-independent.

Although this list of assumptions may seem long, it is satisfied in the vast majority of current metrological protocols, in particular in atomic interferometry and atomic clocks. When these conditions are satisfied, the achievable QFI scales at most quadratically with the time and the number of probes \cite{giovannetti_quantum_2006}
\begin{equation}
    I_x(T)\sim N^2 T^2.
\end{equation} 
This is the so-called \textit{Heisenberg limit}, which is the backbone of most works in quantum metrology.
Ubiquitous as it is, however, the Heisenberg bound only applies when the above list of conditions is satisfied. Several studies have shown how relaxing one or several of these conditions allows to achieve so-called super-Heisenberg scaling. An early example can be found in the work of Boixo et al.~\cite{boixo_generalized_2007}. They considered a situation in which the Hamiltonian $\Hop_x$ acts on several probes at once, and cannot be written as a sum of local contributions. This is the case, for instance, if the parameter to be evaluated is a interaction strength between neighbouring spins on a lattice. Let us consider, for instance, that the Hamiltonian $\Hop_x$ involves two-body interactions, $\Hop_x=x\sum_i \sigma_x^i\sigma_x^{i+1}$. In this case, Boixo et al.~\cite{boixo_generalized_2007} have shown that the QFI can scale as
\begin{equation}
    I_x(T)\sim \frac{N^{4}}{(2!)^2} T^2,
    \label{bound_Boixo}
\end{equation}
which is indeed a super-Heisenberg scaling in $N$. In general, for a Hamiltonian involving $k-$body interactions, the QFI can scale as $N^{2k}T^2$.
Another possibility is to look at time-dependent protocols. This was first studied in details by Pang and Jordan in Ref.~\cite{pang_optimal_2017}. Let us consider that the Hamiltonian is now time-dependent, but we still have $\Hop_x(t)=x\Aop(t)$ and $\Aop(t)$ is bounded. Then we can define its maximum and minimum eigenvalues, which we will call $\lambda_M(t)$ and $\lambda_m(t)$, respectively. In this case, Pang and Jordan showed that the QFI could be bounded by
\begin{equation}
    I_x(T)\leq \left[\int_0^T dt\ \rvert\lambda_M(t)-\lambda_m(t)\lvert\right]^2.
    \label{IntTlimit}
\end{equation}
This limit allows to go beyond quadratic time-scaling. Let us consider, for instance, that the Hamiltonian $\Hop_x(t)$ is local, but increases linearly with time. Then we will have in general $\lambda_{M,m}(t)\propto Nt$, which may result in
\begin{equation}
    I_x(T)\sim N^2T^4.
\end{equation}
and more generally, if $\Hop_x(t)$ scales like $t^{\alpha}$, we can achieve a scaling $T^{2\alpha+2}$. Hence, it is possible to achieve faster-than-quadratic scalings in time by using time-dependent Hamiltonians. Therefore, we see the Heisenberg scaling can be surpassed, provided that the system is non-linear in time, or the Hamiltonian is non-local, which means that the eigenvalues and observables do not scale linearly with the system size. We can summarize this in the following heuristic:

\begin{thm}
 To beat the Heisenberg scaling, go non-linear.
 \label{heur2}
\end{thm}

For time-dependent protocols, controlling the system state may be difficult. A common solution then is to use quantum control techniques, such as counter-diabatic driving or shortcuts to adiabaticity~\cite{Torrontegui:13}. At first sight, quantum control would seem to be very promising in the context of critical quantum metrology. Indeed, let $\Hop=x\Aop$ be a Hamiltonian with a critical point. Near this point, the system will generally be infinitely sensitive to a perturbation. However, preparing the system adiabatically will take infinite time because of the vanishing energy gap. One may want to apply quantum control to quickly bring the system near the critical point, and hence enjoy infinite precision in a finite time. However, the control term must be fully known, which means in particular that it must be independent of the unknown parameter $x$. Hence, the total operation must be of the form

\begin{equation}
    \Hop_x(t)=x\hat{\mathcal{A}}+\hat{\mathcal{B}}(t),
    \label{form_noadv}
\end{equation}
 where $\hat{\mathcal{B}}$, which contains all the control terms, is \textit{independent} of $x$, and $\hat{\mathcal{A}}$ is time-independent. For such a Hamiltonian, it was shown very recently by Gietka et al.~\cite{gietka2021} that the QFI scales at most like $T^2$. Hence, naive control shortcuts does not allow to achieve infinite precision in finite time, or even to achieve super-quadratic scalings in time. However,  this does \textit{not} exclude the possibility of reaching super-Heisenberg scaling; it only shows that, if we want to achieve such a result, the part of the Hamiltonian which encodes the parameter, $\Aop$, needs to be itself time-dependent. In the course of this article, we will show how simple adiabatic ramps or quenches can indeed be used to achieve non-trivial time scalings in a quantum critical system, without using quantum control.

\section{A bound for active interferometry with Gaussian states}
\label{sec:bound}
Many interferometric experiments involve photonic systems, in which the photon number can be both fluctuating and time-dependent. Heisenberg-like scalings do not always hold for these systems, since the number of particles is not uniquely defined. In particular, it is known that by using exotic photon distribution, an infinite precision can in principle be achieved with a finite (even very small) average number of photons~\cite{rivas_sub-heisenberg_2012,GarbePhD}. 
Therefore, one needs to put constraints on the photon statistics in order to dervive meaningful bounds to the achievable precision. 
A common choice is to consider Gaussian states, which are often encountered in experiments~\cite{monras_phase_2013,pinel_quantum_2013,sparaciari_bounds_2015,sparaciari_gaussian-state_2016,safranek_optimal_2016,safranek_gaussian_2016}. We can then find bounds which involve the \textit{average} number of photons, $\langle \Nop\rangle$. For instance, if we use a coherent or squeezed state to evaluate a phase-shift in a Mach-Zehnder interferometer, the QFI obeys, at best, a Heisenberg-like bound $I_\phi\sim \langle\Nop\rangle^2$ \cite{sparaciari_bounds_2015}. In this scenario, the photon number is fluctuating, but the average value $\langle\Nop\rangle$ is constant in time.

In our case we face a different scenario. Even though in the $\eta\rightarrow\infty$ limt, the bosonic mode is indeed in a squeezed, Gaussian, state, its squeezing parameter varies in time, and the average photon number continuously increases as we approach the critical point. Therefore, our system belongs to the more general category of \textit{active} interferometers. For this scenario too, several bounds have been obtained \cite{pinel_quantum_2013,sparaciari_bounds_2015,sparaciari_gaussian-state_2016,safranek_optimal_2016}; the QFI generally scales, at most, like $N_{tot}^2$, with $N_{tot}$ the average photon number \textit{at the end of the evolution}. Nevertheless, to the best of our knowledge, none of these bounds discuss explicitly the dynamics of the active encoding, i.e. the \textit{duration} of the protocol. 
Here, we introduce an important generalization of these bounds, which explicitly takes into account the time-dependence and which is valid for active interferometry with Gaussian states.

Let us consider quadratic Hamiltonians $\Hop_x=(\xop \pop) h_x (\xop \pop)^T$, with $\xop$ and $\pop$ the field quadratures, and $h_x$ a Hermitian two-by-two matrix, which depends on the unknown parameter $x$. Under such Hamiltonian, an initially prepared vacuum state becomes squeezed, expressed by \eqref{squeezedstate}.
 The derivative of the Hamiltonian can be expressed itself as a quadratic field operator,
\begin{equation}
\partial_x H_x(t)=(\xop \pop) M_x(t) (\xop \pop)^T,
\label{quadatform}
\end{equation}
with $M_x=\partial_x h_x$ a (time-dependent) hermitian matrix. Then at all time, this matrix can be diagonalized, and its eigenvalues are denoted by $\phi$ and $\chi$. Then we have shown that the QFI can be bounded by
\begin{align}
I_x(T)\leq 8 \left[\int_0^T dt \sqrt{\chi(t)^2+\phi(t)^2} \Big( 2N(t)+1\Big)\right]^2,
\label{geneboundfulltimedep}
\end{align}
with $N(t)$ the (time-dependent) average number of photons $N=\bra{\psi(t)}\adag\aop\ket{\psi(t)}$. The detailed proof, as well as additional comments on previous bounds, can be found in App.~\ref{App:bounds}. If $\phi$ and $\chi$ are time-independent, we can rewrite
\begin{align}
I_x(T)\leq 8 (\chi^2+\phi^2)\left[\int_0^T dt  \Big( 2N(t)+1\Big)\right]^2
\label{geneboundfinal}
\end{align}
This bound constitutes our first result. It involves only, on the one hand, the eigenvalues $\chi$ and $\phi$, and, on the other end, the average number of photons in time. The former can be deduced from the expression of $\Hop_x$ only, without any reference to the state of the system; the latter is the smallest amount of information we can have about the system state. Hence, this expression allows one to bound the scaling of the quantum Fisher information with very little information about the state. Note also the formal similarity of this expression with Eq. \eqref{IntTlimit}. Our expression includes explicitly the time-dependence of the number of probes, and the ordinary eigenstates of the Hamiltonian has been replaced by the eigenstates of the matrix $M_x$, which are closely related to the notion of symplectic eigenvalues for Gaussian states \cite{safranek_gaussian_2016,GarbePhD}. Although the formula shown here has been derived for vacuum squeezed states and purely quadratic Hamiltonians, we also show in App.~\ref{App:bounds} that similar expressions can be obtained when we allow the state to have non-zero displacement and a linear contribution in the Hamiltonian.

We stress again that, for squeezed states, the QFI can also be explicitly computed using~\eqref{QFI_squeezed}. However, this bound will be most convenient to discuss the time-scaling of the QFI, with no or little actual calculations. We can already see that, in the limit where $N$ is time-independent, we retrieve the Heisenberg prediction $N^2T^2$. On the contrary, it is also immediately apparent from Eq.~\eqref{geneboundfinal} that if the number of photons increases in time, this bound predicts a higher-than-quadratic time scaling; typically, if the photon number increases in time like $t^{\alpha}$, our bound will predict a QFI $\sim T^{2\alpha+2}$. If we define $N_{tot}$ the maximum number of photon during the evolution (which, in our case, is the photon number at the end of the evolution), then we also find that the QFI is always limited by $N_{tot}^2T^2$, which also makes the connection with the previous results for active interferometry with Gaussian states \cite{pinel_quantum_2013,sparaciari_bounds_2015}. Contrary to these previous results, our bound takes explicitly the time-dependence of $N$ into account, and allows to study more complex time scalings, as we show in the following. 

\section{Sudden quench dynamics}
\label{sec:Quench}

Let us now put together all the elements introduced in the previous sections. We will study the precision that can be achieved using the quench protocol sketched in Sec.~\ref{Protocol}. For that, we use the terminology of the QR model, and refer to $\aop$ as a photonic field. However, we stress again that the same results can be directly obtained with other fully-connected models.

\subsection{Thermodynamic limit}

\begin{figure} 
    \centering
    \includegraphics[width=\linewidth]{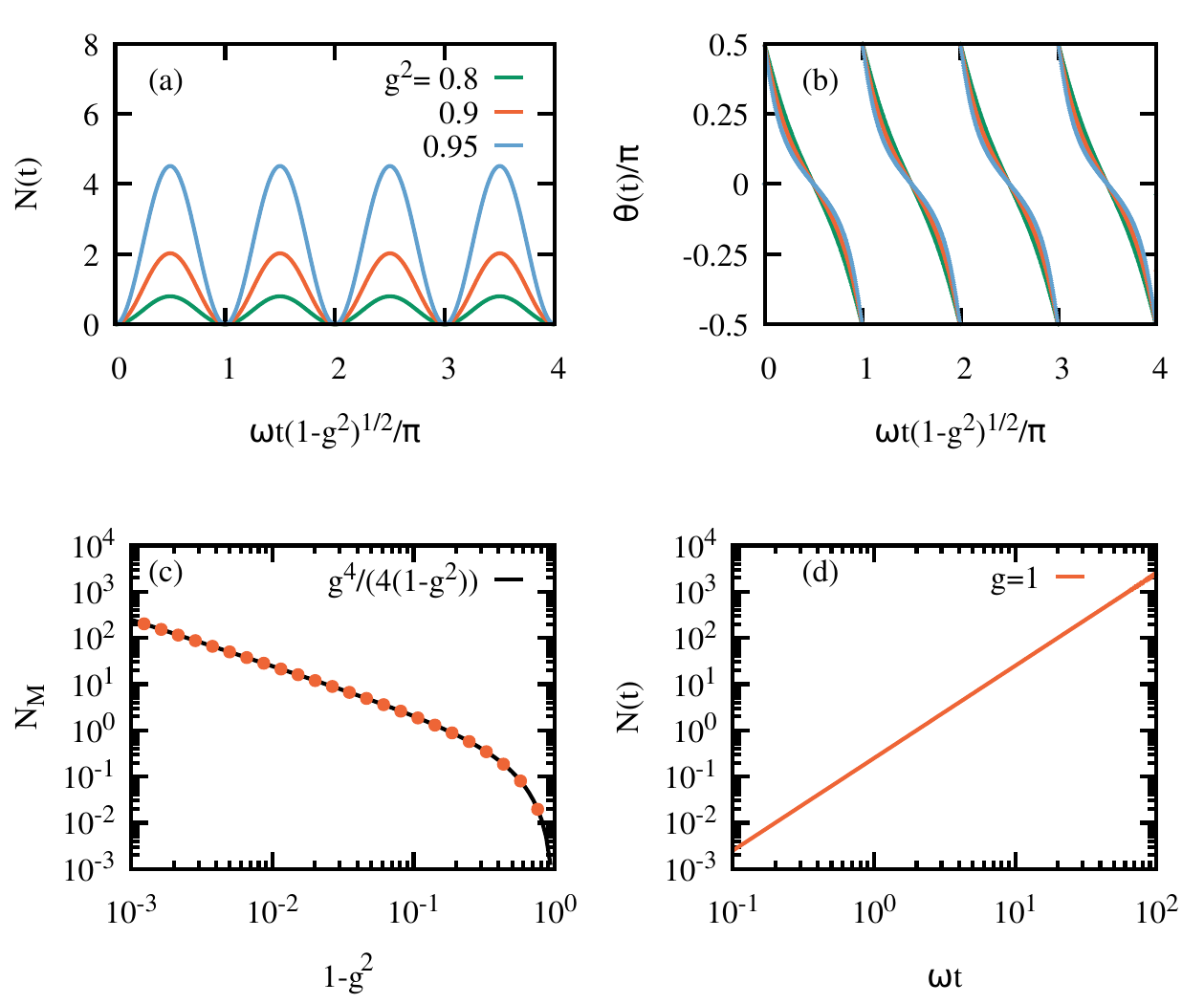}
    \caption{\small{Top panels show the dynamics of the number of bosons $N(t)$ (a) and the squeezing phase $\theta(t)$ (b) upon a sudden quench to $g$ close to the critical point $g_c=1$, namely, $g^2=0.8$, $0.9$ and $0.95$. For $g<1$, $N(t)$ and $\theta(t)$ undergo periodic oscillations, whose period is given by the inverse of the energy gap $\left(\Of \sqrt{1-g^2}\right)^{-1}$. Panels (a) and (b) are plotted against the rescaled time $\Of t\sqrt{1-g^2}/\pi$ to highlight this behavior. Panel (c)
    shows the maximum number of bosons obtain numerically (points), $N_{\rm M}$ versus $1-g^2$, which follows $N_{\rm M}=g^4/(4(1-g^2))$ (solid line). Panel (d) shows the dynamics for a quench to $g=1$. $N(t)$ grows monotonically as $N(t)=(\Of t)^2/4$, while $\theta(t)$ approaches zero as $\theta(t)={\rm arctan}(2/\omega t)$ (not shown here). See main text for further details.}}
    \label{Statequenched}
\end{figure}

Let us first consider that we want to estimate $\Of$, the other parameters being known. We switch instantaneously the physical coupling to a target value $\lambda$, and then let the system evolve freely. We can eliminate the spin degree of freedom, and we are left with the bosonic field evolving under \eqref{eq:H0}-\eqref{potquartic}, with $g=\frac{2\lambda}{\sqrt{\Of\Oq}}$. In the thermodynamic limit $\eta\rightarrow\infty$, the Hamiltonian is purely quadratic, and the state at all time will be a squeezed state of the form \eqref{squeezedstate}. In particular, the squeezing parameter $b$ adopts the following form (see App.~\ref{App:timeev} for the details of the derivation)
\begin{align}
    b(t)&=\frac{2-g^2}{2g^2}+\frac{i\sqrt{1-g^2}}{g^2\tan\left[\sqrt{1-g^2}\Of t-i\text{arctanh}\left(\frac{2\sqrt{1-g^2}}{2-g^2}\right)\right]}.
\label{bsol_quench}
\end{align}
If we quench the system away from the critical point, $g^2<1$, the system will evolve periodically in time, with a period given by the inverse of the gap, i.e. $\tau=\frac{\pi}{\Of\sqrt{1-g^2}}$. 
Moreover, from Eq.~\eqref{bsol_quench} it follows that $b(n\tau)=0$ with $n$ integer, and $b((n+1/2)\tau)=\frac{g^2}{2(2-g^2)}$ its maximum value. The photon number $N(t)=\frac{4\lvert b(t)\rvert^2}{1-4\lvert b(t)\rvert^2}$ has the same periodicity, and its minimum and maximum values are $0$ and $\frac{g^4}{4(1-g^2)}$, respectively. We stress that we have made no approximation here, aside from setting $\eta\rightarrow\infty$. By contrast, if we quench the system directly at the critical point $g=1$, the expression above can be analytically continued, and simplifies to
\begin{align}
b(t)=\frac{\Of t}{2(\Of t -2i)}
\label{bsol_quench_CP}
\end{align}
In this case, we find that the number of photons grows indefinitely in time according to $N(t)=\frac{(\Of t)^2}{4}$. This is only possible at $g=1$ and $\eta\rightarrow\infty$, when there is no stabilizing quartic term in the Hamiltonian, cf. Eq.~\eqref{potquartic}.  The time-evolution for the photon number and the squeezing angle are displayed on Fig.~\ref{Statequenched}, both at and away from the critical point.

\begin{figure} 
    \centering
    \includegraphics[width=\linewidth]{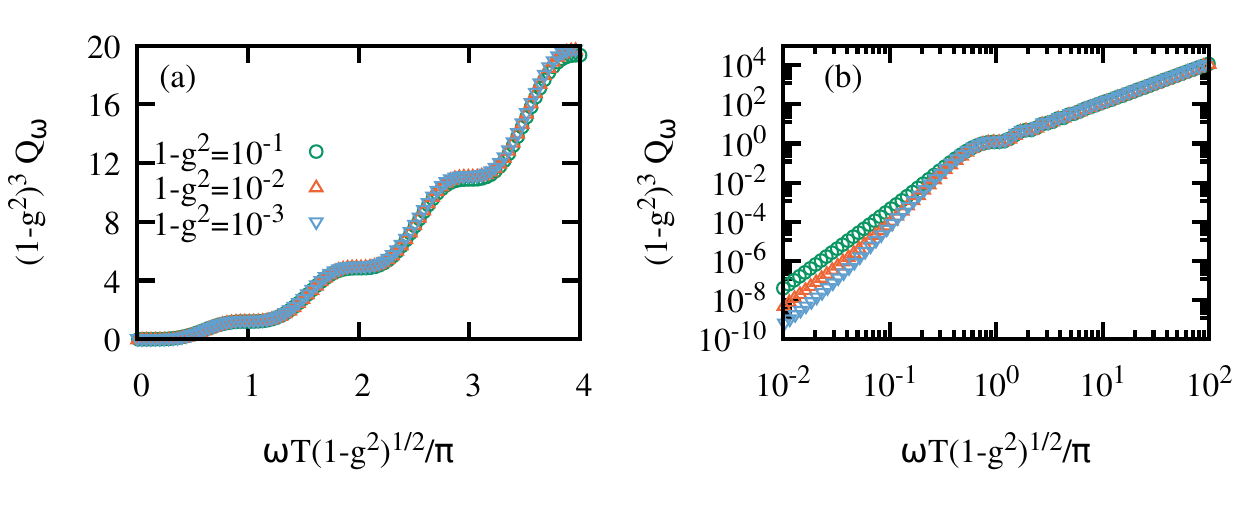}
    \caption{\small{Signal-to-noise ratio for the evaluation of $\omega$ as a function of total time $T$ for different couplings $g^2=0.9$, $0.99$ and $0.999$. In panel (a), the collapse of all the points and the emergence of plateaux separated by $\pi/(\omega\sqrt{1-g^2})$ are evident (cf. Fig.~\ref{Statequenched}(a)). Panel (b) shows the same as (a) but in a log-log scale. }}
    \label{QFIquenchthermo_reallog}
\end{figure}

\begin{figure} 
    \centering
    \includegraphics[width=\linewidth]{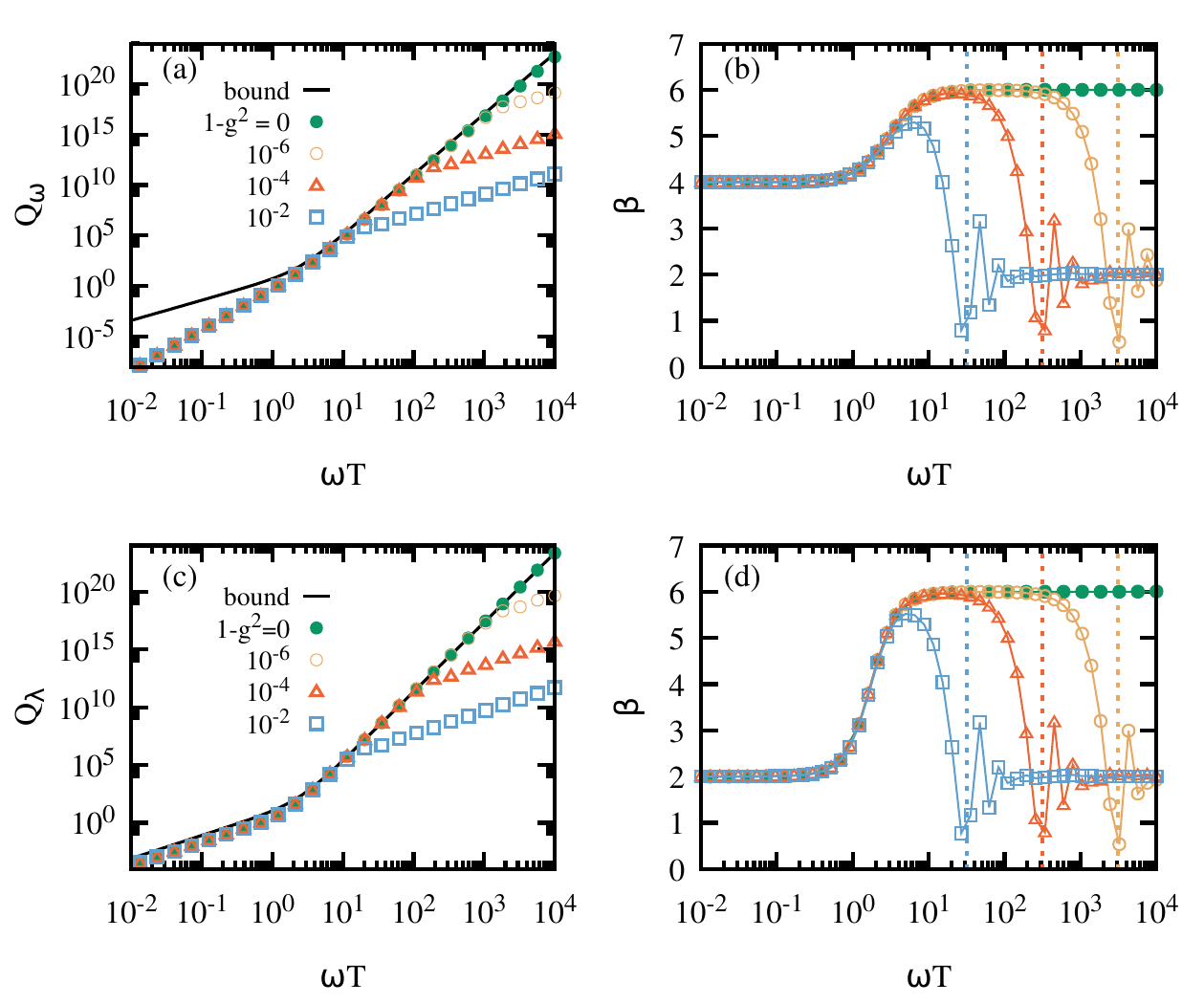}
    \caption{\small{SNR for a sudden quench as a function of the protocol time $T$ and to estimate the parameter $\omega$ (a) and $\lambda$ (c). All the plots are in the thermodynamic limit, $\eta\rightarrow\infty$. Points represent numerical simulations and the solid line is the prediction of the general bound for $g=1$ (cf. Eqs.~\eqref{eq:SQQg}-\eqref{eq:SQQw}). The QFI can be fitted as a polynomial $(\omega T)^\beta$; the fitted value of $\beta$ is plotted on panel (b) and (d) for $Q_\omega$ and $Q_\lambda$, respectively. For short times $(\omega T)\lesssim 10$, the SNR scales like $T^4$ ($Q_\omega$) or $T^2$ ($Q_\lambda$). This corresponds to the regime I sketched in Fig.\ref{fig_sketchmainresults}. For $(\omega T)\gtrsim 10$, the SNR scales like $T^6$, for both parameters (regime II in Fig.\ref{fig_sketchmainresults}). For $g=1$ (full green points), this regime holds until $T\rightarrow\infty$. In contrast, for $g<1$, one enters in regime III for $\omega T\gg\frac{1}{\sqrt{1-g^2}}$, where the SNR scales like $T^2$. The vertical dotted lines in (b) and (d) correspond to $1/\Delta$ to highlight the transition from regime II to III. For $\omega T\gtrsim 10$, the general bound prediction fits well the results.}}
    \label{QFIquenchthermo_scaling}
\end{figure}

By combining \eqref{bsol_quench} and \eqref{QFI_squeezed}, we can now compute the QFI and the signal-to-noise ratio (SNR) $Q_\Of=\Of^2 I_\Of$. The formula \eqref{QFI_squeezed} involves the derivative of $b$ with $\Of$. We have so far expressed $b$ in terms of $g$ and $\Of$. However, if we rewrite everything in terms of the physical parameters, we find that $g$ itself depends on $\Of$, since we have $g=\frac{2\lambda}{\sqrt{\omega\Omega}}$. Therefore, the QFI will involve two components, coming from  $\frac{\partial g}{\partial \Of}\frac{\partial b}{\partial g}$ and $\frac{\partial b}{\partial_{\Of}}$, respectively. As we discuss in App.~\ref{Appendix:physical-effective}, the first contribution actually dominates in most parameter regimes.
On Fig.~\ref{QFIquenchthermo_reallog} we plot $Q_\Of$ with respect to the duration of the protocol $T$ and the distance to the critical point, $1-g^2$. The SNR shows plateaux, separated by intervals $\frac{\pi}{\Of\sqrt{1-g^2}}$. The log-log plot reveals that, for $T>\frac{1}{\Of\sqrt{1-g^2}}$, the QFI has a secular increase $\left(\frac{\Of T}{(1-g^2)}\right)^2$. In the top row of Fig.~\ref{QFIquenchthermo_scaling}, we study more systematically the scaling of the SNR with $T$. We see that the SNR shows actually three different regimes. For short durations $\Of T\lesssim 10$, we obtain a quartic scaling $Q_\Of\propto T^4$. For  $10\lesssim\Of T\lesssim \frac{1}{\sqrt{1-g^2}}$, the SNR scales like $T^6$. Finally, beyond the first plateau, for $\Of T\gtrsim \frac{1}{\sqrt{1-g^2}}$, the SNR settles on a quadratic scaling $T^2$. If we quench the system to values $g$ closer to $1$, the quadratic regime kicks in later. In the limit $g\rightarrow 1$, the first plateau extends to $T\rightarrow \infty$, and the $T^6$ scaling lingers for ever.

The same analysis can be conducted when the unknown parameter is the physical coupling $\lambda$. In the bottom row of Fig.~\ref{QFIquenchthermo_scaling}, we show the scaling of $Q_\lambda=\lambda^2 I_\lambda$ with time. For short time, the SNR scales like $T^2$, instead of $T^4$ for $Q_\Of$. For longer time, the features of $Q_\lambda$ and $Q_\Of$ are exactly the same. As we discuss in details in App.~\ref{Appendix:physical-effective}, this is because a change in either $\Of$ or $\lambda$ have essentially the same effect, which is a renormalization of the effective coupling variable $g$. We can actually make this argument more general. Let us consider \textit{any} model which can be mapped to \eqref{eq:H0}. We want to evaluate a physical parameter $x$. Then \textit{as long as the renormalized coupling $g$ depends on $x$, the SNR $Q_x$ for long $T$ will have the same behavior, independently of $x$ and of the model}. In particular, we will get the same  $T^6$ and $T^2$ scalings if we want to evaluate $h$ or $\Lambda$ in the LMG model, since we have $g=\sqrt{\frac{\Lambda}{h}}$ in this case. A similar dynamical behavior was also obtained recently by Chu et al.~\cite{Chu2021}. Here, we show that these scalings have a broad range of application. We will also show in Sec.~\ref{ss:fsizeSQ} how these results extend to the regime of finite $\eta$ as well.

We can gain a better intuition of this complex interplay of scalings by using the general bound~\eqref{geneboundfinal}. Let us start with the case $g=1$, when the system is quenched at the critical point, and let us consider that we want to evaluate $\lambda$. Then we have (we recall that $g=\lambda/\lambda_c$): 
\begin{align}
\partial_\lambda \Hop= -\Of\frac{\lambda}{\lambda_c^2}\xop^2=-\Of\frac{g}{\lambda_c}\xop^2.\nonumber
\end{align}
This operator is clearly of the form \eqref{quadatform}, with ${M_\lambda=-\omega\frac{g}{\lambda_c}\begin{bmatrix}1&0\\0&0\end{bmatrix}}$. The matrix is already diagonal, with eigenvalues $\phi=-\omega\frac{g}{\lambda_c}$ and $\chi=0$, which are both time-independent. Taking the expression of $N(t)$ and plugging these elements in~\eqref{geneboundfinal}, we find a bound for the signal-to-noise ratio, which reads as
\begin{align}
    Q_\lambda=\lambda^2 I_\lambda\leq 8\Of^2g^4\left[\int_0^T dt  \Big( 2N(t)+1\Big)\right]^2\nonumber \\
    = \left(\frac{2}{9}(\omega T)^6+\frac{8}{3}(\omega T)^4+8(\omega T)^2\right).\label{eq:SQQg}
\end{align}
Similarly, for the bosonic frequency $\omega$, we find $\partial_\Of \Hop=\left(\frac{\pop^2}{2}+(1-g^2)\frac{\xop^2}{2}\right)+\frac{g^2}{2}\xop^2$, and thus $\phi=\chi=\frac{1}{2}$. This leads to a SNR bound 
\begin{align}\label{eq:SQQw}
    Q_\Of\leq \frac{1}{2}\left(\frac{2}{9}(\omega T)^6+\frac{8}{3}(\omega T)^4+8(\omega T)^2\right).
\end{align}

Hence, for short time, the general bound predicts a quadratic scaling. However, as soon as $T$ is large enough (typically for $\Of T>10$), the $T^6$ term dominates. Therefore, our bound correctly predicts that, for a quench at the critical point $g=1$, the SNR features a $T^6$ for long enough time. Fig.~\ref{QFIquenchthermo_scaling}(a) and (b) also show the general bound prediction. For $Q_\lambda$ there is an excellent agreement for all times, i.e., the bound is saturated. For $Q_\Of$, the bound fails to predict the correct scaling for shorter times, but a qualitative agreement is recovered for $\Of T>10$ (in this regime, the bound actually overestimates the actual result by a factor $2$, but captures the scaling). 
Note that one can retrieve the $T^6$ scaling  with hardly any calculation by noting that for long enough times, the photon number is large $N(t)\gg 1$, and the bound becomes essentially proportional to $\left[\int_0^T dt N(t)\right]^2$. The dynamics \eqref{bsol_quench_CP} gives a $N(t)$ which scales quadratically with $t$; therefore, the integral $\int_0^T N(t)dt$ scales like $T^3$, and the bound is $T^6$. Therefore, by simply looking at the scaling of $N$ with time, we can understand the behavior of the SNR for a quench at the critical point.\\

For a quench away from the critical point, the quadratic scaling $Q_\Of\propto T^2$ can also be retrieved from a simple argument. First, let us note that the integral $\int_0^TN(t) dt$ is monotonic in $T$. Therefore, the precision predicted by the general bound can only grow with the protocol duration. Second, although its exact expression is a bit involved, the photon number predicted by \eqref{bsol_quench} is periodic in time, of periodicity $\tau=\frac{\pi}{\Of\sqrt{1-g^2}}$. Let us assume that the average photon number oscillates between zero and its maximum value $N_M$. Let us define $\alpha=\frac{1}{N_M\tau}\int_0^\tau N(t)dt$. Except in very specific cases (for instance, if $N$ increases in very short bursts), $\alpha$ will be typically of order $1$. Then we have $\int_0^t N(t') dt'=N_M(\alpha t + F(t))$, where $F(t)=\int_0^t \frac{N(t)}{N_M}-\alpha dt$ is a periodic function, with $\lvert F \rvert\leq 1$, and $F(n\tau)=0$ for integer $n$. For long $t$, we will have $F(t)\ll \alpha t$. Then, if the QFI saturates the general bound, it follows
\begin{equation}
I_x(T)\sim 32(\chi^2+\phi^2)\alpha^2 N_M^2T^2\sim N_M^2T^2,
\label{Hlimitperiodic2}
\end{equation}
for long $T$, and
\begin{align}
I_x(T=n\tau)&= 8(\chi^2+\phi^2)\left[2n\tau\alpha N_M +n\tau\right]^2\nonumber \\&=n^2 I_x(T=\tau),
\label{Hlimitperiodic1}
\end{align}
for every $n=1,2,\ldots$. Therefore, without even computing the general bound, we can deduce from these simple arguments that it must be monotonic, show a secular quadratic increase in $T$, and be self-similar at intervals $n\tau$. These qualitative features are exactly those of the SNR on Fig.~\ref{QFIquenchthermo_reallog}. 
 Furthermore, the maximum number of bosons, as shown in Fig.~\ref{Statequenched}(c), scales like $N_M\propto \frac{1}{1-g^2}$.  Therefore, the general bound prediction, including the $N_M^2$ prefactor, gives $Q_\Of\propto\left(\frac{T}{1-g^2}\right)^2$, which is what we observe in Fig.~\ref{QFIquenchthermo_reallog}. Combining everything together, we now have the following picture: for short times, the QFI is dominated by non-universal terms, and the general bound is generally not saturated. For times $1\ll \Of T\lesssim \Of\tau$, the effect of the gap is not yet relevant. The system behaves essentially as if it were quenched at the critical point, i.e. the photon number increases like $N(t)\sim t^2$, and the bound scales as $\left[\int_0^TN(t) dt\right]^2\propto T^6$. Finally, for durations $T$ larger than $\tau$, the system undergoes periodic oscillations, and the general bound becomes $N_M^2T^2\sim\left(\frac{T}{1-g^2}\right)^2$. Hence, we have shown that our bound allows to accurately grasp the various scaling regimes of the SNR for $\omega T\gtrsim 1$.

\subsection{Finite-size effects}\label{ss:fsizeSQ}

\begin{figure}
    \centering
    \includegraphics[width=\linewidth]{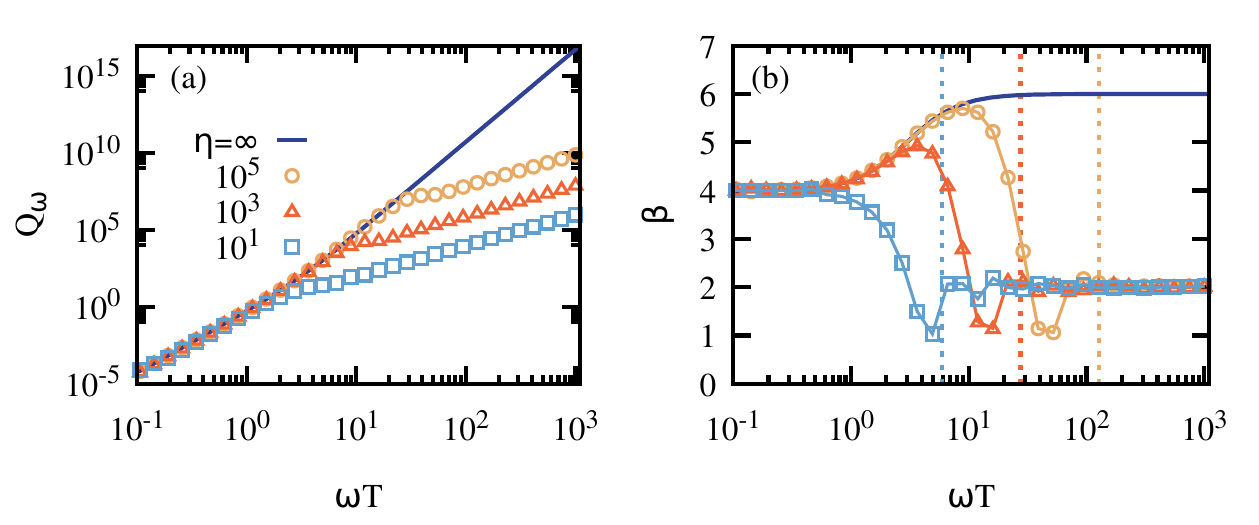}
    \caption{\small{(a) SNR for a sudden quench at $g=1$ versus time, for various values of $\eta$ (points) together with the result for $\eta\rightarrow\infty$ (solid blue line). The SNR scales like $T^4$ for $\omega T \lesssim10$, $T^6$ for $10\ll\omega T\ll \eta^{1/3}$, and $T^2$ for $\omega T\gg \eta^{1/3}$. These results coincide with what we would obtain for $\eta\rightarrow\infty$, doing a quench to $g^*=\sqrt{1-\eta^{-2/3}}$ (see Fig. \ref{QFIquenchthermo_scaling}). Panel (b) shows the fitted exponent $\beta$ to illustrate the different scaling regimes depending on $\omega T$ and $\eta$. The dotted vertical lines indicate the inverse gap at the critical point, $\Of/\Delta=\eta^{1/3}$, for each value of $\eta$. We see that the final $T^2$ scaling regime is established for $\Of T\sim \Of/\Delta$.}} 
    \label{QFIquenchfinite}
\end{figure}

The previous results have been obtained in the thermodynamic limit $\eta\rightarrow\infty$. For $\eta$ finite, the system evolves under the Eqs.~\eqref{eq:H0}-\eqref{potquartic}, and the evolution can no longer be exactly solved. Therefore, we resort to numerical simulations with a converged number of Fock basis to find the QFI. We consider a sudden quench to the critical point with $\eta$ finite. The results are plotted on Fig.~\ref{QFIquenchfinite}. In the limit $\eta\rightarrow\infty$, we have a $T^6$ scaling in the long-time limit, as already discussed. For $\eta$ finite, however, we observe a transition from $T^6$ to $T^2$, which takes place when $\omega T\sim \eta^{1/3}$. 

This behavior can be understood with the following argument. For finite $\eta$, a gap stabilizes around the critical point, of order $\Of\eta^{-1/3}$. For short times, this finite gap is not relevant, and the system evolves as it would in the thermodynamic limit. Another formulation would be to say that, for small time, the number of photons is still small, and the quartic term in Eq.~\eqref{potquartic} is therefore negligible. However, for longer times, the finite gap will create a periodic revival behavior. This resembles the phenomenology in the thermodynamic limit when quenched away from the critical point.  We can here invoke Heuristic \ref{heur1}: \textit{the behavior for a quench at $g=1$ for finite $\eta$ is similar to the behavior one would obtain in the thermodynamic limit by quenching the system at $g^*=\sqrt{1-\eta^{-2/3}}$}. Hence, we see that, although the state is now non-Gaussian and the bound \eqref{geneboundfinal} is no longer applicable, we have the same essential features as in the thermodynamic limit, with a periodic behaviour for $N$, and a SNR showing a secular $T^2$ increase. The difference is that the period of the oscillations, and the boundaries between different scaling regimes, is now given by the parameter $\eta$, instead of the effective coupling $g$.

\section{Adiabatic and finite-time ramps}
\label{sec:Ramp}
 We will now analyze the metrological consequences when, instead of being abruptly quenched, the parameters are slowly tuned to their final values. In this case, the QFI is mostly dominated by the ground-state, equilibrium properties of the Hamiltonian, which corresponds to the \textit{static} paradigm considered in Ref.~\cite{rams_at_2018}.

 \subsection{Adiabatic ramp in the thermodynamic limit}\label{ss:AR}
 
To evolve the state in a controlled way, a possible solution is to tune the parameters very slowly in time, in order to keep the evolution adiabatic. We will consider the thermodynamic limit $\eta\rightarrow\infty$. We take~\eqref{eq:H0}, and slowly increase $g$ towards the critical point. We will define $\epsilon=1-g^2$ for convenience of notation. As long as the time scale introduced by the external driving ($\Delta/\dot{\Delta}$) is much larger than the typical time scale of the system ($1/\Delta$), i.e., as long as
\begin{equation}
\frac{\dot{\Delta}}{\Delta}\ll \Delta,
\label{criterion_adiab}
\end{equation}
the evolution will be adiabatic to very good approximation (see Refs.~\cite{Chandra,Rigolin:08,deGrandi:10} for time-dependent perturbation theory in this context, as well as~\cite{Zurek:05}). Here $\Delta$ is the energy gap during the evolution, and $\dot{\Delta}$ its time derivative. Hence, when initialized in the ground state, the system will remain in it during the evolution. Note that $\Delta\propto\Of\sqrt{\epsilon}$, and $\dot{\Delta}/\Delta\sim\dot{\epsilon}/\epsilon$.
In Ref. \cite{garbe2020}, a time-profile fulfilling these criteria was derived for $\epsilon$, which reads as
\begin{equation}
\epsilon(t)=\frac{1}{1+(t/\tau_Q)^2},
\label{ramp_adiab}
\end{equation}
with $\tau_Q$ some time constant. During this non-linear ramp, we start from $\epsilon(0)=1$, i.e. $g(0)=0$, and then we gradually approach the critical point $\epsilon=0$. The evolution speed $\dot{\epsilon}$ is high at first, then gradually decreases to keep up with the closure of the energy gap. More precisely, we have $\dot{\Delta}/\Delta^2=(t/\omega \tau_Q^2)/\sqrt{1+(t/\tau_Q)^2}$. Therefore, as long as $\tau_Q=\frac{1}{\varphi\Of}$, with $\varphi\ll1$, the criterion \eqref{criterion_adiab} will be satisfied, even in the thermodynamic limit, when the gap becomes exactly zero at the critical point. Note, however, that we only approach asymptotically the critical point, but we never reach it. The corresponding time-profile is schematically plotted in green in the left-hand side of Fig.\ref{fig_sketchmainresults}.

Therefore, if we let the system evolve under this ramp for a time $T$, we expect the system to evolve adiabatically, and to be prepared in the ground-state of the Rabi Hamiltonian. That is, it will be in a squeezed state \eqref{squeezedstate} with $\lvert z\rvert=-\frac{1}{4}\log(\epsilon(T))$ and $\theta=0$. Using \eqref{QFI_squeezed}, the QFI can then be computed exactly as
\begin{equation}
    I_\omega=\frac{1}{8\omega^2}\left(\frac{1}{1-g^2}\right)^2=\frac{1}{8\omega^2\epsilon(T)^2}.
    \label{QFI_adiab}
\end{equation}
Restoring the dependency of $\epsilon$ on $T$, and in the limit of large $T$, we find
\begin{equation}
    Q_\omega=\omega^2 I_\omega\sim\left(\frac{T}{\tau_Q}\right)^4=\left(\varphi\omega T\right)^4.
    \label{QFIadiabramp}
\end{equation}
Therefore, the SNR in this case can scale quartically in time. This scaling can also be understood through the general bound \eqref{geneboundfinal}. For long $t$, $N\sim \frac{1}{\sqrt{\epsilon(t)}}\sim t/\tau_Q$. Since $N$ increases roughly linearly in time, the squared integral in \eqref{geneboundfinal} scales as $T^4$.

\begin{figure}
    \centering
    \includegraphics[width=\linewidth]{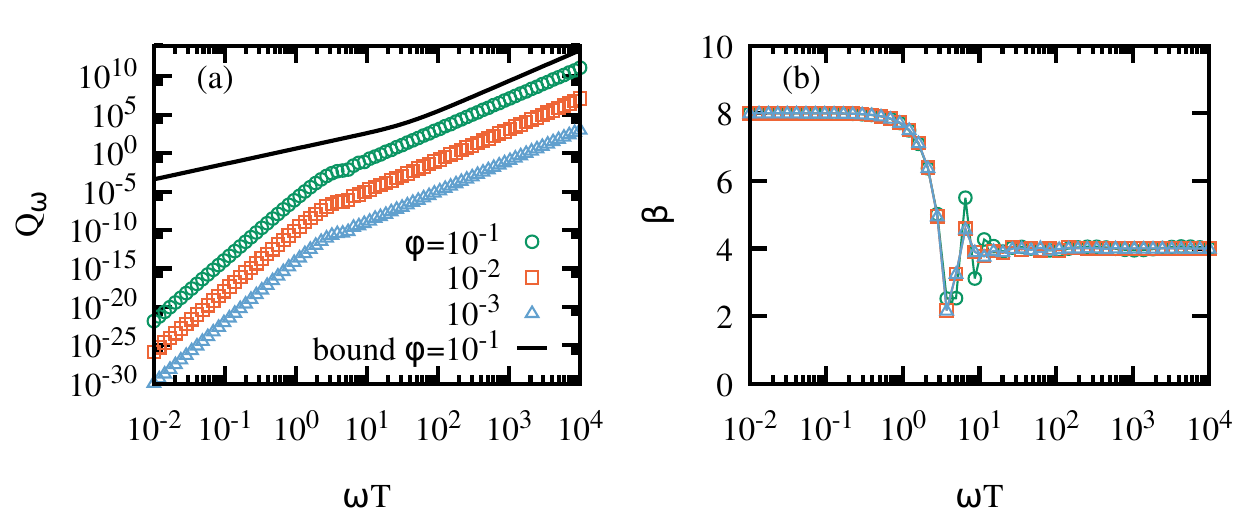}
    \caption{\small{(a) SNR for the parameter $\omega$ under an adiabatic ramp~\eqref{ramp_adiab} with $\tau_Q=1/(\varphi\omega)$, versus $\Of T$, for different values of $\varphi$, together with the bound for $\varphi=10^{-1}$ (solid black line) (cf. Eq.~\eqref{geneboundfinal}).  (b) Scaling exponent $\beta$ for $Q_\omega$ obtained as a best fit to $Q_\omega\sim (\omega T)^\beta$. For short times (regime I) $T\lesssim 1/\omega$ one finds $\beta=8$, where the prediction of~\eqref{QFIadiabramp} does not hold. The general bound is also loose. For longer times, we find $\beta=4$ as predicted in~\eqref{QFIadiabramp} (regime II), and the general bound fits the scaling. Similar results can be found for $Q_\lambda$ but with a scaling $\beta=4$ for $T\lesssim 1/\omega$ (not shown here).}}
    \label{fig6}
\end{figure}
Fig.~\ref{fig6} shows the scaling with $T$ of the SNR, computed with exact simulation (see App.~\ref{App:timeev}). We find that the $T^4$ scaling is indeed recovered for $\Of T\gtrsim 10$. The adiabatic behavior, however, is broken for very small $T$. Indeed, the fulfillment of condition \eqref{criterion_adiab} means that the population of excited states oscillates quickly, with an oscillation rate given by the energy gap, which here is of order $\omega$. For $T\gg 1/\omega$,  the system evolves over several periods, these population will be averaged out, and the system will indeed remain in its ground state. However, if we let the system evolve for a time shorter than the oscillation period $1/\omega$, the system can become excited, and the adiabatic prediction \eqref{QFIadiabramp} is not satisfied anymore. In this regime, we observe that the SNR scales instead as $T^8$.  Since this behavior holds only for short times, and is associated with extremely small SNR, it should be irrelevant in practice.

Let us stress that these results can also be expressed in terms of critical exponents. Let us assume that the gap scales like $\Delta\sim\epsilon^{z\nu}$, and the QFI scales like $I_x\sim\epsilon^{-\tilde{\gamma}}$. Then we can design a ramp $\epsilon(t)=\frac{1}{1+(t/\tau_Q)^{1/(z\nu)}}$, which will satisfy the adiabaticity condition as long as $\tau_Q\gg1/\omega$. Plugging this expression in the QFI, we find $I_x\sim T^{\tilde{\gamma}/(z\nu)}$. In our case, we have $z\nu=1/2$ and $\tilde{\gamma}=2$, which gives the $T^4$ scaling. Finally, we also studied the evaluation of $\lambda$; we found that, for $\Of T\gtrsim 10$, we recover the $T^4$ scaling. As in the sudden quench case, the SNR becomes independent of the parameter being evaluated.

\subsection{Finite-time ramp: Kibble-Zurek mechanism}

The previous ramp, although it optimally keeps the system in the ground state, may be challenging to implement in practice. In general, it might easier to implement ramps according to
\begin{equation}
    g(t)=g_c\left(1-\left(\frac{T-t}{T}\right)^r\right),
    \label{rampprofile}
\end{equation}
with $0<r<\infty$. The corresponding time-profile is plotted in red in the left-hand side of Fig.\ref{fig_sketchmainresults}. In this subsection, we still assume that we are in the thermodynamic limit. Contrary to the previous case, the critical point $g_c=1$ \textit{is} reached in finite time, $g(T)=g_c$. However, the dynamics will cease to be adiabatic in the proximity of the QPT, which is at the core of the Kibble-Zurek mechanism~\cite{Zurek:96,Zurek:05,Dziarmaga:05,Damski:05,delCampo:14,rams_at_2018}. In particular, when $\dot{\Delta}\sim \Delta^2$~\cite{Zurek:05}, the adiabaticity will be broken, which defines the so-called freeze-out time. In our case, this takes place at a time $t_f=T\Big[1-\left(\frac{\omega T}{r}\right)^{-\frac{2}{r+2}}\Big]$, which corresponds to a value
\begin{equation}
    1-g_f=\left(\frac{\omega T}{r}\right)^{-2r/(2+r)}.
    \label{freezout_value}
\end{equation} 

In this case, the standard Kibble-Zurek argument~\cite{Zurek:96,Zurek:05,Dziarmaga:05,Damski:05,delCampo:14} states that the evolution can be decomposed into two parts. First, an adiabatic evolution with the coupling moving from $g(0)=0$ to $g(t_f)=g_f$. Second, an impulse regime in which the system cannot react to external changes imposed by $g(t)$, and thus the system is effectively quenched from $g_f$ to the final value $g(T)=g_c=1$. Moreover, the QFI at the end of the evolution becomes approximately equal to the QFI at the freeze-out instant. 

\begin{figure}
    \centering
    \includegraphics[width=\linewidth]{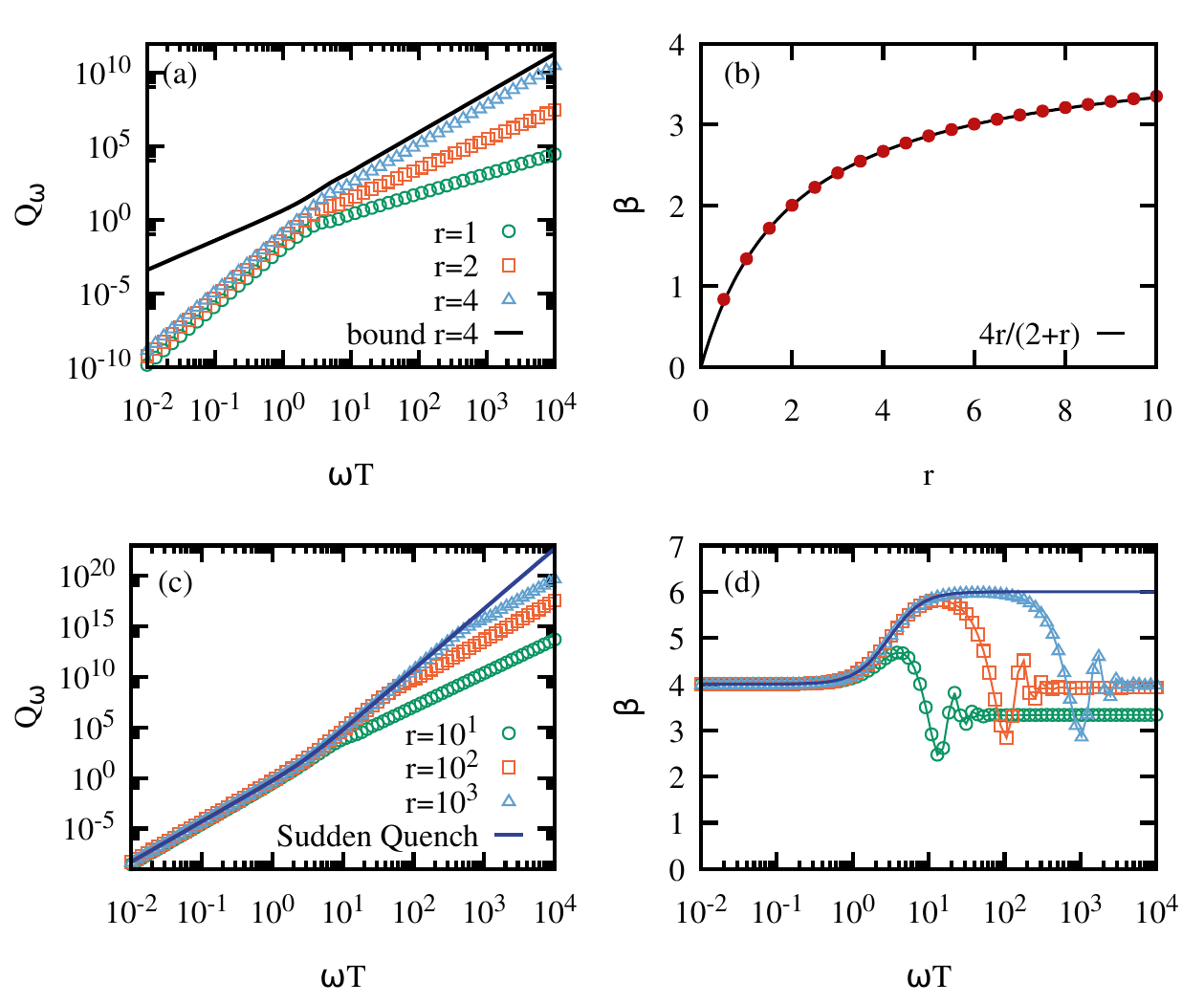}
    \caption{\small{(a) SNR $Q_\Of$ for a finite-time ramp \eqref{rampprofile} versus $\Of T$, for various values of the non-linear exponent $r$. We also show the resulting bound from Eq.~\eqref{geneboundfinal} for $r=4$ (solid black line). (b) Fitted exponent $\beta$ in the interval $\omega T\in[10^3,10^4]$, which agrees very well with the KZ predicted exponent, $4r/(2+r)$. (c) Illustration of the behavior of $Q_\Of$ for $r\gg 1$, compared to the sudden quench results (solid blue line). Panel (d) shows the fitted exponent $\beta$, which clearly reveals a $T^6$ scaling as for a sudden quench protocol, but which holds only for $10\lesssim \omega T\lesssim r$.}}
    \label{fig7}
\end{figure}

Combining this with Eq.~\eqref{QFI_adiab}, the KZ mechanism predicts a QFI proportional to $\frac{1}{\omega^2(1-g_f^2)^2}$, which results in
\begin{equation}
    Q_\omega=\left(\frac{\omega T}{r}\right)^{4r/(2+r)}.
    \label{KZprediction}
\end{equation}
We computed the SNR under~\eqref{rampprofile} (see App.~\ref{App:timeev}) and compared it to this prediction. The results are plotted in Fig.~\ref{fig7}. On the top panel, we show the scaling of the SNR for relatively small $r$. We observe that, for $10\lesssim \Of T$, the SNR scales indeed according to the KZ value. Note that the bound~\eqref{geneboundfinal} is able to capture this scaling behavior. A closer examination, with higher values of $r$ (bottom row of Fig.~\ref{fig7}) reveals that there are actually three scaling regimes. For very short times $\Of T \lesssim 10$, $Q_\Of$ scales as $T^4$. For $10<\Of T <r$, one obtains $Q_\Of\propto T^6$, while for $\Of T>r$, we recover the Kibble-Zurek scaling (cf. Eq.~\eqref{KZprediction}). This can be interpreted as follows. For $\Of T < r$, the freeze-out value $1-g_f$ is larger than 1. Since $g$ must be positive, this is not possible, which means that the adiabaticity is actually broken from the very beginning of the evolution. Therefore, the entire evolution can be deemed as a sudden quench, in which the coupling is instantaneously brought from $g(0)=0$ to $g_c=1$, and left to evolve for a time $T$ (cf. Sec.~\ref{sec:Quench}); we then recover the $T^6$ scaling we observed in Sec.\ref{sec:Quench}. For $\Of T > r$, $0\leq g_f\leq 1$, so that the evolution can be decomposed according to the adiabatic-impulse approximation and the KZ argument holds (cf. Fig.~\ref{fig7}).

In the limit $r\rightarrow\infty$, the KZ prediction \eqref{KZprediction} leads to $Q_\Of\propto T^4$. However, this scaling can only be attained for very long protocols, for $\Of T>r\rightarrow\infty$. Furthermore, the SNR carries  a very small prefactor $\frac{1}{r^4}$.  We can connect this to the results for the fully adiabatic ramp which we introduced in Sec.~\ref{ss:AR}, where $Q_\Of\propto T^4$, with a prefactor $\varphi^4\ll 1$. The critical point is not reached in the adiabatic ramp, but asymptotically approached for very long $T$. Therefore, we see that in the limit $r\gg 1$ and $\omega T\gg 1$, the finite-time ramp and the fully adiabatic ones become similar. By contrast, for $r\gg 1$ and $1\ll\omega T\ll r$, the finite-time ramp has the same behavior as the sudden quench, as shown in the lower panel of Fig.~\ref{fig7}. Hence, depending on the tuning of $r$ and $T$, the non-linear ramp provides an interpolation between the simple linear ramp, the fully adiabatic evolution, and the sudden quench.

\begin{figure}
    \centering
    \includegraphics[width=\linewidth]{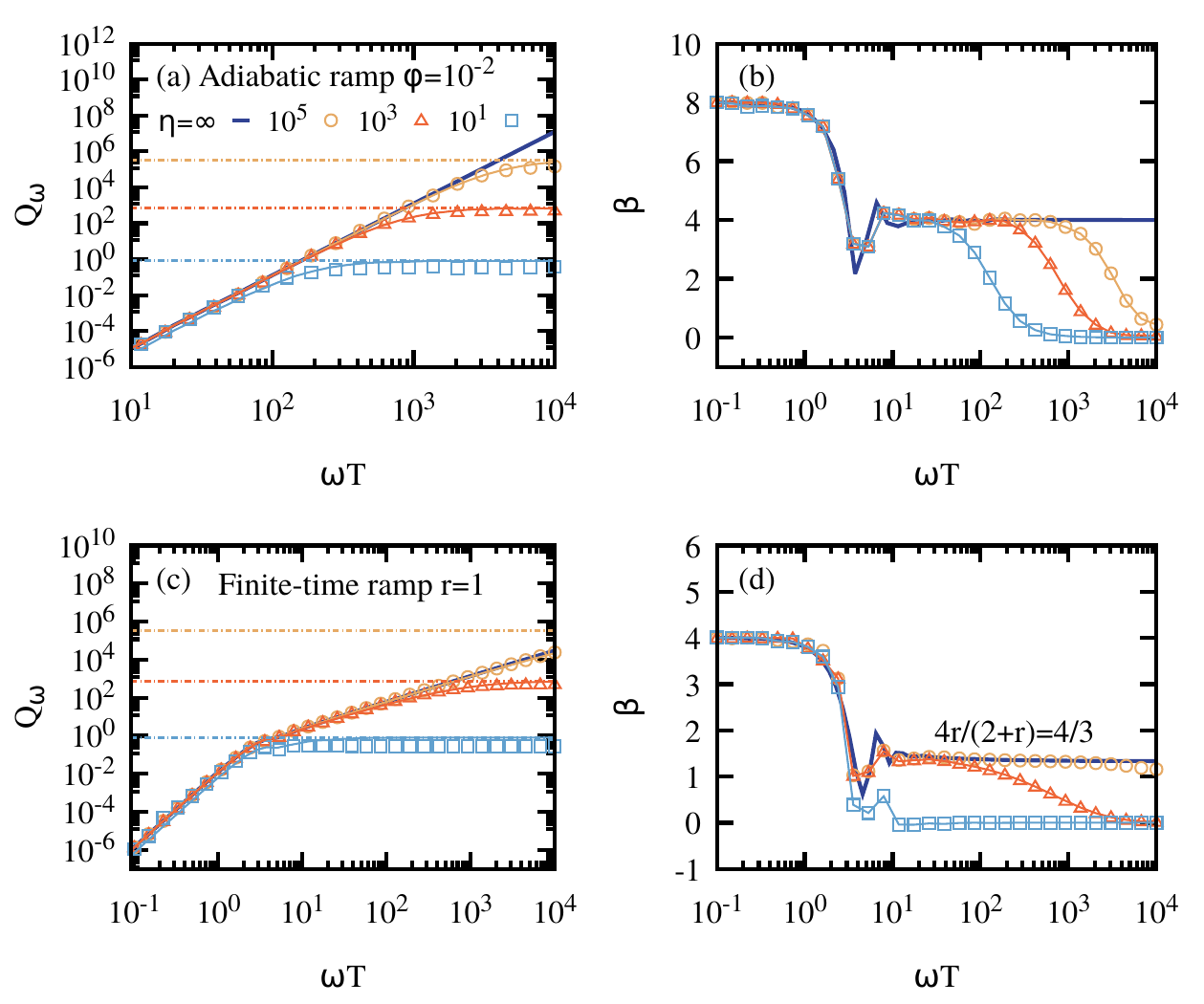}
    \caption{\small{Panels (a) and (b): SNR $Q_\omega$ and its fitted exponent $\beta$, respectively, for an adiabatic ramp with $\varphi=10^{-2}$ and different finite values of $\eta$, together with the case  $\eta\rightarrow \infty$ (solid blue line). The evolution for finite $\eta$ follows the thermodynamic limit prediction for small $\Of T$, then saturates at a value $Q_\Of\sim \eta^{4/3}$. Horizontal dashed lines show the SNR of the ground state at $g_c$ for different $\eta$, which also corresponds to the saturation of the SNR  obtained by performing the adiabatic ramp for $\eta\rightarrow\infty$, but only up to $g^*\sim\sqrt{1-\eta^{-2/3}}$ (solid color lines)  (see main text for details).  Panels (c) and (d) show the same but for a finite-time ramp with $r=1$. In the long-time limit, the SNR saturates to the same value (note that the curve for $\eta=10^5$ does not show the saturation because the display time is too short.)}}
    \label{fig8}
\end{figure}

\subsection{Finite-size effects}

Finally, we study the adiabatic \eqref{ramp_adiab} and finite-time ramps \eqref{rampprofile} for finite-size system. The results are displayed in Fig.~\ref{fig8}. For the adiabatic ramp at short times $\omega T\ll \eta^{1/3}$, we observe the same behavior as in the adiabatic limit, with a scaling going from $T^8$ to $T^4$. This can understood as follows: for finite $\eta$, the gap saturates around its minimum value in the critical region, of width $\Gamma\sim\eta^{-2/3}$. Now, if we apply the evolution \eqref{ramp_adiab} for a total time $\omega T\ll \eta^{1/3}$, we will obtain at the end of the evolution $\epsilon(T)\sim\left(\frac{\tau_Q}{T}\right)^2\gg\eta^{-2/3}=\Gamma$. In other words, at the end of the evolution, we will have $1-g^2\gg\Gamma$. Therefore, we remain safely out of the critical zone; the finite-size effects play a negligible role, and we recover the $\eta\rightarrow\infty$ results. To the contrary, if $\omega T\gg \eta^{1/3}$, the adiabatic ramp brings us within the critical zone. In this region, the QFI saturates at a value $I_\omega\sim\frac{1}{\Of^2\Gamma^2}=\frac{1}{\Of^2}\eta^{4/3}$. This result can be retrieved using \eqref{QFI_adiab} and applying Heuristic \ref{heur1}. This is depicted in the upper panels of Fig.~\ref{fig8}. For $\omega T\gg \eta^{1/3}$, the SNR saturates at a value $\propto \eta^{4/3}$, and becomes independent of $T$. Note that this equivalent to perform the adiabatic ramp to get only up to $g^*=\sqrt{1-\Gamma}\sim\sqrt{1-\eta^{-2/3}}$. We stress that the same results hold for $Q_\lambda$. 

The results for the finite-time ramp are very similar. Let us consider again the freeze-out value \eqref{freezout_value}. For $\frac{\Of T}{r}\ll\eta^{(2+r)/3r}$, it follows $1-g_f\gg\Gamma$, i.e. the freeze-out occurs outside of the critical zone, and thus  the thermodynamic limit results are recovered. For $\frac{\Of T}{r}\gg\eta^{(2+r)/3r}$, the SNR saturates at $\eta^{4/3}$. Those are the features we observe in the lower panels of Fig.~\ref{fig8}, where we plotted the evolution for $r=1$. First, we obtain the short-time scaling $T^4$, then the KZ scaling $T^{4r/(2+r)}=T^{4/3}$, and finally the saturation with a $T^0$ scaling. For larger $r$, there exists an intermediate region for $1\ll\Of T\ll r$, where the SNR scales like $T^6$.

To sum up, finite-size effects for a sudden quench (cf. Fig.~\ref{QFIquenchfinite}) and in adiabatic and finite-time ramps (cf. Fig.\ref{fig8}) have a similar impact. For short durations, the finite-size effect are negligible, and we recover the thermodynamic limit scalings. For long time, the finite-size effects are dominant. The transition between these two regimes is governed by the minimum gap value, $\Of \eta^{1/3}$. These various regimes are summarized in Fig.\ref{fig_sketchmainresults}.\\

To conclude this section, let us discuss how the performances of the different strategies compare against each other. From Fig.~\ref{QFIquenchthermo_scaling} and Fig.~\ref{fig7}, on the one hand, and Fig.~\ref{QFIquenchfinite} and Fig.~\ref{fig8}, on the other hand, we can see that a sudden quench at $g=1$ is always the optimal strategy, in the sense that it always gives the highest QFI for a given duration $T$~\footnote{Note however that the quench achieves a higher precision by increasing the number of excitations in the system. Hence, it may not necessarily be the optimal strategy in a context where both $T$ and $N$ are fixed.}. This is particularly visible when $\eta$ becomes large, in which case the precision achievable with the ramp quickly saturates when $T$ increases, while the performances of the quench keep improving. Fully exploiting these improved performances, however, requires a somewhat more complex measurement strategy, as we will see in the next section.

\begin{figure}
    \centering
    \includegraphics[width=\linewidth]{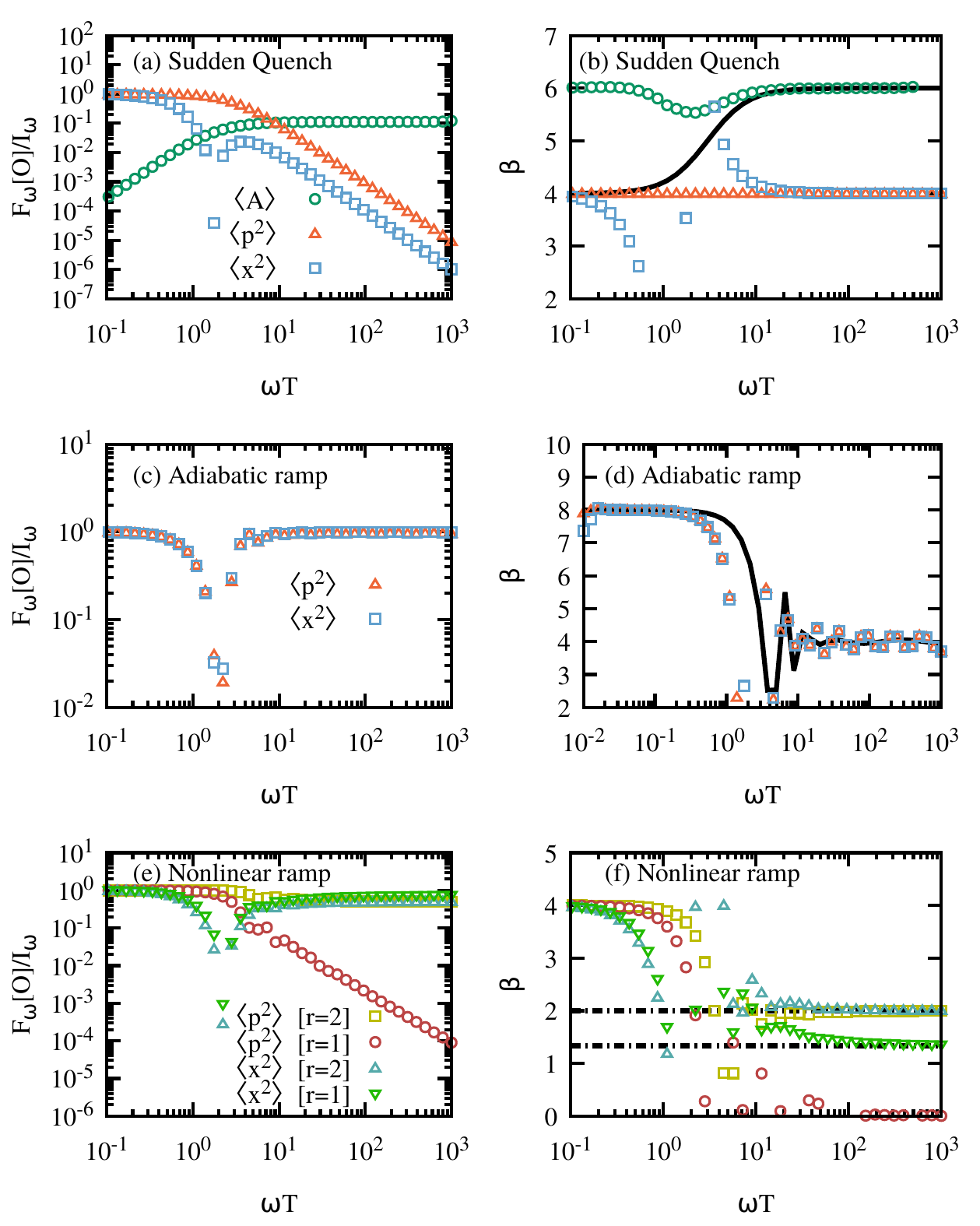}
    \caption{\small{Left: Ratio between the Fisher information $F_\omega[\Oop]$ for an observable $\Oop$ and the QFI $I_\omega$. Right: Corresponding scaling exponent $\beta$ such that $F_\omega[\Oop]\sim (\omega T)^\beta$ (the solid black line represents the scaling of the QFI). Top row, panels (a) and (b), correspond to a sudden quench protocol where $\Oop=\hat{x}^2$, $\hat{p}^2$ and $\hat{A}$ as introduced in Eq.~\eqref{eq:Oopt}. Note that the observable $\hat{A}$ yields a $T^6$ scaling as the QFI, while other quadradures provide $T^4$. The results for an adiabatic ramp with $\varphi=10^{-1}$ are plotted in the middle row, panels (c) and (d), where both quadratures are sufficient to saturate the QFI and thus its $T^4$ scaling (solid black line for the QFI in (d)). The results for non-linear ramps with $r=1$ (full symbols) and $r=2$ (open symbols) are shown in panels (e) and (f), which indicate that it is possible to saturate the QFI and thus reproduce the corresponding Kibble-Zurek scaling (horizontal dashed lines in (f)). If we set $r$ to be very large (not shown here), we observe first a $T^4$ scaling which corresponds to the sudden quench regime, and then a KZ scaling for $\frac{r}{\omega}\ll T\ll \frac{1}{\kappa}$, followed by a saturation.}}
    \label{fig9}
\end{figure}

\section{Saturability of the QFI}
In this section, we will comment on how the QFI limits the actual precision that could be reached in an experimental implementation of our protocol. We will focus on the thermodynamic limit. Let us first recall how our protocol would be exploited to measure the frequency $\omega$. Starting from $\lambda=0$, we increase the coupling following a quench or a ramp of duration $T$, producing some state $\ket{\psi_f}$. We then measure some observable $\hat{O}$, and use the measurement results to infer the unknown value $\omega$. The error $\delta \omega$ is then limited by the Fisher information (FI), given by
$$\delta\omega\geq \frac{1}{\sqrt{F_\omega(\Oop)}},\hspace{10 pt} F_\omega(\Oop)=\frac{\left( \frac{d\langle\hat{O}\rangle}{d\omega}\right)^2}{\text{Var}(\hat{O})},$$
where both the average and the variance of $\Oop$ are taken over the state $\ket{\psi_f}$. The choice of a different observable $\Oop$ may render the estimation more or less precise. The QFI then gives us a lower bound on the error $\delta \omega$, when the choice of $\Oop$ is optimized. This bound can generally be saturated (see \cite{paris_quantum_2011} for the formal conditions), in which case we have $\text{Max}_{\Oop} F_\omega(\Oop)=I_\omega$, or equivalently,  in terms of the squared signal-to-noise ratio, o.e. $\omega^2F_\omega(\Oop)$, which gives $Q_\omega$ when $\Oop$ is optimized.
The optimal observable may however be extremely complicated as a combination of higher-order moments. However, since the states in our case are Gaussian or close to Gaussian, we can expect that measuring the second-order correlations will be sufficient to reach the QFI. In this section, we will show that this is indeed the case. For the ramp, the QFI scaling can be reached by measuring the fluctuations of one quadrature through homodyne detection. For the quench, a more complex quadratic operator is needed, which requires measurements of the fluctuations along both quadratures.

\subsection{Finite-time ramp}

Let us consider first the adiabatic ramp, in the thermodynamic limit. The system is prepared in its ground state, with a coupling value $g$ very close to one. The natural observables in a bosonic system are the quadratures, which can be accessed by homodyne measurement. However, the quadratures always have zero average value in our case. Therefore, measuring $\Oop=\xop$ or $\Oop=\pop$ will always lead to $\langle\Oop\rangle=0$, and hence zero precision. Instead, we need to look at the \textit{fluctuations} of the quadratures, by measuring $\xop^2$ and $\pop^2$. In practice, this could be made by performing an homodyne detection of one quadrature, and integrating the noise of the homodyne current. A natural choice is to measure the squeezed quadrature, by setting $\Oop=\pop^2$, this gives $\langle\Oop\rangle=\sqrt{1-g^2}$ (see Eq.~ \eqref{Eq:quadflu_GS}). 
We then get $\omega \frac{d\langle\Oop\rangle}{ d\omega}=(\omega\frac{d g}{d\omega}) \frac{d\sqrt{1-g^2}}{dg}\sim \frac{1}{\omega\sqrt{1-g^2}}$. \footnote{The quantity $(\omega \frac{d g}{d\omega})$ depends on the underlying physical model, but is always dimensionless and of order $1$}. The variance is derived using Wick's theorem, we get $\text{Var}(\Oop)=3\langle \pop^2\rangle^2\sim 1-g^2$. Putting everything together, we get $F_\omega(\pop^2)\sim \left(\frac{1}{\omega(1-g^2)}\right)^2$. The precision has then exactly the same scaling in $1-g^2$ than the QFI obtained in Eq.~\eqref{QFI_adiab}, eventually, restoring the dependency of $1-g^2=\epsilon(T)$ on $T$, as per \eqref{rampprofile}, we find the same $(\omega T)^4$ scaling as the QFI. This argument indicates that measuring the noise in one quadrature allows to saturate the QFI. Our exact numerical simulations confirm this insight, as shown in Fig.~\ref{fig9}. Here, the high precision has two origins: On the one hand, the derivative $\frac{d\langle\Oop\rangle}{d\omega}$ scales like $(1-g^2)^{-1/2}$ and diverges as the critical point is approached, thus the observable becomes very sensitive. On the other hand, the noise $\text{Var}(\Oop)\sim 1-g^2$ is suppressed. However, $\Oop=\pop^2$ is not the only observable which saturates the QFI. Indeed, let us see what happens if we measure the \textit{anti-squeezed} quadrature instead, and set $\Oop=\xop^2$. Then we find $\langle\Oop\rangle=(1-g^2)^{-1/2}$, $\omega\frac{d\langle\Oop\rangle}{d\omega}\sim(1-g^2)^{-3/2}$, and $\text{Var}(\Oop)\sim (1-g^2)^{-1}$. We then obtain $F_\omega(\xop^2)\sim \left(\frac{1}{\omega(1-g^2)}\right)^2$. Hence, the measurement along the anti-squeezed quadrature gives the \textit{same} scaling of precision as the measurement along the squeezed one, and also saturates the QFI. This is because, although the variance is much larger (and actually divergent) in this case, the derivative $\frac{d\langle\Oop\rangle}{d\omega}$ is correspondingly higher. Instead of homodyne detection, we may also measure the number of photons $\adag\aop=\frac{\xop^2+\pop^2-1}{2}$. The photon number is essentially dominated by the anti-squeezed quadrature, and give the same FI. This is actually a more general property, coming from the critical scaling behavior. If an observable $\Oop$ scales like some power $(1-g^2)^\alpha$ near the critical point, the derivative $\omega\frac{d\langle\Oop\rangle}{d\omega}\sim\frac{d\langle\Oop\rangle}{dg}$ will scale like $(1-g^2)^{\alpha-1}$, and if the dynamics is (at least approximately) Gaussian, we will also have $\text{Var}(\Oop)\sim \langle\Oop\rangle^2\sim (1-g^2)^{2\alpha}$. Combining these quantities,  one finds a signal-to-noise ratio which scales \textit{always} like $(1-g^2)^{-2}$, independently of the value of $\alpha$. 

If we now consider the finite-time ramp, the above arguments remain essentially unchanged. As it can be seen in Fig.\ref{fig9}, homodyne measurement along both $\xop$ and $\pop$ quadratures gives a FI which increases with the ramp time $T$, and saturates the QFI in almost any case. We checked other quadratures and obtained every time the same scaling, with only a change in the prefactor. We found a single exception to that rule: For a linear ramp $r=1$, measuring the squeezed quadrature $\pop^2$ gives a precision which saturates to a finite value, and fail to reach the QFI. By contrast, measuring the anti-squeezed $\xop^2$ always allows to saturate the QFI. Hence, we have a situation in which measuring the \textit{noisy} quadrature is not just a good strategy, but actually a much better one than measuring the quadrature where noise is suppressed. Note that this is a very isolated case; if a measurement is performed along a slightly tilted direction, or if the ramp is slightly non-linear, the QFI scaling will be restored very fast. 
These results indicate, on the one end, that homodyne detection is almost always optimal when the ramp is used. On the other end, and counter-intuitively, they show that it is \textit{not} necessary, or even counter-productive, to measure observables whose fluctuations are suppressed. Several works have considered using critical point as a tool to produce states with reduced quantum noise~\cite{frerot_quantum_2018}, which can then be used in standard phase-shift sensing protocols. Here, instead, the relevant parameters are encoded in the noise signature itself. Our protocol \textit{amplifies} the noise, but with an amplification coefficient which depends critically on the parameter to be estimated.

\subsection{Sudden quench}

To understand the FI achievable in the sudden quench case, we need to describe the dynamics of the quantity to be measured. In App. \ref{App:FIsuddenquench} we provide the analytical expression for the time-evolution of $\xop^2$, $\pop^2$ and $\adag\aop$, and the associated FI. The key findings are the following: Measuring the photon number, or the noise along any quadrature, always yields a precision scaling with the quench duration like $(\omega T)^4$. We recall that the QFI scales like $(\omega T)^6$, and theferore, homodyne and photon-number measurement are always suboptimal. However, we also identified a family of observables which allows saturating the QFI. An example of observable saturating the QFI can be expressed as
\begin{align}\label{eq:Oopt}
\hat{A}=-\frac{\xop^2}{(\omega T)^2}+\pop^2.
\end{align}
This observable can be accessed via Gaussian operations and homodyne measurements, but it clearly requires a non-trivial measurement setup.
These analytical results are confirmed by our numerical simulations. For example, in Fig.\ref{fig9} we show the comparison between one- and multi-quadratures measurement strategies, and how the latter can saturate the QFI.

Let us conclude this section with two comments. First, although the sudden quench gives the best QFI, fully reaching this precision comes at the cost of a more complex measurement procedure. If, experimentally, only single-quadrature measurement or photon counting are available, the precision will fall short of the QFI. However, the sudden quench will still give a quartic scaling, while the finite-time ramp will give a KZ scaling $T^{\frac{2r}{r+2}}\leq T^4$. Therefore, even when the measurements are constrained, the quench is \textit{still} the optimal strategy. The advantage will only be less important than in the absence of measurement constraints.

Second, we have focused the discussion on the measurement of $\omega$. The results are unchanged if we want to measure $\lambda$ instead. Once again, this is because a change in either $\omega$ or $\lambda$ have the same effect, namely a renormalization of the effective coupling $g$ (see Appendix \ref{App:FIsuddenquench} for more details).

\resub{\section{Decoherence effects}
\label{sec:decoh}}

\begin{figure}
    \centering
    \includegraphics[width=\linewidth]{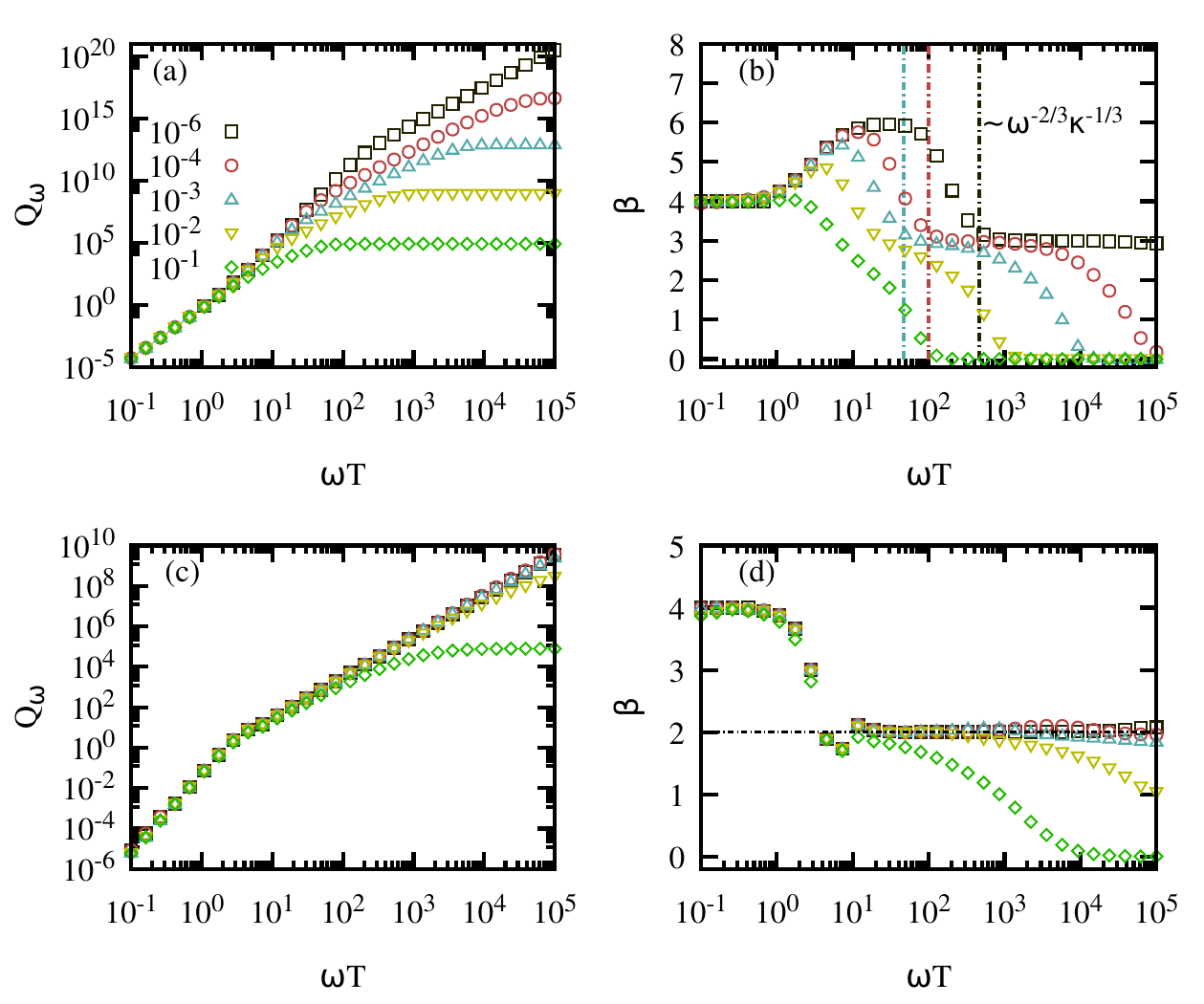}
    \caption{\small{Impact of decoherence on the QFI for a sudden quench (top row) and a non-linear ramp with $r=2$ (bottom row). Panels (a) and (c) show the behavior of the SNR $Q_\omega$ for increasing values of $\kappa/\omega$, from $10^{-6}$ to $10^{-1}$, while panels (b) and (d) show the resulting fitted-exponent $\beta$ such that $Q_\omega\sim (\omega T)^\beta$. Vertical dashed lines in (b) indicate the transition from coherent to a noisy regime for $\kappa/\omega=10^{-3}$, $10^{-4}$ and $10^{-6}$, that takes place at a time $T_0\sim \omega^{-2/3}\kappa^{-1/3}$ (the lines correspond to $(\kappa/100)^{-1/3}$), while the horizontal dashed line in (d) shows the predicted Kibble-Zurek scaling for $r=2$.}}
    \label{fig10}
\end{figure}

Finally, let us look at the effect of decoherence in our model to assess the robustness of the reported QFI scalings~\cite{huelga_improvement_1997,Alipour:14,Beau:17}. We will focus on the effect of photon loss in the thermodynamic limit. The density matrix $\rop$ then evolves according to the following Lindblad equation,
\begin{equation}
    \frac{d\rop}{dt}=-i[\Hop_0,\rop]+\kappa(2\aop\rop\adag -\adag\aop\rop-\rop\adag\aop).
    \label{Lindblad_quench}
\end{equation}
We consider both sudden quench and finite ramp to $g=1$, in the regime $\kappa\leq\omega$. The behavior of the QFI is displayed in Fig.~\ref{fig10}. Let us first look at the non-linear ramp. If the ramp duration $T$ is short enough, the system does not have time to relax to the steady-state, and the QFI still obeys first the non-universal $T^4$ scaling and then the KZ scaling. For very long $T$, however, the system decays to the steady-state before the ramp is completed. In this latter case, we find that the QFI saturates and stops increasing with $T$. The smaller $\kappa$, the longer the KZ scaling holds, and the larger the final QFI. A fit for various $\kappa$ reveals that the steady-state QFI scales like $\frac{\omega^4}{\kappa^4}$. In addition, it is worth mentioning that the impact of decoherence in this context may lead to an anti-Kibble-Zurek scaling~\cite{Dutta:16,Gao:17,Puebla:20,Weinberg:20,Ai:21}. Although not evident from the results shown in Fig.~\ref{fig10}, the corrections to the noiseless QFI may follow a universal anti-Kibble-Zurek scaling in the weak decay regime $\kappa/\omega\ll 1$ as discussed in Ref.~\cite{Puebla:16}.

For the sudden quench, the situation is more intricate. We see the appearance of a new time-scale $T_0\sim \omega^{-2/3}\kappa^{-1/3}$. The evolution shows four regimes: For $T\lesssim\frac{1}{\omega}$, the QFI follows the non-universal $T^4$ scaling. For $\frac{1}{\omega}\lesssim T\lesssim T_0$, the QFI follows a $T^6$ scaling, as it does in the absence of noise. For $T_0\lesssim T\lesssim \frac{1}{\kappa}$, the QFI shows a new regime, with a scaling law $T^3$. Finally, for $T\gtrsim \frac{10}{\kappa}$, the steady-state is reached, and the QFI saturates again at a value that scales as $\frac{\omega^4}{\kappa^4}$. As we show in Appendix \ref{App:dissipation}, this time-scale $T_0$ can be understood by looking at the noise structure of the state. During the evolution, the system is in a squeezed thermal state, 
\begin{equation}
    \Sop(z)e^{-\log(\frac{\upsilon+1}{\upsilon-1})\adag\aop}\Sop(z)^\dag,
    \label{thermalsqueezedstate}
\end{equation}
where $\Sop(z)={\rm exp}\left[\frac{1}{2}\left(|z|e^{-i\theta}\aop^2-|z|e^{i\theta}a^{\dagger 2}\right)\right]$ is the squeezing operator, and $\upsilon$ is the so-called symplectic eigenvalue~\cite{safranek_gaussian_2016}, which describes the purity of the state (or, equivalently, its effective temperature).
 For $\upsilon=1$, we have a pure state with zero temperature, and for $\upsilon\rightarrow\infty$, we have a completely mixed state with infinite effective temperature. During the quench, both the squeezing $\lvert z\rvert$ and the thermal parameter $\upsilon$ increase. However, for $T\lesssim T_0$, the thermal noise is still very small, and the state is effectively a pure squeezed state, with the same properties as in the noiseless case. For $T\gtrsim T_0$, the thermal noise becomes important, and the system enters the new scaling regime. As a last remark, let us also emphasize that the total photon number (including both squeezing and thermal noise) always grows quadratically in time, both for $T\lesssim T_0$ and $T\gtrsim T_0$. For $T\lesssim T_0$, the QFI scales like $T^6$, which can be obtained using our Heisenberg-like bound $I_\omega\propto \left(\int dt (2N(t)+1)\right)^2$. By contrast, for $T\gtrsim T_0$, the QFI scales  like $T^3$, which could be obtained by a bound of the form $I_\omega\propto \left(\int dt (2N(t)+1)\right)$. Hence, this time scale $T_0$ represents also a transition from a Heisenberg-like to a shot-noise metrological behavior.


\section{Conclusion and outlook}
\label{sec:Conclusion}
We have presented a thorough analysis of the quantum Fisher information achievable in the context of critical quantum metrology with fully-connected models.
In particular, we considered three families of critical quantum metrology protocols: sudden quenches, adiabatic sweeps and finite-time ramps. Sudden quenches and adiabatic sweeps are limiting cases which make use of dynamical and static properties of critical systems, respectively. Alternatively, in finite-time ramps the adiabaticity of the protocol can be continuously tuned, providing a formal connection between the first two cases. We have shown the existence of different regimes, characterized by different scalings of the QFI with the protocol duration. Most of these scaling regimes are universal, in the sense that they are independent of both the specific model and the parameter being evaluated. In particular, our protocols can achieve $T^6$, $T^4$ or Kibble-Zurek $T^{4r/(2+r)}$ scaling, which go beyond the paradigmatic quadratic time scaling. This is because our protocols belong to the category of active interferometric protocols, in which the number of probes is allowed to change in time.
We have also provided a general upper bound for the quantum Fisher information achievable using time-dependent quadratic Hamiltonians. We showed that such bound provides an accurate estimate of the scaling of the achievable precision in most regimes, as a function of the time-dependent average number of photons generated during the implementation of the protocol. In general, this bound proved to be a valuable resource to extract the most important features of each scaling regime. We showed that the optimal precision predicted by the QFI could be reached using standard homodyne or photon-number measurements. Finally, we studied how the losses affect our protocols, and showed that non-trivial scaling regimes survive as long as the protocol duration is smaller than the dissipation rate.

The interest in this analysis is of both practical and theoretical nature. Let us briefly discuss these aspects separately. 
From a practical point of view, our study can guide the implementation of critical quantum sensors with atomic and solid-state quantum optical devices. For example, fully-connected models include finite-component critical systems, which can present critical properties without the complexity of many-body quantum systems. The identification of different scaling regimes, and the analysis of the precision achievable with finite-time ramps, make it possible to identify the optimal working point given the characteristics of the considered quantum technology. In the manuscript we focus mostly on the quantum Rabi model as a case study, but the introduced method can be directly applied to any instance of fully-connected models, such as the LMG or Dicke model. 

From a purely theoretical point of view, our results stress out the link between critical quantum metrology and the emergence of  universal scaling laws such as the Kibble-Zurek mechanism. Besides its significance for sensing applications, the QFI is itself an interesting physical quantity, as it gives a measure of the system response to external perturbations. Our analysis provides then a characterization of the static and dynamical susceptibility of quantum critical systems in proximity of the critical point.

Finally, our study suggests possible future directions to improve critical quantum metrology protocols from a fundamental perspective. In general, the Heisenberg scaling is defined under a specific set of assumptions and it is expressed as a function of fundamental resources, such as time or number of probes. Our general upper bound for the QFI shows how the scaling in time could be improved by generalizing the class of considered protocols. Furthermore, we have shown that the sudden quench strategy performs generally better than the finite-time ramp. Both features indicate that a dynamical modulation of the Hamiltonian could be a promising way to improve the performances of such protocols (see for example the recent work~\cite{Yang:22b}). 
An example could be the introduction of a periodic time-dependence in the Hamiltonian term that encodes the parameter to be estimated. However, we have also shown that, to fully exploit this dynamical behavior, it is necessary to go beyond single-quadrature measurement. It is therefore of importance to study which measurement strategies would allow to exploit these features while remaining experimentally accessible. In addition, another interesting avenue within critical quantum metrology would be determining the role of the range of the interactions within a quantum many-body system~\cite{Yang:22} in this dynamical scenario.

Note added: Ref. \cite{gietka_exponentially_2021} appeared simultaneously in arXiv, where a different metrological protocol in fully-connected systems was considered. The system is brought across the critical point, a complementary approach to the present study, where the system remains in the normal phase. This other protocol leads to a transient exponential scaling of the QFI.

\begin{acknowledgements}
 This work was supported by the Austrian Academy of Sciences (\"OAW) and by the Austrian Science Fund (FWF) through Grant No. P32299 (PHONED). R. P. acknowledges support from the European Union's Horizon 2020 FET-Open project SuperQuLAN (899354). O. A. acknowledges support from the UK EPSRC EP/S02994X/1 and Newcastle University (Newcastle University Academic Track fellowship). 
\end{acknowledgements}

\appendix

\section{Scale-invariance and critical exponents}
\label{App_criexp}

Let us consider the following parameter transformation:

\begin{equation}\label{Eq:Symmetry-transform_one}
\left\lbrace
\begin{split}
\pop \to &\pop'=\alpha\pop, \\
\xop \to &\xop'=\frac{1}{\alpha}\xop, \\
\eta \to &\eta'=\frac{1}{\alpha^6}\eta \hspace{10pt}(\text{therefore }\Gamma \to \Gamma'=\alpha^4\Gamma)\\
\omega \to &\omega'=\frac{1}{\alpha^2} \omega\\
1-g^2 \to &(1-g'^2)=\alpha^4(1-g^2) 
\end{split}
\right.
\end{equation}

This corresponds simply to a Bogoliubov transformation on the quadrature, accompanied by a rescaling of the parameters. \\

Now, let us assume we start from some large but finite value of $\eta$, with some value of $g$; then we apply the rescaling with $\alpha<1$, which makes the system evolve towards larger and larger $\eta$, while maintaining $g$ in the interval $[0,1]$.  Under this transformation, the quadratic term of the Hamiltonian \eqref{potquartic} remain exactly invariant. The quartic term can be decomposed into two contributions; the factor $\omega\frac{1}{4\eta}\xop^4$, which is scale-invariant, and the term $f(g)$, which is not. Let us assume here we derive the effective model starting from the QR model, i.e. $f(g)=g^4$. Then the full Hamiltonian transforms as
\begin{align}
\Hop'=\Hop+\omega\frac{(g'^4-g^4)}{4\eta}\xop^4=\Hop+\omega(g'^2+g^2)\frac{(1-g^2)(1-\alpha^4)}{4\eta}\xop^4.
\end{align}
Both terms $g'^2+g^2$ and $1-\alpha^2$ are of order $1$ at most. Hence,  we can neglect them and the correction is bouded by $\omega\frac{(g'^4-g^4)}{4\eta}\xop^4\sim\omega\frac{(1-g^2)}{\eta}\xop^4$. Now, this correction is of the same form as the original quartic term, but with an additional $(1-g^2)$ term. This is because the rescaling is a dilatation transformation on $g$; the amount by which $g$ changes is proportional to its distance to the critical point. If originally we have $g=1$, the scaling transformation just gives $(1-g'^2)=\alpha^4(1-g^2)=0\rightarrow g'=1$; in other words, the critical point $g=1$ is a fixed point of the renormalization flow. Therefore, for $g=1$, the correction to the Hamiltonian must exactly vanish, which is indeed the case. For $g\neq1$, the coupling constant will change; the longer the initial distance $1-g^2$ to the critical point, the bigger the change, and the larger the correction.

Now, let us compare the correction with the quadratic term of the Hamiltonian; the two are of comparable strength when $\frac{\langle\xop^4\rangle}{\eta}\sim\langle\xop^2\rangle$, with an average taken over the ground state of the unperturbed Hamiltonian $\Hop$. We can see, however, that this is never the case, since the ratio $\frac{\langle\xop^4\rangle}{\langle\xop^2\rangle}\sim\langle\xop^2\rangle$ is $\eta^{1/3}$ at most. We can also express it so: either we have $1-g^2\gg\Gamma$ originally, and in this case the term $\frac{\xop^4}{\eta}$ is small; or we have $1-g^2\leq\Gamma\ll 1$, and in this case it is the term $1-g^2$ in the correction which is small.

The same reasoning can be made if we start from another model than the QR model. The quartic term is $f(g)\frac{\xop^4}{\eta}$. $\frac{\xop^4}{\eta}$ is exactly invariant under the rescaling. Then either we start from $1-g^2\leq\Gamma$, in which case $g$ is almost left unchanged by the flow, and therefore $f(g)$ is also invariant; or we start from $1-g^2\gg\Gamma$, in which case $f(g)$ does changes, but then the term $\frac{\xop^4}{\eta}$ is so small that the correction is negligible. 

Therefore, no matter the value of $g$ from which we start, as long as $\eta$ is large, the Hamiltonian remains approximately invariant under the scaling transformation, with a correction which is always small. At the critical point $g\rightarrow1$, the system becomes exactly scale-invariant. 
Therefore, we find here the usual scaling features of a critical system, despite the absence of spatial structure. 

In standard quantum many-body systems one defines scaling relations in terms of correlation length and system size. The correlation length in the thermodynamic limit scales as the coupling according to the critical  exponent $\nu$, e.g. $\xi\sim(1-g^2)^{-\nu}$. A correlation length, of course, cannot be defined directly for a fully-connected system~\cite{Botet:82,Botet:83}. Yet, the frequency ratio $\eta$ can be interpreted as an effective system size, even in the zero-dimensional QR model. As for a candidate of correlation length, we need a quantity that scales as the system size under the scaling transformation. $\langle\Delta x\rangle^6$, with $\Delta x$ the standard deviation of $\xop$, satisfies this property. Therefore, we can now define the notations $L=\eta$, $\xi=\langle\Delta x\rangle^6$. In the critical region, the quadrature variance equals the system "size", or more precisely, it reaches the maximum value allowed by the quartic term. We have the critical exponents for $\xi$ and the gap $\Delta$, $\xi\sim(1-g^2)^{-\nu}$ in the thermodynamic limit, $\xi\sim L$ at the critical point and $\Delta\sim (1-g^2)^{z\nu}$. In the main text, we have shown that the critical exponents are $z\nu=1/2$ and $\nu=3/2$. 

We can now state the scaling argument. Let us take the gap as an example. In the thermodynamic limit, the gap scales with the coupling as $\Delta\sim \omega(1-g^2)^{1/2}$. If we now consider a finite $\eta$ and arbitrary $g$, we should have $\Delta=\omega(1-g^2)^{1/2}f(\omega,\eta,g)$, with $f$ an \textit{a priori} arbitrary function. However, we can constrain this function by noting that $\Delta$ must be homogeneous to $\omega$, and invariant under the scaling transformation (since the Hamiltonian itself is scale-invariant). Since $\omega(1-g^2)^{1/2}$ has already the good dimension and is scale-invariant, this means that $f$ must be scale-invariant \textit{and} dimensionless. Therefore, it can only depend on combinations of $\omega$, $g$ and $\eta$ that satisfy this property. The only combination that works results in $L/\xi=\eta(1-g^2)^\nu$, or equivalently, $\frac{1-g^2}{\Gamma}=\eta^{1/\nu}(1-g^2)$. Therefore, we deduce that  $\Delta\sim\omega(1-g^2)^{1/2}f(\frac{1-g^2}{\Gamma})$. The further requirement that $\Delta$ depends only $g$ in the thermodynamic limit and on $\eta$ near the critical point brings the condition $f(x)\rightarrow1$ for $x\gg1$, and $f(x)\rightarrow x^{-1/2}$ for $x\ll1$. Therefore, near the critical point, this argument predicts that we will have $\Delta\sim \omega \Gamma^{-1/2}=\omega \eta^{-1/3}$. We can also do the same for the quadrature variance, $\langle\xop^2\rangle\sim(1-g^2)^{-1/2}$.  In the general case, we would have $\langle\xop^2\rangle\sim(1-g^2)^{-1/2}f'(\omega,\eta,g)$. This time, the quantity $\langle\xop^2\rangle$ is \textit{not} scale-invariant, since the operator $\xop$ changes explicitly during the rescaling; therefore, the variance will evolve like $\langle\xop^2\rangle\rightarrow\alpha^{-2}\langle\xop^2\rangle$. However, the quantity $(1-g^2)^{-1/2}$ has already the same scaling behavior; therefore, we find again that the function $f'$ must be scale-invariant and dimensionless. To summarize, the scaling argument can be seen as a way to extend the usual dimensional analysis to dimensionless quantities; we must find parameter combinations which have not only zero physical dimension, but also a zero \textit{scaling dimension}. 

\section{QFI for squeezed states}
\label{App:QFI_squeezed}

Here we recall the derivation of the QFI for squeezed states \eqref{QFI_squeezed}. Taking a vacuum squeezed state $\ket{\psi}$, it is  convenient to introduce
\begin{align}
\ket{\psi}=\frac{1}{\sqrt{\cosh|z|}}{\rm exp}\left[-b a^{\dagger,2} \right]\ket{0},
\end{align}
with $b=\frac{\tanh(|z|)}{2}e^{i\theta}$. Then,
\begin{align*}
\ket{\partial_x\psi}&=-(\partial_x\lvert b\rvert)\frac{\sinh(2|z|)}{2}\ket{\psi}-(\partial_x b) \aop^{\dag 2}\ket{\psi}.
\end{align*}
To compute the QFI, we make use of the following relation:
$$\aop^2\aop^{\dag,2}=\frac{1}{4}[\xop^4+\pop^4+2:\xop^2\pop^2:+4(\xop^2+\pop^2)+2],$$
where $::$ means we are taking the symmetric combination $:\hat{a}\hat{b}:=\frac{\hat{a}\hat{b}+\hat{b}\hat{a}}{2}$. For the squeezed vacuum state, we have $\langle\xop^2\rangle=\frac{e^{2\lvert z\rvert}}{2}$ and $\langle\pop^2\rangle=\frac{e^{-2\lvert z\rvert}}{2}$. Then Wick's theorem gives us $\langle\xop^4\rangle=\langle\xop^2\rangle^2$,  $\langle\pop^4\rangle=\langle\pop^2\rangle^2$,  $\langle:\xop^2\pop^2:\rangle=\langle\xop^2\rangle\langle\pop^2\rangle+2\langle:\xop\pop:\rangle^2=\langle\xop^2\rangle\langle\pop^2\rangle$, and we find
\begin{align*}
  \langle\psi|\aop^2\aop^{\dag 2}|\psi\rangle&=\frac{1}{16}[3\cosh(4\lvert z\rvert)+8\cosh(2|z|)+5]\\
  &=(3\cosh^2(|z|)-1)\cosh^2(|z|).
\end{align*}
We can now compute the products involving the derivative of the state, 
\begin{widetext}
\begin{align*}
 \langle\psi|\partial_x\psi\rangle&=i(\partial_x\theta)\sinh(|z|)\cosh(|z|)|b|=i(\partial_x\theta)\frac{\sinh^2(|z|)}{2},\\
\langle\partial_x\psi|\partial_x\psi\rangle&=|\partial_x b|^2(3\cosh^2(|z|-1)\cosh^2(|z|)-(\partial_x|b|)^2\cosh^2(|z|)\sinh^2(|z|).
\end{align*}
\end{widetext}
Finally, the QFI reads as
\begin{align}
\nonumber
I_x&=4\left[\langle\partial_x \psi|\partial_x\psi\rangle-(\langle \psi|\partial_x\psi\rangle)^2\right]\\ \nonumber
&=8(\partial_x |b|^2+|b|^2(\partial_x\theta)^2)\cosh^4(|z|)\\ \nonumber
	&=2\left(\left(\frac{\partial|z|}{\partial x}\right)^2+\cosh^2(|z|)\sinh^2(|z|)\left(\frac{\partial\theta}{\partial x}\right)^2\right)\\
	&=\frac{8|\partial_x b|^2}{(1-4|b|^2)^2}.
	\label{QFI_corrected}
\end{align}

\section{QFI bounds}
\label{App:bounds}
In this section, we make comments on the existing bounds of the QFI, we present the derivation of Eq.~\eqref{geneboundfinal}, and we also show how it can be extended to displaced states.

\subsection{Comments on previous bounds}
The Hamiltonian $\Hop_x$ generates a unitary evolution $\Uop$. The state evolves towards $\ket{\psi_t}=\Uop\ket{\psi_0}$.

The QFI can be rewritten as
\begin{align}
I_x(t)=4\Big(\bra{\psi_0}F_x(t)^2\ket{\psi_0}-\bra{\psi_0}F_x(t)\ket{\psi_0}^2\Big),
\end{align}
where we have defined
\begin{align}
F_x(t)=i\Uop^\dag\partial_x\Big(\Uop\Big).
\end{align}
Now, it is straigthforward to show that the derivative of $F_x$ can be expressed as: $\partial_t F_x(t)=\Uop^\dag(\partial_xH_x(t))\Uop$. Therefore, 
\begin{equation}
F_x(T)=\int_0^T \Uop^\dag(\partial_xH_x(t))\Uop dt.
\end{equation}
We want to compute the variance of $F_x$ on the initial state $\ket{\psi_0}$. For this, we can use the convexity of the standard deviation , i.e.  $\sqrt{{\rm Var}(A+B)}\leq \sqrt{{\rm Var}(A)}+\sqrt{{\rm Var}(B)}$. From this, we can deduce
\begin{align}\nonumber
I_x(T)\leq &4\left(\int_0^T dt \sqrt{{\rm Var}_{\psi_0}\left(\hat{U}^\dag_x(0\rightarrow t) (\partial_xH_x(t)) \hat{U}_x(0\rightarrow t)\right)}\right)^2\\
&=4\left(\int_0^T dt \sqrt{{\rm Var}_{\psi_t}\left(\partial_xH_x(t) \right)}\right)^2
\label{geneboundvariance}
\end{align}
In the first line, the variance is taken over the initial state $\ket{\psi_0}$; in the second, it is taken over the time-evolved state $\ket{\psi_t}$. This equation is the most general bound we can set on the precision for a Hamiltonian evolution with a pure initial state. Again, we emphasize that this expression is valid for any Hamiltonian $\Hop_x$, even time-dependent. Note that the term under the integral is time-dependent in two different ways. First, the derivative of the Hamiltonian, $\partial_x\Hop_x(t)$, can be intrinsically time-dependent. Second, the variance is taken over the state $\ket{\psi_t}$, which evolves in time.

This expression provides a unified view of all the bounds discussed in the main text. Let us assume that we have $\Hop_x(t)=x\Aop(t)$, with $\Aop(t)=\partial_x\Hop_x(t)$ independent of $x$, and $\Hop_x(t)$ is bounded. Then we can define the (time-dependent) extremal eigenvalues $\lambda_M(t)$ and $\lambda_m(t)$ of $\Aop(t)$. The variance can always be bounded by the difference between these eigenvalues: ${\rm Var}_{\chi}(\Aop(t))\leq\frac{1}{4}\lvert\lambda_M(t)-\lambda_m(t)\rvert$, for \textit{any} state $\ket{\chi}$, and all $t$. Plugging this into \eqref{geneboundvariance}, we retrieve the bound \eqref{IntTlimit}
\begin{equation}
    I_x(T)\leq \left[\int_0^T\rvert\lambda_M(t)-\lambda_m(t)\lvert\right]^2.
\end{equation}
This bound can be saturated when the state is in a coherent superposition of the two extremal values: $\ket{\chi(t)}=\frac{\ket{\lambda_M(t)}+\ket{\lambda_m(t)}}{\sqrt{2}}$ \cite{giovannetti_quantum_2006,pang_optimal_2017}. This condition can be challenging to implement in practice, especially if the Hamiltonian is time-dependent; however, it was shown that this could be done for small systems, using quantum control \cite{pang_optimal_2017}.

Let us now assume that all the conditions 1)-4) discussed in the main text are satisfied. Then the extremal eigenvalues become time-independent; and because $\Hop_x$ is local, they can scale at most with the number of components. This means we have $\lvert\lambda_M(t)-\lambda_m(t)\rvert=\alpha N$, with $\alpha$ a time-independent, non-universal constant which depends on the precise expression of $\Hop_x$. This finally gives us back the Heisenberg limit
\begin{align}
\label{IntHlimit}
    I_x(T)&\leq \left[\int_0^T \alpha N dt\right]^2\\\nonumber
    &\leq \alpha^2 N^2 T^2.
\end{align}
Conversely, if $\Hop_x$ acts on $k$ particles at the same time, then the eigenvalues can scale like $N^k$, and we retrieve the bounds \eqref{bound_Boixo}. Finally, if we let $\Hop_x$ be time-dependent, but still bounded, we fall back to the discussion of Pang and Jordan \cite{pang_optimal_2017}.

\subsection{Derivation of our bound}
We will now show derive our bound \eqref{geneboundfinal}. The proof contains essentially two parts: the first part is to use \eqref{geneboundvariance}, thus we reduce the estimation of the QFI to evaluating the variance of $\partial_xH_x$ at each time. The second step is to notice, using Wick's theorem, that the variance of a quadratic Hamiltonian on a Gaussian state is essentially equal to $N^2=\langle\Nop\rangle^2$. Combining the two, we find that the QFI should be bounded by a quantity like $\left(\int_0^T N(t)\right)^2$.
More precisely, we assume that both $\Hop_x$ and its derivative $\partial_x \Hop_x$ are purely quadratic, with no linear part,
\begin{equation}
\partial_x H_x=(\xop \pop) M_x (\xop \pop)^T.
\end{equation}
We also assume that, in the initial state, we have $\langle\xop\rangle=\langle\pop\rangle=0$, which is the case for a vacuum state. We will first assume that $\Hop_x$ is time-independent, and relax this assumption at the end. Starting with \eqref{geneboundvariance}, we need to compute the variance of $\hat{O}(t)=\Uop^\dag\partial_x H_x\Uop$ over the \textit{initial} state. We can rewrite: $$\hat{O}(t)=[\xop(t) \pop(t)] M_x \begin{bmatrix}\xop(t)\\ \pop(t) \end{bmatrix}$$
Here $M_x$ is a two-by-two hermitian matrix, and $\xop(t)=\Uop^\dag \xop\Uop$ is the time-evolved quadrature in Heisenberg picture. Because the Hamiltonian is purely quadratic, the quadratures stay centered around zero at all times: $\langle \xop(t)\rangle=\langle \pop(t) \rangle=0$ for all  $t$. Now, at any given $t$, we can define two quadratures $(\xop_m(t), \pop_m(t))^T=R(\xop(t),\pop(t))^T$, which are obtained from the original quadratures by a (time-dependent) rotation $R(t)$, and which satisfy: $\langle :\xop_m \pop_m:\rangle=0$ (where $::$ means we are taking the symmetric combination $:\hat{a}\hat{b}:=\frac{\hat{a}\hat{b}+\hat{b}\hat{a}}{2}$). $\xop_m$ and $\pop_m$ simply correspond to the directions of maximal and minimum squeezing, respectively. Note that these two quadratures can always be defined, even when the state is mixed. Then we can absorb the rotation in the matrix $M_x$, and we obtain
$$\hat{O}(t)=[\xop(t) \pop(t)] M_x \begin{bmatrix}\xop(t)\\ \pop(t) \end{bmatrix}=[\xop_m(t) \pop_m(t)] R(t) M_x R(t)^T \begin{bmatrix}\xop_m(t)\\ \pop_m(t), \end{bmatrix}$$ with the rotated matrix
$$R(t)M_xR(t)^T=\begin{bmatrix}A_1(t)&b(t)+ic(t)\\b(t)-ic(t)&A_2(t). \end{bmatrix}$$
In other words, the time-evolution has two different components. It rotates the direction of squeezing, and changes its amount. The second operation is now encoded in the quadrature $\xop_m$, while the first is contained in the matrix $R(t)M_xR(t)^T$. Now, the coefficients $A_i$, $b$ and $c$ can have a very complicated time-dependence in general. However, since $R$ is just a rotation, the eigenstates of $M_x$, which we call $\phi$ and $\chi$, remain the same at all time. The goal is now to find a bound on ${\rm Var}(\hat{O})$ which depends only on this time-independent factors. Let us expand $\hat{O}(t)$
 \begin{equation}
 \hat{O}(t)=A_1 \xop_m^2 + A_2 \pop_m^2+ 2b(:\xop_m \pop_m:)-c.
 \label{devO1}
 \end{equation}
 The coefficient $c$ is a constant with no effect on the variance, and can be dropped. We can develop
\begin{widetext}
\begin{align} \label{devO2}
\hat{O}^2&=A_1^2 \xop_m^4+A_2^2\pop_m^4  + A_1A_2(\xop_m^2\pop_m^2+\pop_m^2\xop_m^2)+ b^2(\xop_m\pop_m+\pop_m\xop_m)^2+2b(2A_1 :\xop_m^3\pop_m:+2A_2 :\xop_m\pop_m^3:)\\ \nonumber
&=A_1^2 \xop_m^4+A_2^2\pop_m^4  + (2A_1A_2+4b^2):\xop_m^2\pop_m^2:+ (b^2-A_1A_2)+2b(2A_1 :\xop_m^3\pop_m:+2A_2 :\xop_m\pop_m^3:),
\end{align}
\end{widetext}
where we have dropped the explicit time-dependence to lighten notation. In the second line, we have used the handy relation: $\xop_m^2\pop_m^2+\pop_m^2\xop_m^2=2:\xop_m^2\pop_m^2:-1$. Now, when we take the average over the initial state, we may use Wick's theorem $\langle \xop_m(t)^4\rangle=3\langle \xop_m(t)^2\rangle^2$, $\langle :\xop_m(t)^2\pop_m(t)^2:\rangle=\langle \xop_m(t)^2\rangle \langle \pop_m(t)^2\rangle+2\langle :\xop_m(t) \pop_m(t):\rangle^2=\langle \xop_m(t)^2\rangle \langle \pop_m(t)^2\rangle$, and $\langle :\xop_m(t)^3\pop_m(t):\rangle=\langle :\xop_m(t)\pop_m(t)^3:\rangle=0$, at all times. Then after straightforward manipulation, we find
\begin{align}
    \langle \hat{O}^2(t) \rangle-\langle \hat{O}(t) \rangle^2=&2(A_1^2 \langle \xop_m^2\rangle^2+A_2^2 \langle \pop_m^2\rangle^2)\nonumber\\&+4\langle \xop_m^2\rangle \langle \pop_m^2\rangle b^2- A_1A_2 +b^2,
    \label{VarO}
\end{align}
where all the averages are taken over the \textit{initial} state. Now, we want to replace the time-dependent components $A_i$, $b$ and $c$ by the time-independent $\chi$, $\phi$. We have the following relations:
\begin{align*}
 \chi\phi=A_1 A_2-(b^2+c^2)\leq A_1 A_2-b^2\\
\chi+\phi=A_1+A_2\\
(\phi-\chi)^2=(A_1-A_2)^2+4(b^2+c^2)\geq (A_1-A_2)^2\\
\phi^2+\chi^2\geq A_1^2+A_2^2.
\end{align*}
which gives
\begin{widetext}
\begin{align*}
    \langle \hat{O}^2(t) \rangle-\langle \hat{O}(t) \rangle^2&=2(A_1^2 \langle \xop_m^2\rangle^2+A_2^2 \langle \pop_m^2\rangle^2)+4\langle \xop_m^2\rangle \langle \pop_m^2\rangle b^2- A_1A_2 +b^2\\
    &\leq 2(A_1^2 \langle \xop_m^2\rangle^2+A_2^2 \langle \pop_m^2\rangle^2)+4\langle \xop_m^2\rangle \langle \pop_m^2\rangle (b^2+c^2)- A_1A_2 +(b^2+c^2)\\
    &=2\Big(A_1\langle \xop_m^2\rangle+A_2\langle \pop_m^2\rangle\Big)^2+(1+4\langle \xop_m^2\rangle \langle \pop_m^2\rangle)(b^2+c^2-A_1A_2)\\
    &=2\Big(A_1\langle \xop_m^2\rangle+A_2\langle \pop_m^2\rangle\Big)^2-\chi\phi(1+4\langle \xop_m^2\rangle \langle \pop_m^2\rangle)
\end{align*}
\end{widetext}
This bound is saturated whenever $c=0$; that is, when $\partial_x H_x$ has no constant term. Now, we can write in general
\begin{align*}
    (ac+bd)^2-(ad+bc)^2\leq (ac+bd)^2\leq (ac+bd)^2+(ad+bc)^2,
    \end{align*}
    so that
    $$(a^2-b^2)(c^2-d^2)\leq(ac+bd)^2\leq \frac{(a-b)^2(c-d)^2+(a+b)^2(c+d)^2}{2},$$
for all $a,b,c,d\in \mathbb{R}$.  
We can apply this to the first term in the previous equation, and we get
\begin{widetext}
\begin{align*}
     \langle \hat{O}^2(t) \rangle-\langle \hat{O}(t) \rangle^2&\leq (A_1-A_2)^2\Big(\langle \xop_m^2\rangle-\langle \pop_m^2\rangle\Big)^2+(A_1+A_2)^2\Big(\langle \xop_m^2\rangle+\langle \pop_m^2\rangle)^2-\chi\phi(1+4\langle \xop_m^2\rangle \langle \pop_m^2\rangle\Big)\\
     &= (A_1-A_2)^2\Big(\langle \xop_m^2\rangle-\langle \pop_m^2\rangle\Big)^2+(\phi+\chi)^2\Big(\langle \xop_m^2\rangle+\langle \pop_m^2\rangle\Big)^2-\chi\phi(1+4\langle \xop_m^2\rangle \langle \pop_m^2\rangle\Big)\\
&\leq (\chi-\phi)^2\Big(\langle \xop_m^2\rangle-\langle \pop_m^2\rangle\Big)^2+(\phi+\chi)^2\Big(\langle \xop_m^2\rangle+\langle \pop_m^2\rangle\Big)^2-\chi\phi(1+4\langle \xop_m^2\rangle \langle \pop_m^2\rangle\Big).
\end{align*}
\end{widetext}
Now, we have successfully eliminated all the time-dependent coefficients $A_i$, $b$ and $c$, and we are left only with the time-independent eigenvalues $\phi$ and $\chi$. There is still, however, some time-dependence left in the expectation values of the quadratures. We will now find how this bound can be amended to a more elegant form, involving only the photon number at a given time. The expression above can be massaged as
\begin{widetext}
\begin{align*}
     \langle O^2(t) \rangle-\langle O(t) \rangle^2&\leq (\chi^2+\phi^2)\Big[(\langle \xop_m^2\rangle-\langle \pop_m^2\rangle)^2+(\langle \xop_m^2\rangle+\langle \pop_m^2\rangle)^2\Big]+\chi\phi\Big(2(\langle \xop_m^2\rangle+\langle \pop_m^2\rangle)^2-2(\langle \xop_m^2\rangle-\langle \pop_m^2\rangle)^2-1-4\langle \xop_m^2\rangle \langle \pop_m^2\rangle\Big)\\
     &=2(\chi^2+\phi^2)[\langle \xop_m^2\rangle^2+\langle \pop_m^2\rangle^2]+\chi\phi[4\langle \xop_m^2\rangle\langle \pop_m^2\rangle-1].\\
\end{align*}
\end{widetext}
The term $4\langle \xop_m^2\rangle\langle \pop_m^2\rangle-1$ is always positive, due to Heisenberg's inequality. Now, we can distinguish two cases: if $\chi\phi\geq0$, we use the triangle inequality $\phi\chi\leq \frac{\chi^2+\phi^2}{2}$, and we find a new bound
     \begin{align*}
     \langle \hat{O}^2(t) \rangle-\langle \hat{O}(t) \rangle^2&\leq 2(\phi^2+\chi^2)\Big[\left(\langle \xop_m^2\rangle+\langle \pop_m^2\rangle\right)^2-\langle \xop_m^2\rangle\langle \pop_m^2\rangle-\frac{1}{4}\Big]\\
     &\leq 2(\phi^2+\chi^2)\Big[\left(\langle \xop_m^2\rangle+\langle \pop_m^2\rangle\right)^2-\frac{1}{2}\Big].
     \end{align*}
where, in the last line, we have used Heisenberg's inequality $\langle \xop_m^2\rangle\langle \pop_m^2\rangle\geq\frac{1}{4}$. Recall that $\langle\xop\rangle=\langle\pop\rangle=0$).  If $\chi\phi\leq0$ instead, we have directly
\begin{align*}
     \langle \hat{O}^2(t) \rangle-\langle \hat{O}(t) \rangle^2&\leq2(\chi^2+\phi^2)[\langle x_m^2\rangle^2+\langle p_m^2\rangle^2]\\
     &= 2(\phi^2+\chi^2)\Big[\left(\langle x_m^2\rangle+\langle p_m^2\rangle\right)^2-2\langle x_m^2\rangle\langle p_m^2\rangle\Big]\\
     &\leq 2(\phi^2+\chi^2)\Big[\left(\langle x_m^2\rangle+\langle p_m^2\rangle\right)^2-\frac{1}{2}\Big].
     \end{align*}
Hence, in both cases, we find the same bound. It is now  easy to rewrite everything in terms of the photon number; at each time, we have $N(t)=\langle\frac{\xop^2_m(t)+\pop^2_m(t)-1}{2}\rangle$, and therefore
\begin{align*}
     \langle \hat{O}^2(t) \rangle-\langle \hat{O}(t) \rangle^2 &\leq 2(\phi^2+\chi^2)\Big[ \left(2N(t)+1\right)^2-\frac{1}{2}\Big].
     \end{align*}
Finally, combining this with the expression \eqref{geneboundvariance}, we find
\begin{align}
I_x(t)&\leq 8 (\chi^2+\phi^2)\left(\int_0^T dt \sqrt{( 2N(t)+1)^2-\frac{1}{2}}\right)^2\\ \nonumber
&\leq 8 (\chi^2+\phi^2)\left[\int_0^T dt \Big( 2N(t)+1\Big)\right]^2.
\end{align}
This derivation was performed assuming that $\Hop_x$ was time-independent. Now, if $\Hop_x$ is explicitely time-dependent, we can still define a matrix $M(t)$ and its two eigenvalues $\phi(t)$ and $\chi(t)$. Those are also the eigenvalues of the rotated matrix $R(t)M(t)R(t)^T$, at any given time $t$. The steps of the derivation remain exactly the same, and we find instead
\begin{align}
I_x(T)\leq 8 \left[\int_0^T dt \sqrt{\chi(t)^2+\phi(t)^2} \Big( 2N(t)+1\Big)\right]^2.
\end{align}
Note that although $\chi$ and $\phi$ are now time-dependent, there are still entirely determined by $\Hop_x$. Thus, in this case too, we have successfully separated the QFI into a contribution which depends on the Hamiltonian only, and one which depends only on the average photon number.

\subsection{Extension to non-zero displacements}

We will now show how the bound above can be extended when the state is still Gaussian, but with a non-zero displacement. We now assume that the initial state is an arbitrary pure Gaussian state, and the Hamiltonian now has the form
\begin{equation}
\Hop_x=(\xop \pop) h_x (\xop \pop)^T +u(\xop \pop)^T.
\end{equation}
For now, we will assume that $u$ is \textit{independent} of $x$. Then the derivative $\partial_x\Hop_x$ is still purely quadratic, but the state has now a non-zero displacement which may change in time. We can apply once again apply some time-dependent rotation $(\xop_m,\pop_m)^T=R(t)(\xop,\pop)^T$, so that the rotated quadratures verify the following property: $\xop_m=\Xop+\alpha$ and $\pop_m=\Pop+\beta$, with $\alpha$ and $\beta$ scalars, and we have the following properties: $\langle\Xop^n\rangle=\langle\Pop^n\rangle=0$ for odd $n$, and $\langle:\Xop\Pop:\rangle=0$. In other words, what we have done is apply first a rotation, then a displacement, to obtain quadratures centered around zero and aligned with the squeezing. Then we can again absorb the rotation inside $M_x$, and we obtain again Eq. \eqref{devO1}. If we now develop $\xop_m=\Xop+\alpha$, we find
 \begin{align}
 \nonumber
 \hat{O}(t)&=A_1 \Xop^2 + A_2 \Pop^2+ 2b(:\Xop \Pop:)+2(A_1\alpha+b\beta)\Xop\\&+2(A_2\beta+b\alpha)\Pop
 +(A_1\alpha^2+A_2\beta^2+2b\alpha\beta-c).
 \label{devO1bis}
 \end{align}
 The last term is again a scalar (albeit time-dependent) with no effect on the variance, so it can be safely neglected. Let us write $\mu_1=2(A_1\alpha+b\beta)$ and $\mu_2=2(A_2\beta+b\alpha)$. Then we can develop
\begin{widetext}
\begin{align}
\hat{O}^2&=A_1^2 \Xop^4+A_2^2\Pop^4  + (2A_1A_2+4b^2):\Xop^2\Pop^2:+ (b^2-A_1A_2)+2b(2A_1 :\Xop^3\Pop:+2A_2 :\Xop\Pop^3:)\\\nonumber
&+\mu_1^2\Xop^2+\mu_2^2\Pop^2+2\mu_1\mu_2:\Xop\Pop:+2\mu_1A_1\Xop^3+2\mu_1A_2:\Xop\Pop^2:+4\mu_1b:\Xop^2\Pop:+2\mu_2A_2\Pop^3+2\mu_2A_1:\Xop^2\Pop:+4\mu_2b:\Xop\Pop^2:.
\label{devO2bis}
\end{align}
\end{widetext}
The first line is just the expression \eqref{devO2} where we replaced $\xop_m$ and $\pop_m$ by $\Xop$ and $\Pop$, and the second line is an extra term which disappears when $\alpha=\beta=0$. Now we can apply Wick's theorem and exploit the relations $:\Xop\Pop:=0$ etc. Restoring the expression of $\mu_1$ and $\mu_2$, we finally obtain the expression of the variance
\begin{align}\nonumber
    \langle \hat{O}^2(t) \rangle-\langle \hat{O}(t) \rangle^2=&2(A_1^2 \langle \Xop^2\rangle^2+A_2^2 \langle \Pop^2\rangle^2)+4\langle \Xop^2\rangle \langle \Pop^2\rangle b^2\nonumber\\&- A_1A_2 +b^2
    +4(A_1\alpha+b\beta)^2\langle \Xop^2\rangle\nonumber\\&+4(A_2\beta+b\alpha)^2\langle \Pop^2\rangle.
\end{align}
This expression is just \eqref{VarO}, plus an extra term which couples the variance of the quadratures and the displacement. The first term captures the effect of squeezing, the second the interplay between squeezing and displacement. We now make the following manipulation: We integrate again the displacement in the first term (noting that $\langle\xop_m^2\rangle=\langle\Xop^2\rangle+\alpha^2$, and the same for $\pop_m$), but \textit{not} in the second term. We obtain
\begin{align} \nonumber
    \langle \hat{O}^2(t) \rangle-\langle \hat{O}(t) \rangle^2&=2(A_1^2 \langle \xop_m^2\rangle^2+A_2^2 \langle \pop_m^2\rangle^2)+4\langle \xop_m^2\rangle \langle \pop_m^2\rangle b^2\\  \nonumber
    &- A_1A_2 +b^2+8b\alpha\beta(A_1\langle\Xop^2\rangle+A_2\langle\Pop^2\rangle)\\
    &-(4\alpha^2\beta^2b^2+2A_1^2\alpha^4+2A_2^2\beta^4).
\end{align}

The first term can now be bounded, in exactly the same way as in the previous subsection, by $2(\phi^2+\chi^2)[(2N(t)+1)^2-\frac{1}{2}]$ (noting that $N=\frac{\langle\xop_m^2+\pop_m^2\rangle-1}{2}$. For the second term, we can bound it by applying the Cauchy-Schwarz and triangle inequalities

\begin{align*}
8b\alpha\beta(A_1\langle\Xop^2\rangle+A_2\langle\Pop^2\rangle)&\leq 8\alpha\beta b\sqrt{A_1^2+A_2^2}\sqrt{\langle\Xop^2\rangle^2+\langle\Pop^2\rangle^2}\\
&\leq 8\alpha\beta b\sqrt{A_1^2+A_2^2}(\langle\Xop^2\rangle+\langle\Pop^2\rangle)\\
&\leq2 (b^2+A_1^2+A_2^2)(\langle\Xop^2\rangle+\langle\Pop^2\rangle)(\alpha^2+\beta^2)\\
&\leq \frac{1}{2} (b^2+A_1^2+A_2^2) (\langle\Xop^2\rangle+\langle\Pop^2\rangle+\alpha^2+\beta^2)^2\\
&=\frac{1}{2} (b^2+A_1^2+A_2^2) (\langle\xop_m^2\rangle+\langle\pop_m^2\rangle)^2.
\end{align*}

\begin{widetext}
And finally, noting that $b^2+A_1^2+A_2^2\leq 2(b^2+c^2)+A_1^2+A_2^2=\chi^2+\phi^2$, we can bound the variance by
\begin{align}\nonumber
    \langle \hat{O}^2(t) \rangle-\langle \hat{O}(t)\rangle^2&\leq  2(\phi^2+\chi^2)[(2N(t)+1)^2-\frac{1}{2}]+\frac{1}{2}(\phi^2+\chi^2)[(2N(t)+1)^2]-(4\alpha^2\beta^2b^2+2A_1^2\alpha^4+2A_2^2\beta^4)\\
    &\leq \frac{5}{2}(\phi^2+\chi^2)[(2N(t)+1)^2],
    \label{Bound_dispnoH}
\end{align}
\end{widetext}
which is very similar to our previous bound, but with a different prefactor. 

Finally, let us consider an even more general case, when the linear term in $\Hop_x$, $u$, can now depend on $x$. We define $v=\partial_x u$. We can again apply a quadrature rotation, and we get $\partial_x\Hop_x=(\xop_m \pop_m) M_x (\xop_m \pop_m)^T +vR^{-1}(\xop_m \pop_m)^T$. The rotated vector $vR^{-1}$ can be written as $(d_1,d_2)$; although the expression of these two terms is unknown, we know we have $d_1^2+d_2^2=\lvert v\rvert^2$, since the rotation preserves the norm. This term adds additional linear $\Xop$ and $\Pop$ term. We find that the expressions \eqref{devO1bis} and \eqref{devO2bis} will remain the same, but we now have $\mu_1=2(A_1\alpha+b\beta+d_1)$ and $\mu_2=2(A_2\beta+b\alpha+d_2)$. Integrating the displacement in the first, but not the second term, we get
\begin{widetext}
\begin{align}\nonumber
    \langle \hat{O}^2(t) \rangle-\langle \hat{O}(t) \rangle^2&=2(A_1^2 \langle \xop_m^2\rangle^2+A_2^2 \langle \pop_m^2\rangle^2)+4\langle \xop_m^2\rangle \langle \pop_m^2\rangle b^2- A_1A_2 +b^2\\ \nonumber 
    &+8b\alpha\beta(A_1\langle\Xop^2\rangle+A_2\langle\Pop^2\rangle)+4d_1(A_1\alpha+b\beta)\langle\Xop^2\rangle+4d_2(A_2\beta+b\alpha)\langle\Pop^2\rangle+d_1^2\langle\Xop^2\rangle +d_2^2\langle\Pop^2\rangle\\ &-(4\alpha^2\beta^2b^2+2A_1^2\alpha^4+2A_2^2\beta^4).
    \label{devO3}
\end{align}
The first terms can be bounded as before. For the other terms, we find using Cauchy-Schwarz and triangle inequalities
\begin{align*}
    d_1(A_1\alpha+b\beta)\langle\Xop^2\rangle+d_2(A_2\beta+b\alpha)\langle\Pop^2\rangle&\leq\sqrt{d_1^2+d_2^2}\sqrt{(A_1\alpha+b\beta)^2\langle\Xop^2\rangle^2+(A_2\beta+b\alpha)^2\langle\Pop^2\rangle^2}\\ \nonumber
    &\leq\sqrt{d_1^2+d_2^2}\sqrt{(A_1\alpha+b\beta)^2+(A_2\beta+b\alpha)^2}\sqrt{\langle\Xop^2\rangle^2+\langle\Pop^2\rangle^2}\\ \nonumber
    &\leq\sqrt{d_1^2+d_2^2}\sqrt{(A_1\alpha+b\beta)^2+(A_2\beta+b\alpha)^2}(\langle\Xop^2\rangle+\langle\Pop^2\rangle)\\
    &\leq\sqrt{d_1^2+d_2^2}\sqrt{(\alpha^2+\beta^2)}\sqrt{(A_1^2+2b^2+A_2^2)}(\langle\Xop^2\rangle+\langle\Pop^2\rangle)\\
    &\leq\sqrt{d_1^2+d_2^2}\sqrt{(\alpha^2+\beta^2)}\sqrt{(\chi^2+\phi^2)}(\langle\Xop^2\rangle+\langle\Pop^2\rangle)
\end{align*}
$$d_1^2\langle\Xop^2\rangle +d_2^2\langle\Pop^2\rangle\leq(d_1^2+d_2^2)(\langle\Xop^2\rangle+\langle\Pop^2\rangle)$$
\begin{align*}
8b\alpha\beta(A_1\langle\Xop^2\rangle+A_2\langle\Pop^2\rangle)&\leq2 (b^2+A_1^2+A_2^2)(\langle\Xop^2\rangle+\langle\Pop^2\rangle)(\alpha^2+\beta^2)\\
&\leq 2 (\chi^2+\phi^2)(\langle\Xop^2\rangle+\langle\Pop^2\rangle)(\alpha^2+\beta^2).
\end{align*}
We have now successfully isolated the $d$ factors, so that the bound only depends on $d_1^2+d_2^2=v^2$. Putting everything together, the second term in \eqref{devO3} can now be bounded by
\begin{align*}
    (\langle\Xop^2\rangle+\langle\Pop^2\rangle)\Big[2(\chi^2+\phi^2)(\alpha^2+\beta^2)+4\lvert v\rvert\sqrt{\chi^2+\phi^2}\sqrt{\alpha^2+\beta^2}+\lvert v\rvert^2\Big]&\leq 2(\langle\Xop^2\rangle+\langle\Pop^2\rangle)\Big[\sqrt{(\chi^2+\phi^2)(\alpha^2+\beta^2)}+\lvert v\rvert\Big]^2\\
    &\leq 2(\langle\Xop^2\rangle+\langle\Pop^2\rangle)(\chi^2+\phi^2+\lvert v\rvert^2)(\alpha^2+\beta^2+1)\\
    &\leq\frac{1}{2}(\chi^2+\phi^2+\lvert v\rvert^2)(\langle\Xop^2\rangle+\langle\Pop^2\rangle+\alpha^2+\beta^2+1)^2\\
    &=\frac{1}{2}(\chi^2+\phi^2+\lvert v\rvert^2)(2N+2)^2.
\end{align*}
And finally, we can bound the variance of $\hat{O}$ as
\begin{align}
    \langle \hat{O}^2(t) \rangle-\langle \hat{O}(t)\rangle^2&\leq  2(\phi^2+\chi^2)\left[(2N(t)+1)^2-\frac{1}{2}\right]+2(\phi^2+\chi^2+\lvert v\rvert^2)[(N(t)+1)^2].
    \label{Bound_dispH}
\end{align}
\end{widetext}
The bounds \eqref{Bound_dispnoH} and \eqref{Bound_dispH} are most likely not tight, and do not explicitly give back the previous bound \eqref{geneboundfinal} in the limit $\alpha,\beta,v\rightarrow0$. However, they show that, even in the best possible case, adding a non-zero displacement to the state should not allow to achieve better scaling that what we obtained with vacuum squeezed states. Most importantly, these bounds differ from \eqref{geneboundfinal} only by some prefactors. This indicates that much of the insight we have obtained about scaling regimes should apply also when we have displacement.

\section{Dynamics under $\Hop_{0}$}
\label{App:timeev}
Here we give further details about the evolution of the state under $\Hop_0$ given in Eq.~\eqref{eq:H0} and any protocol $g(t)$, the expression of the squeezing \eqref{bsol_quench} and the ground-state value of $b$ and $\theta$. We consider the thermodynamic limit $\eta\rightarrow\infty$ of $\Hop$, namely, $\Hop_0$, which can be rewritten as 
\begin{align}
    H(t)=\omega a^\dagger a-\frac{g^2(t)\omega}{4}(a+a^\dagger)^2,
    \label{QRM_projected}
\end{align}
with $g^2\leq 1$. The initially prepared state $\ket{\psi(0)}=\ket{0}$, i.e. the ground state at $g=0$ evolves following the time-evolution operator $\hat{U}(t)$ such that $\ket{\psi(t)}=\hat{U}(t)\ket{0}$.  The operator $\hat{U}(t)$ itself evolves according to $\dot{\hat{U}}(t)=-i\Hop(t)\hat{U}(t)$. Up to an irrelevant global phase, this results in
\begin{align}
    \dot{\hat{U}}(t)=-i\omega \left( \left(1-\frac{g^2(t)}{2}\right)a^\dagger a-\frac{g^2(t)}{4}(a^2+a^{\dagger 2})\right)U(t),
\end{align}
To find the equation of motion for $b(t)$, we use Q-space representation, where $U_\alpha(t)=\bra{\alpha}U(t)\ket{\alpha}$,
with $\ket{\alpha}$ a coherent state. Using the substitution $\aop\hat{U}\rightarrow(\alpha+\partial_{\alpha^*})U_{\alpha}$ and $\adag \Uop\rightarrow \alpha^*U_\alpha$, we obtain
\begin{align}\label{eq:Ut}
    \dot{U}_{\alpha}(t)=-i\omega &\left( \left(1-\frac{g^2(t)}{2}\right)\alpha^*(\alpha+\partial_{\alpha^*})\right. \nonumber \\&\left.-\frac{g^2(t)}{4}\left(\left(\alpha+\partial_{\alpha^*}\right)^2+\alpha^{* 2}\right) \right)U_{\alpha}(t),
\end{align}
which suggests a solution of the form of 
\begin{align}
    U_\alpha(t)=e^{k(t)+b(t)\alpha^{*,2}+c(t)\alpha^*\alpha+d(t)\alpha^2}.
\end{align}
The equations of motion for the coefficients $k(t)$, $b(t)$, $c(t)$ and $d(t)$ follow from Eq.~\eqref{eq:Ut}. In particular, $b(t)$ is decoupled from the rest
\begin{align}\label{eq:bt}
    \dot{b}(t)=-i\omega\left(-\frac{g^2(t)}{4}+2\left(1-\frac{g^2(t)}{2}\right)b(t)-g^2(t)b^2(t)\right).
\end{align}
We can now recast the operator $U_\alpha(t)$ in its original form by replacing $\alpha$ and $\alpha^*$ by $\aop$ and $\adag$, and requiring normal ordering,
\begin{align}
    \hat{U}(t)=e^{k(t)}e^{b(t)a^{\dagger,2}}N[e^{c(t)a^\dagger a}]e^{d(t)a^2}
\end{align}
 In this manner, an evolved state under the protocol $g(t)$ is given by
\begin{align}
    \ket{\psi(t)}=U(t)\ket{\psi(0)},
\end{align}
where we have assumed $t=0$ as initial time. Now it is easy to see that if the initial state is the vacuum, $\ket{\psi(0)}=\ket{0}$, then 
\begin{align}
    \ket{\psi(t)}=e^{k(t)}e^{b(t)a^{\dagger 2}}\ket{0}=(1-4\lvert b\rvert^2)^{1/4} e^{b(t)a^{\dagger 2}}\ket{0},
\end{align}
the second equality being obtained by imposing normalization. This is just a squeezed state of the form~\eqref{squeezedstate}, with the squeezing parameter $b(t)$ described by the equation of evolution Eq.~\eqref{eq:bt} with the initial condition $b(0)=0$. 
As discussed in the main text, Eq.~\eqref{eq:bt} admits a exact solution in certain cases, such as in the sudden quench scenario.

Finally we comment that the ground-state properties of $\Hop_0$ can be easily obtained by setting $\dot{b}=0$ in~\eqref{eq:bt}. Indeed, this leads to $b=-1/2 +(1+\sqrt{1-g^2})^{-1}$ so that $\theta=0$, which reproduces the ground-state squeezing of $\Hop_0$~\cite{Hwang:15}.

\section{Connection between the QFI of the physical and effective models}
\label{Appendix:physical-effective}

In Sec. \ref{Protocol}, we have discussed how to express the QFI and how, in the thermodynamic limit, it could be computed through the derivative of the squeezing parameter \eqref{QFI_squeezed}. Here, we discuss a subtle, but important point concerning this derivative. The model \eqref{potquartic} is entirely described by three independent parameters, $g$, $\Of$ and $\eta$. Let us first consider the limit $\eta\rightarrow\infty$, in which only the first two are relevant. Let us consider that we suddenly quench the system from zero coupling to some value $g\sim1$. The state evolves as a squeezed state, with a squeezing parameter described by \eqref{bsol_quench}. We can compute the partial derivatives $\partial b/\partial g$ and $\partial b/\partial \Of$. In the limit $g\rightarrow1$, we can compute explicitly
$$g\frac{\partial b(T)}{\partial_g}=-2+i(\omega T -2i)+\frac{2i(4i-\omega T)}{(\omega T-2i)^2}$$
$$\omega\frac{\partial b(T)}{\partial_\omega}=\frac{-i\omega T}{(\omega T-2i)^2}$$
$$1-4\lvert b\rvert^2=\frac{4}{(\omega T)^2+4}.$$
For long time (more precisely, for $1\ll\omega T\ll 1/\sqrt{1-g^2}$, which corresponds precisely to the region II in Fig.\ref{fig_sketchmainresults}), this reduces to
\begin{align}
    g\frac{\partial b(T)}{\partial_g}\sim i\omega T\\
\omega\frac{\partial b(T)}{\partial_\omega}\sim-\frac{i}{\omega T}\\
1-4\lvert b\rvert^2\sim\frac{4}{(\omega T)^2}.
\end{align}
Now, let us assume that $g$ and $\omega$ are completely independent parameters, which we want to estimate. Let us define the corresponding SNR $Q_g^0$ and $Q_\omega^0$. Then using \eqref{QFI_squeezed}, we have
\begin{align}
    Q_g^0= \frac{8}{1-4\lvert b\rvert^2} \bigg| g\frac{\partial b}{\partial g}\bigg|^2\rightarrow (\omega T)^6\\
    Q_\omega^0=\frac{8}{1-4\lvert b\rvert^2} \bigg| \omega\frac{\partial b}{\partial \omega}\bigg|^2\rightarrow (\omega T)^2
\end{align}
 Hence, for large $T$, $Q_g^0$ will show a critical $T^6$ scaling, while $Q_\omega^0$ has a normal $T^2$ scaling. This is not surprising, since only a change of $g$ changes the position relative to the critical point, and can thus exploit the critical sensitivity of the system. In general, for long $T$, we have $ g\frac{\partial b(T)}{\partial_g}\gg\omega\frac{\partial b(T)}{\partial_\omega}$.

Now, when we consider the physical QR model, the parameters $g$ and $\Of$ are not independent anymore. Therefore, we have to consider the derivatives $\partial_\omega g\partial_g b$ and $\partial_\omega b$, instead. But we have shown that the first contribution will dominate for long $T$. Therefore, for $\Of T\gg1$, we will have
\begin{align}
    Q_\omega&=\frac{8}{1-4\lvert b\rvert^2} \bigg| \omega\frac{\partial b}{\partial \omega}+\omega \partial_\omega g\frac{\partial b}{\partial g}\bigg|^2\\\nonumber
    &\sim \frac{8}{1-4\lvert b\rvert^2} \frac{1}{4} \bigg|\omega \partial_\omega g\frac{\partial b}{\partial g}\bigg|^2= \frac{1}{4} Q_g^0,
\end{align}
where we have used $\partial_\Of g=\frac{-\lambda}{\Of^{3/2}\Oq^{1/2}}=-\frac{1}{2}\frac{g}{\Of}$. Similarly, if we want to evaluate $\lambda$, we must compute $\partial_\lambda g\partial_g b$; using $\partial_\lambda g=\frac{g}{\lambda}$, we find
\begin{equation}
    Q_\lambda=Q_g^0.
\end{equation}
Therefore, we see that for $\Of T\gg 1$, the estimation of $\lambda$ and $\omega$ will give the same SNR, up to a constant factor. This is precisely what we observe in Sec.\ref{sec:Quench}, with a common scaling behavior for $Q_\lambda$ and $Q_\omega$. This shows that, from the perspective of critical metrology, a  parameter change is only pertinent insofar as it induces a change in $g$, and thus moves the system towards or away from the critical point. 

This argument is, however, only valid for $\Of T>1$. For shorter times, $Q_g$ and $Q_{\Of}$ are actually of the same order of magnitude, and different scalings can be achieved for $Q_\lambda$ and $Q_\Of$. The same conclusion can be reached if we consider the adiabatic scenario rather than the sudden quench. The expression of the squeezing parameter is different in this case, but we still find that $\partial_g b$ dominates $\partial_\Of b$.
Importantly, this conclusion can be straightforwardly carried to \textit{any} model which can be effectively described by \eqref{eq:H0}. If we want to estimate any parameter $x$, we will obtain the same universal profile for $Q_x$; the only exception is when the effective coupling $g$ is totally independent of $x$.

\section{Fisher Information for the sudden quench}
\label{App:FIsuddenquench}
To study the dynamics during a sudden quench, it is most convenient to move to phase space. The state is entirely described by its covariance matrix \begin{equation}
    \sigma=\begin{bmatrix}\langle\xop^2\rangle & \langle:\xop\pop:\rangle \\ \langle:\xop\pop:\rangle &\langle\pop^2\rangle.\end{bmatrix}
    \label{covmat_def}
\end{equation}
For a pure squeezed state \eqref{squeezedstate}, $\sigma$ can be decomposed as: $$\sigma=\begin{bmatrix}\cos(\theta)& \sin(\theta)\\ -\sin(\theta) & \cos(\theta)\end{bmatrix} \begin{bmatrix} \frac{e^{2\lvert z\rvert}}{2}& 0\\ 0 & \frac{e^{-2\lvert z\rvert}}{2}\end{bmatrix} \begin{bmatrix}\cos(\theta)& -\sin(\theta)\\ \sin(\theta) & \cos(\theta)\end{bmatrix}$$.

If we quench the system, the evolution is governed by the Hamiltonian \eqref{eq:H0}, and  the covariance matrix follows a Lyapunov equation
\begin{align*}
\partial_t \sigma=B \sigma + \sigma B^T,\\
B=\begin{bmatrix}0 &\omega \\\omega(g^2-1)&0,\end{bmatrix}
\end{align*}
and $\sigma$ can be decomposed as $\sigma(t)=\sum_i c_i(t) M_i$, with eigenmatrices obeying
$B M_i+M_i B^T=\lambda_i M_i$. The eigenmatrices and eigenvalues are:
\begin{align}
M_0&=\begin{bmatrix}0 &1 \\-1&0\end{bmatrix},\\  M_1&=\begin{bmatrix}\frac{1}{\sqrt{1-g^2}} &0 \\0&\sqrt{1-g^2}\end{bmatrix},\\
M_\pm&=\begin{bmatrix}\frac{1}{\sqrt{1-g^2}} &\pm i \\\pm i&-\sqrt{1-g^2}\end{bmatrix},
\end{align}
with $\lambda_0=\lambda_1=0$ and $\lambda_\pm=\pm 2i\omega\sqrt{1-g^2}$. 
And the covariance matrix will be expressed as 
\begin{equation}
\sigma(t)=\begin{bmatrix}
\frac{c_1+c_++c_-}{\sqrt{1-g^2}} & i(c_+-c_-)\\i(c_+-c_-) & \sqrt{1-g^2}(c_1-(c_++c_-)).
\end{bmatrix} 
\label{covmatcoeff}
\end{equation}
Assuming we start from the vacuum, we find
$c_1(t)=\text{cst}=\frac{1}{4}\left(\frac{1}{\sqrt{1-g^2}}+\sqrt{1-g^2}\right)$, and  $c_\pm(t)=-\frac{1}{8}\left(\frac{1}{\sqrt{1-g^2}}-\sqrt{1-g^2}\right)e^{\lambda_\pm t}$. 
Combining these expressions, we find the quadrature fluctuations at the end of the evolution, for $t=T$:
\begin{widetext}
\begin{align}
\nonumber
    \langle\xop^2\rangle&=\frac{1}{4(1-g^2)}\left[1-\cos\left(2\omega\sqrt{1-g^2}T\right)\right]+\frac{1}{4}\left[1+\cos\left(2\omega\sqrt{1-g^2}T\right)\right].\\ \nonumber
    \langle\pop^2\rangle&=\frac{1}{4}\left[1+\cos\left(2\omega\sqrt{1-g^2}T\right)\right]+\frac{1-g^2}{4}\left[1-\cos\left(2\omega\sqrt{1-g^2}T\right)\right].\\
    \langle:\xop\pop:\rangle&=\frac{1}{4}\left[\frac{1}{\sqrt{1-g^2}}-\sqrt{1-g^2}\right]\sin\left(2\omega\sqrt{1-g^2}T\right).
    \label{quadquench_exact}
\end{align}
\end{widetext}
Now, we will consider a quench very close to the critical point, such that $\sqrt{1-g^2}\omega T\ll 1$; we can then expand everything in powers of $1-g^2$, and we find:
\begin{align}
 \label{quadnoise_expanded}
\nonumber
 \langle x^2\rangle&=\frac{1}{2}+\frac{(\omega T)^2}{2}-(1-g^2)\left(\frac{(\omega T)^2}{2}+\frac{(\omega T)^4}{6}\right),\\ 
 \langle p^2\rangle&=\frac{1}{2}-(1-g^2)\left(\frac{(\omega T)^2}{2}\right).\\ \nonumber
 \langle :xp:\rangle &=\frac{\omega T}{2}-(1-g^2)\left(\frac{\omega T}{2}+\frac{(\omega T)^3}{3}\right).
 \end{align}
 
 We now want to evaluate how accurately $\omega$ can be evaluated by performing an homodyne measurement. For this, we need the derivative of the above expressions with respect to $\omega$, which gives
 \begin{align*}
 \omega\frac{d{\langle x^2\rangle}}{d\omega}&=\omega\partial_\omega \langle \xop^2\rangle+\partial_\omega( g^2)\partial_{g^2}\langle \xop^2\rangle,\\
 &=(\omega T)^2+\alpha\left(\frac{(\omega T)^2}{2}+\frac{(\omega T)^4}{6}\right),\\
  \omega\frac{d{\langle p^2\rangle}}{d\omega}&=\alpha \frac{(\omega T)^2}{2},\\
  \omega\frac{d{\langle :x p:\rangle}}{d\omega} &=\frac{\omega T}{2}+\alpha\left(\frac{\omega T}{2}+\frac{(\omega T)^3}{3}\right).
  \end{align*}
 Here we have defined $\alpha=\partial_\omega( g^2)$, although its precise value depends on the specific model, it is always dimensionless and of order $1$. Note that, alternatively, we could also have started from the exact expressions~\eqref{quadquench_exact} and then perform the derivative setting $g$ to $1$. 
 Now, in the regime $1\ll\omega T\ll \frac{1}{\sqrt{1-g^2}}$, the dominant terms for each expressions are
 \begin{align}
     \label{dominant_quadterm}
     \langle\xop^2\rangle\sim \frac{(\omega T)^2}{2},\\ \nonumber
     \langle\pop^2\rangle\sim \frac{1}{2},\\ \nonumber
     \langle:\xop\pop:\rangle\sim \frac{\omega T}{2},\\ \nonumber
     \omega \frac{d\langle \xop^2\rangle}{d\omega}\sim \frac{\alpha}{6}(\omega T)^4,\\ \nonumber
     \omega \frac{d\langle \pop^2\rangle}{d\omega}\sim \frac{\alpha}{2}(\omega T)^2,\\ \nonumber
     \omega \frac{d\langle :\xop\pop:\rangle}{d\omega}\sim \frac{\alpha}{3}(\omega T)^3.
 \end{align}
These expressions have two important features: First, for the derivative, the dominant term comes from the dependency of $g$ on $\omega$. We find again that, near the critical point, the sensitivity with respect to a parameter $x$ will depend above all on how the distance to the critical point changes with $x$. It also means that we will obtain essentially the same results if we want to estimate the coupling $\lambda$ rather than $\omega$.

Second, the ratio between the derivative and the average value is always of order $(\omega T)^2$, for every quantity. Therefore, a choice of any of these quantities as observable will yield a similar FI. For instance, let us assume that we measure $\xop^2$. Since the state is Gaussian, we will have $\text{Var}(\xop^2)\sim 3\langle\xop^2\rangle^2$ as per Wick's theorem. Then the estimation of $\omega$ will yield a squared signal-to-noise ratio $\frac{(\omega d{\langle x^2\rangle/d\omega)^2}}{\text{Var}(\xop^2)}\sim \frac{\alpha^2 (\omega T)^8}{(\omega T)^4}\sim (\omega T)^4$. We can apply the same reasoning to $\pop^2$  instead, and we find the exact same scaling. Because the ratio between the variance and its derivative is fixed for both quadratures, they will both yield the same precision.
 
 Let us now see how we can go beyond this quartic scaling. In general, if we have access to both quadratures, we can define arbitrary combinations of $\xop^2$, $\pop^2$ and $:\xop\pop:$. Let us consider the following combination
 $$\Oop=\frac{\gamma}{(\omega T)^2}\xop^2 + \pop^2,$$ 
 with $\gamma$ some real factor. Using \eqref{dominant_quadterm}, we find that the dominant term of the derivative is given by $$\omega\frac{d\Oop}{d\omega}=\alpha(\omega T)^2 \left(\frac{\gamma}{6}+\frac{1}{2}\right)+O(\omega T).$$
The dominant term of the variance is:
\begin{align*}
\text{Var}(\Oop)&=2\frac{\gamma^2}{(\omega T)^4}\langle\xop^2\rangle^2+2\langle\pop^2\rangle^2+\frac{\gamma}{(\omega T)^2}\left(4\langle:\xop\pop:\rangle^2-1\right)\\
&=\frac{\gamma^2}{2}+\frac{1}{2}+\gamma +O\left(\frac{1}{\omega T}\right).
\end{align*}
For a generic $\gamma$, we find that the derivative scales like $(\omega T)^2$, while the square-root of the variance is of order $1$. Hence, we find again the same ratio $(\omega T)^2$, which gives a FI scaling like $(\omega T)^4$. However, if we set $\gamma=-1$, we can see that the dominant term of the variance will cancel out, but $\textit{not}$ the one in the derivative. Restoring the complete expressions, we find that $\text{Var}(\Oop)=\frac{2}{(\omega T)^2}$ and $\omega\frac{d\Oop}{d\omega}=\frac{\alpha}{3}(\omega T)^2$ in this case, which eventually yields $$\frac{\left(\omega\frac{d\Oop}{d\omega}\right)^2}{\text{Var}(\Oop)}\propto(\omega T)^6$$
Hence, we can achieve a FI scaling like $T^6$, which is precisely the QFI scaling.

More generally, we can define a family of operators, of the form $\Oop=\frac{\gamma}{(\omega T)^2}\xop^2+\frac{\chi}{\omega T}:\xop\pop:+\pop^2$, and tune the parameters $\gamma$ and $\chi$ to cancel the dominant term of the variance, while preserving the derivative. We can thus obtain a family of observables saturating the QFI. Note, however, that the homodyne measurement of a single quadrature does \textit{not} belong to this family. Indeed, a single quadrature measurement gives an operator of the form $\Oop=\xop_\phi^2=(\cos(\phi)\xop+\sin(\phi)\pop)^2$. Although this operator is also a combination of $\xop^2$, $\pop^2$ and $:\xop\pop:^2$, we have access to a single tuning parameter $\phi$. The dominant terms in the derivative and standard deviation are $$\omega\frac{d\Oop}{d\omega}\sim\frac{(\omega T)^2}{2}+\frac{\cos^2(\phi)}{6}(\omega T)^4+\frac{2\cos(\phi)\sin(\phi)}{3}(\omega T)^3,$$ $$\sqrt{\text{Var}(\Oop)}\sim\frac{1}{2}+\frac{\cos^2(\phi)}{2}(\omega T)^2+\cos(\phi)\sin(\phi)(\omega T).$$
Now, one can follow the same reasoning and cancel the dominant term in the variance. This is indeed possible by setting $\cos(\phi)=-\frac{1}{(\omega T)}$ and $\sin(\phi)=\sqrt{1-\frac{1}{(\omega T)^2}}\sim 1$. However, if we do this, the dominant term of the derivative will \textit{also} cancel out. Working with the full expressions, we find that we get $\omega\frac{d\Oop}{d\omega}\sim \frac{1}{\omega T}$ and $\sqrt{\text{Var}(\Oop)}\sim \frac{1}{(\omega T)^3}$ in this case, which eventually will again yield a precision scaling like $T^2$, and a quartic FI. Hence, the measurement of a single quadrature gives us a family of state described by a single parameter $\phi$; however, to reach the QFI, we need to tune two parameters independently, which requires measurement of a more complex Gaussian observable.

In summary, for sudden quench, the derivative and variance of most quantities have generically a ratio of $(\omega T)^2$. However, this connection can be broken by a specifically chosen combination of observables, allowing to reach the QFI scaling of $T^6$. In general, this may require a fine-tuning of the coefficient $\gamma$, which involves being able to measure the protocol duration with great accuracy. A possible way to circumvent this difficulty is to perform an adaptative measurement, performing quenches of increasing duration and improving gradually our estimate of the parameter.

\section{Dynamics under dissipation}
\label{App:dissipation}
Under dissipation, the system will be in a squeezed thermal state. It is still fully described by its covariance matrix $\sigma$. For a thermal squeezed state \eqref{thermalsqueezedstate}, $\sigma$ can be decomposed as
$$\sigma=\begin{bmatrix}\cos(\theta)& \sin(\theta)\\ -\sin(\theta) & \cos(\theta)\end{bmatrix} \begin{bmatrix} \upsilon\frac{ e^{2\lvert z\rvert}}{2}& 0\\ 0 & \upsilon \frac{e^{-2\lvert z\rvert}}{2}\end{bmatrix} \begin{bmatrix}\cos(\theta)& -\sin(\theta)\\ \sin(\theta) & \cos(\theta)\end{bmatrix}.$$
We may still define two directions with minimum and maximum noise, with fluctuations $\upsilon e^{-2\lvert z\rvert}$ and $\upsilon e^{2\lvert z\rvert}$, respectively. The squeezing parameter hence gives the asymmetry between the maximal and minimum noise, while the thermal noise increases the variance of both quadratures. Using the expression above, and the original definition \eqref{covmat_def}, we can also express $\upsilon$, $z$ and $\theta$ in terms of the noise in each quadrature $$\tan(2\theta)=\frac{2\langle:\xop\pop:\rangle}{\langle\pop^2\rangle-\langle\xop^2\rangle},$$
$$\upsilon^2=\text{Det}[\sigma]=4(\langle\xop^2\rangle\langle\pop^2\rangle-\langle:\xop\pop:\rangle^2),$$
$$\upsilon\cosh(2\lvert z\rvert)={\rm Tr}[\sigma]=\langle\xop^2\rangle+\langle\pop^2\rangle.$$
These expressions will be particularly useful to compute the QFI later on.
Before we dive into the time-evolution of $\sigma$, we need to comment on the expression for the QFI. The expression \eqref{QFI_general} is valid only for pure state. For mixed state, the QFI can have a generally much more complex form. However, for a one-mode Gaussian state, it can be expressed in a compact form as~\cite{pinel_quantum_2013}
\begin{align}
 I_\omega=\frac{2\upsilon^2}{1+\upsilon^2}&\left[\frac{1}{2}\sinh^2(2\lvert z\rvert)\left(\frac{d\theta}{d\omega}\right)^2+2\left(\frac{d\lvert z\rvert}{d\omega}\right)^2\right]\nonumber\\ &+\frac{1}{\upsilon^2-1}\left(\frac{d\upsilon}{d\omega}\right)^2,
 \label{QFI_noisy}
 \end{align}
and the SNR is simply $Q_\omega=\omega^2 I_\omega$. The goal will then to express the angle $\theta$, squeezing parameter $\lvert z\rvert$, and noise parameter $\upsilon$, as well as their derivative with respect to $\omega$.

When we perform a quench in the presence of noise, the density matrix evolves under the Lindblad equation~\eqref{Lindblad_quench}. Moving to phase space, this translates into a Lyapunov equation for $\sigma$,  $\partial_t \sigma=B \sigma + \sigma B^T+ D$, with $B=\begin{bmatrix}-\kappa &\omega \\\omega(g^2-1)&-\kappa\end{bmatrix}$ and $D=\kappa\mathbb{I}$. We may again solve this equation by defining the eigenmatrices $B M_i + M_i B^T=\lambda_i M_i$. We find that the eigenmatrices $M_i$ are the same as in the absence of dissipation, while the eigenvalues now read as
$$\lambda_0=\lambda_1=-2\kappa, \hspace{10pt} \lambda_\pm=-2\kappa\pm 2i\omega\sqrt{1-g^2}.$$
Decomposing the covariance matrix as $\sigma=\sum_i c_i M_i$ and the diffusion matrix as $D=\sum_i c^D_i M_i$, 
and assuming we start from the vacuum, we can find the expression for the coefficients $c_i$ upon a sudden quench. At the end of the evolution, we find
  \begin{align*}
 c_+(T)+c_-(T)&=\frac{2c_+(0)}{1+u^2}\left[1+e^{-2\kappa T}(u\sin(2\kappa uT)+u^2\cos(2\kappa uT))\right],\\
 -i(c_+(T)-c_-(T))&=\frac{2c_+(0)}{1+u^2}\left[u+e^{-2\kappa T}(u^2\sin(2\kappa uT)-u\cos(2\kappa uT))\right],\\
 c_+(0)=c_-(0)&=-\frac{1}{8}\left(\frac{1}{\sqrt{1-g^2}}-\sqrt{1-g^2}\right),\\
 c_1(T)&=\frac{1}{4}\left(\frac{1}{\sqrt{1-g^2}}+\sqrt{1-g^2}\right),\\
 c_0&=0,
 \end{align*}
 where we defined $u=\frac{\omega\sqrt{1-g^2}}{\kappa}$. If $u$ is a small parameter, we have
 \begin{align}
 \label{quad_thermodev}
 \nonumber
\langle\xop^2\rangle&=\frac{1}{2}+ \frac{\omega^2}{4\kappa^2}\left(1-C_1\right)+u^2\left(\frac{\omega^2}{4\kappa^2}(C_3-1)+\frac{C_1-1}{4}\right)\\
\langle\pop^2\rangle&=\frac{1}{2} +u^2\left(\frac{C_1-1}{4}\right)\\ \nonumber
\langle:\xop\pop:\rangle&= \frac{\omega }{4\kappa}(1-e^{-2\kappa T})+\frac{u^2}{4}\left(-\frac{\kappa}{\omega}(1-e^{-2\kappa T})+\frac{\omega}{\kappa}(C_2-1)\right)
\end{align}
$C_1=e^{-2\kappa T}(1+2\kappa T)$, $C_2=e^{-2\kappa T}(1+2\kappa T+2\kappa^2T^2)$ $C_3=e^{-2\kappa T}(1+2\kappa T+2\kappa^2T^2+\frac{4}{3}\kappa^3T^3)$.

 \subsubsection{Transient regime}
Let us study what happens when $\kappa T\ll 1$. In this regime, the quench is too short for the system to reach the steady-state. All the quantities will depend on $T$. We can develop the expressions above in terms of $\kappa T$. For instance, we will have $C_1\sim 1-2(\kappa T)^2+\frac{(2\kappa T)^3}{3}+O((\kappa T)^4)$, $C_2\sim 1-\frac{4}{3}(\kappa T)^3+\frac{(2\kappa T)^4}{8}+O((\kappa T)^5)$, and $C_3\sim 1-\frac{2}{3}(\kappa T)^4+\frac{(2\kappa T)^5}{30}+O((\kappa T)^6)$ 
After a tedious but straightforward development of the quadratures, we arrive to
\begin{widetext}
\begin{align}
\nonumber
\langle\xop^2\rangle&=\frac{1}{2} +\frac{(\omega T)^2}{2}+ \frac{2\omega^2\kappa T^3}{3} +(1-g^2)\left(-\frac{(\omega T)^2}{2}-\frac{(\omega T)^4}{6}+\frac{2}{3}\omega^2\kappa T^3+\frac{4}{15}\omega^4\kappa T^5\right),\\
\langle\pop^2\rangle&=\frac{1}{2} +(1-g^2)\left(-\frac{(\omega T)^2}{2}+\frac{2}{3}\omega^2\kappa T^3\right),\\ \nonumber
\langle:\xop\pop:\rangle&= \frac{\omega T}{2}-\frac{\omega\kappa T^2}{2}+\frac{1-g^2}{2}\left(-\omega T-\frac{2}{3}(\omega T)^3+\omega\kappa T+\omega^3\kappa T^4\right).
\end{align}
\end{widetext}
Note that if we set $\kappa=0$, we recover the expressions \eqref{quadnoise_expanded}.

Equipped with the expressions above, we can now extract the relevant parameters $\upsilon$, $\theta$ and $\lvert z\rvert$. Keeping only the largest coefficients, we find
$$\tan(2\theta)=\frac{2\langle:\xop\pop:\rangle}{\langle\pop^2\rangle-\langle\xop^2\rangle}\sim -\frac{2}{\omega T}+(1-g^2)\frac{2\omega T}{3},$$
$$\upsilon^2=4(\langle\xop^2\rangle\langle\pop^2\rangle-\langle:\xop\pop:\rangle^2)\sim 1+\frac{10}{3}\omega^2\kappa T^3-(1-g^2)\left(\frac{7}{10}\omega^4\kappa T^5\right),$$
and 
$${\rm Tr}[\sigma]=\upsilon\cosh(2\lvert z\rvert)=\frac{(\omega T)^2}{2}-(1-g^2)\frac{(\omega T)^4}{6}.$$

Now, we must distinguish two important cases. We can define the time-scale $T_0=\omega^{-2/3}\kappa^{-1/3}$. As long as $T\ll T_0$, we have $\upsilon^2\sim 1$, and the state is almost pure. In this regime, we can work out the dominant terms for all observables, and their derivative. Up to prefactors, the results are  $\cosh(2\lvert z\rvert)\sim\sinh(2\lvert z\rvert)\sim (\omega T)^2$, 
    $\upsilon^2-1\sim \omega^2\kappa T^3$,  $\omega\frac{d\lvert z\rvert}{d\omega}\sim (\omega T)^2$, $\omega\frac{d\theta}{d\omega}\sim \omega T$, 
    $\omega\frac{d\upsilon}{d\omega}\sim\omega^4\kappa T^5$.
The various terms in~\eqref{QFI_noisy} scale therefore as $\sinh^2(2\lvert z\rvert)\left(\frac{d\theta}{d\omega}\right)^2\propto (\omega T)^6$, $\left(\frac{d\lvert z\rvert}{d\omega}\right)^2\propto (\omega T)^4$, and $\frac{1}{\upsilon^2-1}\left(\frac{d\upsilon}{d\omega}\right)^2\propto \omega^6\kappa T^7$. 
Recall that the term $\frac{\upsilon^2}{1+\upsilon^2}$ is always bounded between $1/2$ and $1$ and can be ignored. Since we are in the regime $\frac{1}{\omega}\ll T\ll \frac{1}{\kappa}$, the dominant term is the first one, that is, we recover the $T^6$ scaling we obtained in the absence of decoherence. Hence, the short-time dynamics for the noisy quench is essentially identical to the quench without boson loss, and gives the same QFI scaling. Note that the QFI arises from the interplay between $d\theta/d\omega$ and $\sinh(2\lvert z\rvert)$. During the evolution, the system experiences both phase-shift and amplification, and both effects contribute to the final precision.

In the case $T\gg T_0$, by contrast, $\upsilon$ becomes large, and the thermal noise cannot be neglected anymore. In this case, we obtain modified scalings for the observables $\cosh(2\lvert z\rvert)\sim\sinh(2\lvert z\rvert)\sim\frac{\omega}{\sqrt{\kappa}} T^{1/2}$, $\upsilon^2-1\sim \omega^2\kappa T^3$, $\omega\frac{d\lvert z\rvert}{d\omega}\sim (\omega T)^2$, $\omega\frac{d\theta}{\omega}\sim \omega T$,    $\omega\frac{d\upsilon}{\omega}\sim(\omega T)^2$.
The terms in the QFI scale now as
$\sinh^2(2\lvert z\rvert)\left(\frac{d\theta}{d\omega}\right)^2\propto \frac{\omega^4}{\kappa} T^3=\frac{\omega^2}{\kappa^2}\frac{T^3}{T_0^3}$,
$\left(\frac{d\lvert z\rvert}{d\omega}\right)^2\propto (\omega T)^4$, and
$\frac{1}{\upsilon^2-1}\left(\frac{d\upsilon}{d\omega}\right)^2\propto \frac{\omega^2}{\kappa} T$. 
The dominant term is again the first one (note that the second term gives a larger scaling, but is smaller in absolute value, because $T\ll \frac{1}{\kappa}$). Hence, we see that we now achieve a cubic scaling with $T$.

\subsubsection{Steady-state}
In the regime $\kappa T\gg 1$, the quench is long enough for the system to reach its steady-state. All observables, as well as the QFI, saturate at a constant value. Working out the expressions \eqref{quad_thermodev} leads to
 \begin{align*}
\langle\xop^2\rangle&=\frac{1}{2}+ \frac{\omega^2}{4\kappa^2}-(1-g^2)\left(\frac{\omega^2}{4\kappa^2}+\frac{\omega^4}{4\kappa^4}\right),\\
\langle\pop^2\rangle&=\frac{1}{2} -(1-g^2)\left(\frac{\omega^2}{4\kappa^2}\right),\\
\langle:\xop\pop:\rangle&= \frac{\omega }{4\kappa}-(1-g^2)\left(\frac{\omega}{2\kappa}\right),
\end{align*}
from which we can obtain the dominant terms for the observables and their derivatives, $\cosh(2\lvert z\rvert)\sim\sinh(2\lvert z\rvert)\sim\frac{\omega}{\kappa}$, $\tan(2\theta)\sim \frac{\kappa}{\omega}$, 
    $\upsilon^2-1\sim \frac{\omega^2}{\kappa^2}$, 
    $\omega\frac{d\lvert z\rvert}{d\omega}\sim \frac{\omega^2}{\kappa^2}$, 
      $\omega\frac{d\theta}{d\omega}\sim \frac{\omega}{\kappa}$, 
     $\omega\frac{d\upsilon}{d\omega}\sim\frac{\omega^3}{T^3}$. 
Plugging everything in the QFI, we find that $Q_\omega\sim \frac{\omega^4}{\kappa^4}$.

To summarize, the various regimes for the quench in the presence of boson loss are the following: for $T\ll\frac{1}{\omega}$, the dynamics follow non-universal behavior. For $\frac{1}{\omega}\ll T\ll\frac{1}{\omega^{2/3}\kappa^{1/3}}$, the thermal noise is still negligible; the system behaves as in the absence of dissipation, and the QFI scales like $T^6$. For $\frac{1}{\omega^{2/3}\kappa^{1/3}}\ll T\ll\frac{1}{\kappa}$, the thermal noise becomes important; the system enters a thermal squeezed state, with both squeezing and thermal noise increasing with $T$. In this regime, the QFI achieves a modified scaling $T^3$. Finally, for $T\gg\frac{1}{\kappa}$, the system reaches its steady-state, and the QFI saturates at a value $\frac{\omega^4}{\kappa^4}$. All of these findings are corroborated by the simulations presented in the main text.

\bibliographystyle{apsrev4-1.bst}
\bibliography{QFI_fullyconnected_resub.bib}


\end{document}